\documentclass[
    aps,
    prx,
    letterpaper,
    nobalancelastpage,
    twocolumn,
    superscriptaddress,
    nofootinbib,
    longbibliography
]{revtex4-2}

\usepackage[colorlinks, citecolor=blue]{hyperref}
\usepackage{graphicx}
\usepackage{amsmath}
\usepackage{amssymb}
\usepackage[english]{babel}
\usepackage{color}
\usepackage[version=4]{mhchem}
\usepackage{dsfont}
\usepackage{multirow}
\usepackage{booktabs}
\usepackage{array}
\usepackage{stmaryrd}
\usepackage{physics}
\usepackage{bm}
\usepackage{comment}
\usepackage{braket}

\usepackage{soul}

\usepackage{textcomp}

\renewcommand{\t}[1]{\text{#1}}

\newcommand{\UIUC}{
    Department of Physics,
    The University of Illinois at Urbana-Champaign,
    Urbana, IL 61801, USA
}

\renewcommand{\cite}[1]{\mbox{\citep{#1}}}
\newcommand{\edited}{\textcolor{blue}}

\addto\captionsenglish{}
\addto\captionsenglish{\renewcommand\thefigure{\arabic{figure}}}

\newcommand{\clockpiprobraw}{0.978(5)}

\newcommand{\clocktwopiprobraw}{0.979(5)}

\newcommand{\ramanpiprobraw}{0.987(4)}

\newcommand{\bellfidelityraw}{{0.90(1)}$\pm${0.014(3)}$_\text{bound}$}
\newcommand{\clockpiprob}{0.985(4)}

\newcommand{\clocktwopiprob}{0.996(4)}

\newcommand{\ramanpiprob}{0.996(3)}

\newcommand{\ramantwopiprob}{0.988(8)}

\newcommand{\regpiprob}{0.94(5)}

\newcommand{\regtwopiprob}{0.96(3)}

\newcommand{\depumpprob}{0.04(1)}
\newcommand{\repumpprob}{0.992(6)}
\newcommand{\bellfidelity}{{0.950(9)}$\pm${0.005(3)}$_\text{bound}$}

\begin{document}

\title{Parallelized telecom quantum networking with a ytterbium-171 atom array}
\author{Lintao Li}\email{ltli@illinois.edu}
\affiliation{\UIUC}
\author{Xiye Hu}
\affiliation{\UIUC}
\author{Zhubing Jia}
\affiliation{\UIUC}
\author{William Huie}
\affiliation{\UIUC}
\author{Won Kyu Calvin Sun}
\affiliation{\UIUC}
\author{Aakash}
\affiliation{\UIUC}
\author{Yuhao Dong}
\affiliation{\UIUC}
\author{Narisak Hiri-O-Tuppa}
\affiliation{\UIUC}
\author{Jacob P. Covey}\email{jcovey@illinois.edu}
\affiliation{\UIUC}

\begin{abstract}
The integration of quantum computers and sensors into a quantum network opens a new frontier for quantum information science. We demonstrate high-fidelity entanglement between ytterbium-171 atoms -- the basis for state-of-the-art atomic quantum processors and optical atomic clocks -- and optical photons directly generated in the telecommunication wavelength band where loss in optical fiber is minimal. We entangle the nuclear spin of the atom with a single photon in the time bin basis, and find an atom measurement-corrected (raw) atom-photon Bell state fidelity of \bellfidelity ~(\bellfidelityraw). Photon measurement errors contribute $\approx0.037$ to our infidelity and can be removed with straightforward upgrades. Additionally, by imaging our atom array onto an optical fiber array, we demonstrate a parallelized networking protocol that can provide an $N$-fold boost in the remote entanglement rate. Finally, we demonstrate the ability to preserve coherence on a memory qubit while performing networking operations on communication qubits. Our work is a major step towards the integration of atomic processors and optical clocks into a high-rate or long-distance quantum network.
\end{abstract}
\maketitle

\section{Introduction}\label{Intro}
Quantum networks utilize entanglement shared among many nodes to enable new opportunities for cryptographically secured communication~\cite{Gisin2002,Pirandola2019,Pirandola2020}, quantum sensing~\cite{Gottesman2012,Malia2022} and timekeeping~\cite{Komar2014,Nichol2022}, and blind~\cite{Wootters1982,Barz2012,Fitzsimons2017} or distributed~\cite{Jiang2007,Monroe2014} quantum computing. Entanglement between two network nodes is typically generated probabilistically via a Bell state measurement of two photons that are each entangled with a matter-based, atom-like qubit in a node. This technique has been used to realize high fidelity~\cite{Stephenson2020,Saha2024} and long-distance entanglement~\cite{Hensen2015,vanLeent2022,Krutyanskiy2023,Uysal2024} between atom-like qubits. The rate and fidelity of entanglement between the nodes is primarily determined by the properties of the atom-like qubits and the photonic interface that collects photons into optical fibers~\cite{Covey2023}.

\begin{figure}[t!]
	\includegraphics[width=\linewidth]{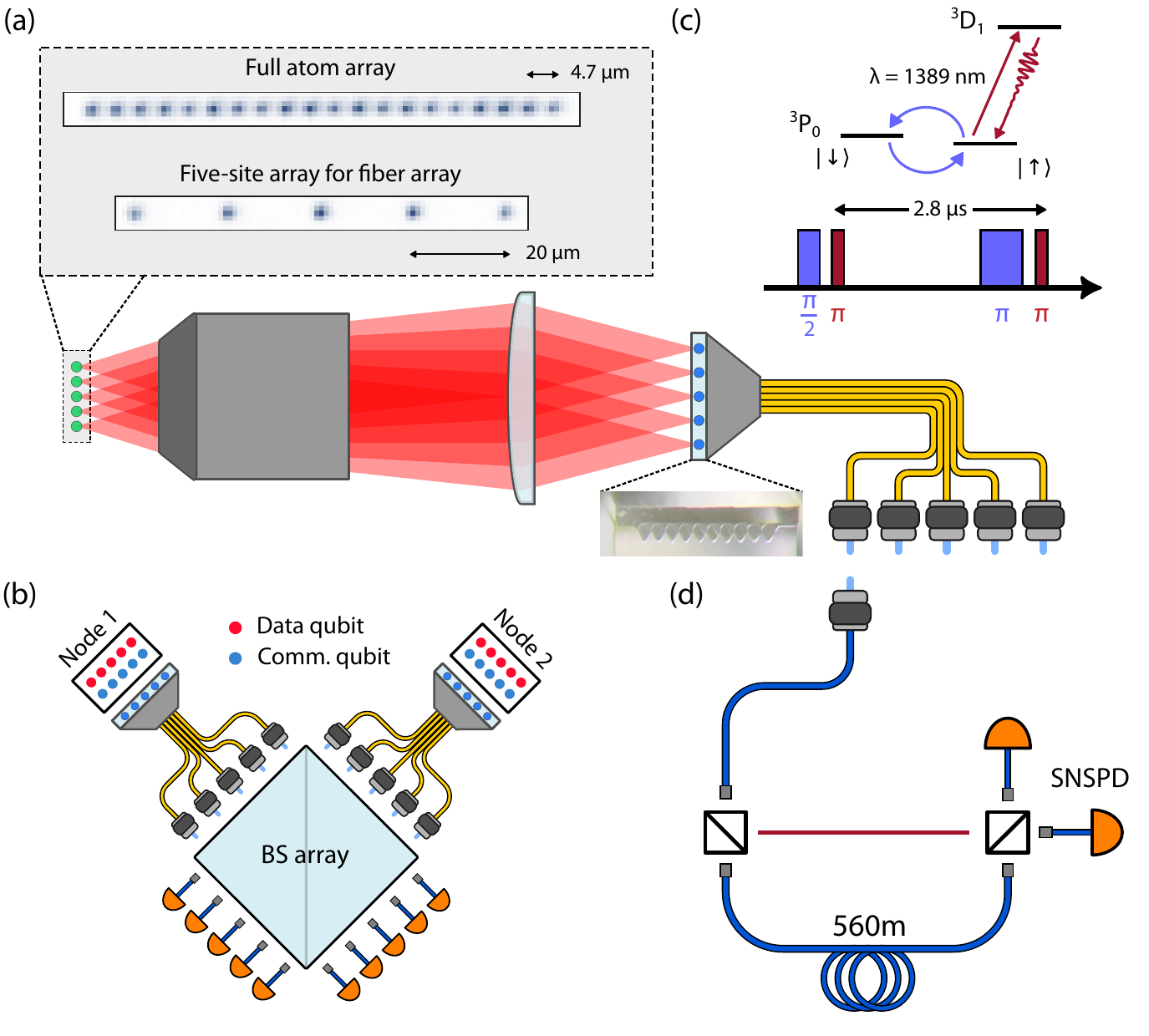}
	\caption{
        \textbf{Overview of the platform}.
        (a) An imaging system with a high-NA objective maps an array of atoms in optical tweezers to an array of single-mode optical fibers. We nominally use an array of 20 tweezers with spacing $\approx4.7$ $\mu$m, but an array of 5 tweezers with spacing $\approx20$ $\mu$m is employed for optimal matching with the mode-field diameter of the fiber array. The inset shows the image of a typical v-grooved fiber array with 10 fibers in a row. (b) Our vision for parallelized networking with atom array processors using fiber, detector, and beam splitter arrays. (c) We utilize time bin encoding to entangle the metastable nuclear spin of ytterbium-171 atoms (blue pulses) with individual photons with wavelength 1389 nm (red pulses). (d) After sending the photons through a 40-m fiber, we utilize a time-delay interferometer (TDI) and superconducting nanowire single photon detectors (SNSPDs) to characterize the atom-photon Bell state.
    }
    \label{fig:schematic}
\end{figure}

Following the development of classical communication technology, quantum networks will benefit from the use of photons in the telecommunication (telecom) wavelength band (approximately $1.25-1.65$ $\mu$m), both because of the minimal attenuation in optical fiber and because of the associated growth of silicon, CMOS-compatible photonic technologies such as fast and efficient switches, modulators, and detectors. However, most quantum networks with atom-like qubits operate at visible or near-ultraviolet wavelengths~\cite{Moehring2007,Bernien2013,Bhaskar2020,Saha2024} and require conversion to the telecom band for long-distance networking~\cite{vanLeent2022,Stolk2022,Krutyanskiy2023,Bersin2024} which adds complexity and footprint, reduces efficiency, and may add noise and/or extra photons. Although some rare earth ions embedded in solid-state hosts, for instance, operate directly at telecommunication wavelengths, their transitions are weak and require substantial Purcell enhancement~\cite{Zhong2017,Dibos2018,Kindem2020,Uysal2024}. A platform that combines \textit{both} highly coherent qubits capable of scalable deterministic quantum logic \textit{and} a high emission bandwidth directly in the telecom wavelength band has yet to be developed. 

Here, we demonstrate quantum networking with highly coherent nuclear spin qubits that are directly connected to a strong, telecom-band transition at a wavelength 1389 nm, where the attenuation is $\approx0.3$ dB/km in modern telecom fiber (see Fig.~\ref{fig:schematic}). Using a time-delay interferometer (TDI) with a 560-m fiber on one arm, we perform tomography of time bin-encoded photonic qubits. We observe an atom measurement-corrected (raw) Bell pair fidelity of \bellfidelity~(\bellfidelityraw) between the metastable nuclear spin qubit of a ytterbium-171 atom and a single photon. To our knowledge, this fidelity matches the record for entanglement between atom-like emitters and telecom-band photons~\cite{vanLeent2022}. It is also the first direct entanglement with telecom-band photons for trapped atom/ion systems, the first demonstration of time-bin encoding for neutral atoms, and the first quantum networking demonstration with neutral alkaline earth(-like) atoms which host optical clock transitions.

Moreover, many applications of quantum networks require fast entanglement generation. The coherence time of the qubits represents a minimal timescale for useful entanglement generation~\cite{Hucul2015}. However, a more stringent requirement for distributed quantum computing, for instance, is the gate rate in a quantum processor. Ideally, remote entanglement would be established at least this fast, but the attempt rate is fundamentally limited by atomic properties such as the decay rate. Hence, the optimal strategy for increasing the entanglement generation rate is via parallelization. With an array of $N$ qubits in each node, remote entanglement generation could be attempted in parallel across $N$ channels to boost the communication rate by a factor of $N$ [see Fig.~\ref{fig:schematic}(b)]. Such \textit{spatial} multiplexing outperforms \textit{temporal} multiplexing~\cite{Huie2021,LiThompson2024,Canteri2024,Hartung2024} in short network links for which latency from two-way communication is negligible.

Therefore, we additionally demonstrate a platform for parallelized networking by coupling our one-dimensional atom array to a lensed fiber array. We show that the atom-photon entanglement fidelities are uniform across the atom/fiber array, with negligible crosstalk between channels. Finally, we demonstrate the ability to preserve coherence on a memory qubit while performing networking operations on communication qubits. Our work sets the stage for high-rate, high-fidelity remote entanglement between large-scale neutral atom quantum processors, opening the door to new opportunities for atomic clock networks~\cite{Komar2014,Nichol2022} and distributed quantum computing~\cite{Huie2021,LiThompson2024,Canteri2024,Hartung2024,Monroe2014,Jiang2007,Sunami2024}.

\section{Multi-level control of a ytterbium-171 atom array}

We use a one-dimensional array of up to 20 optical tweezers with wavelength $\approx760$ nm and depth $U/k_B\approx500$ $\mu$K [see Fig.~\ref{fig:schematic}(a)]. After loading atoms into optical tweezers, we image the array to learn its occupancy, cool the atoms, and perform optical pumping to the $m_F=-1/2$ ground nuclear spin state. Many details of our system can be found elsewhere~\cite{Huie2023}, but we have since added gray molasses cooling~\cite{Jenkins2022} and electromagnetically induced transparency (EIT) cooling~\cite{Morigi2000,Lis2023} (see Appendix \ref{app:gm-eit} for further details). EIT cooling brings the atomic temperature close to the motional ground state, and offers a simpler alternative to sideband cooling for achieving moderately low temperatures.

\begin{figure}[t!]
	\includegraphics[width=\linewidth]{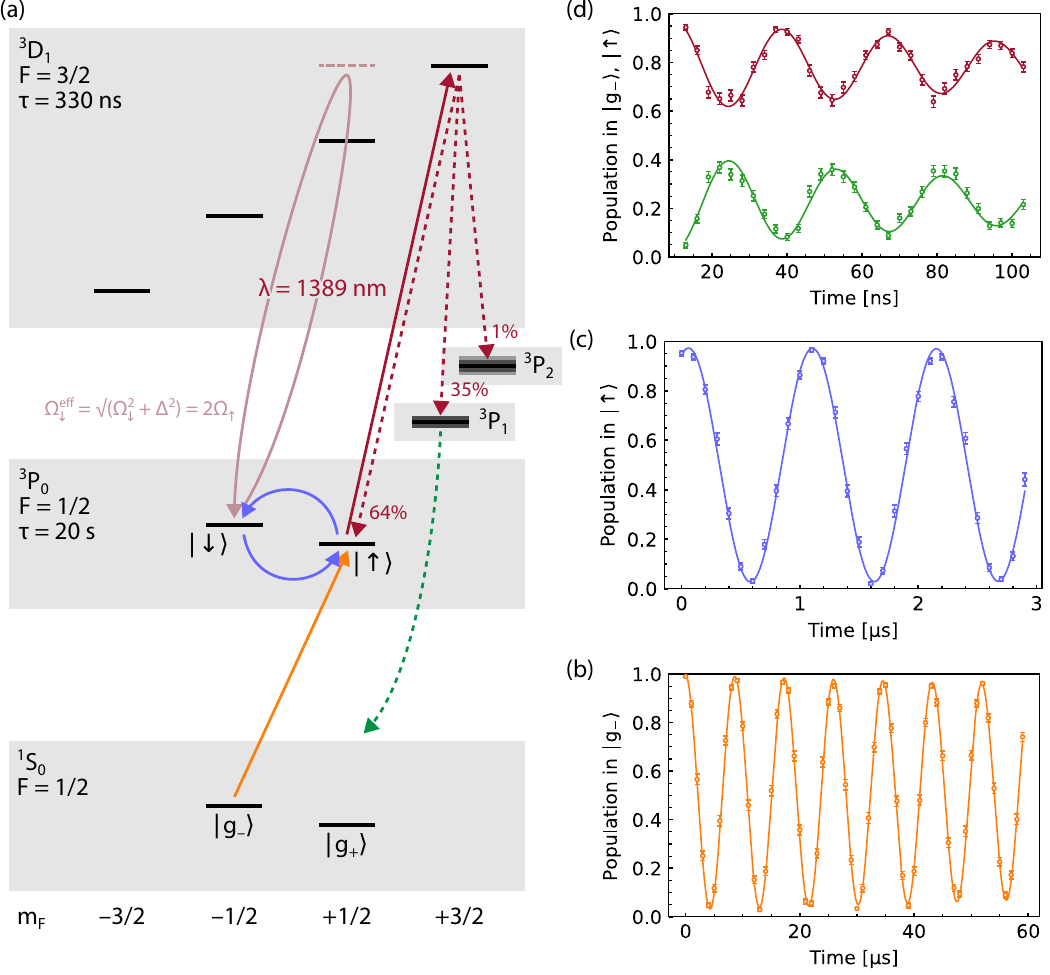}
	\caption{
        \textbf{Multi-level control of the $^{171}$Yb array}.
        (a) The relevant level structure that shows the optical ``clock" transition (yellow), the metastable nuclear spin qubit (blue), and the telecom transition for atom-photon entanglement (red). The decay pathways from the $^3$D$_1$ state are shown, where the pathway to $^3$P$_1$ subsequently leads to the ground state in $\approx1$ $\mu$s. (b) Rabi oscillations on the clock transition with $\pi$-pulse time of $\tau_\pi^\text{o}\approx4$ $\mu$s and corrected $\pi$- ($2\pi$-)pulse fidelity of \clockpiprob~(\clocktwopiprob). (c) Rabi oscillations of the metastable nuclear spin qubit with $\pi$-pulse time of $\tau_\pi^\text{m}\approx0.6$ $\mu$s and corrected $\pi$-pulse fidelity of \ramanpiprob. (d) Rabi oscillations of the telecom transition with $\pi$-pulse time of $\tau_\pi^\text{t}\approx16$ ns and corrected $\pi$-pulse fidelity of \regpiprob. $\tau_\pi^\text{t}$ was chosen to minimize population in the $^3$D$_1$ $m_F=1/2$ state (light red loop; see text). We detect the population in both $^1$S$_0$ (green) and $^3$P$_0$ (red) after spontaneous emission.
    }
    \label{fig:control}
\end{figure}

The strong telecom transition in ytterbium is from the metastable ``clock" state $^3$P$_0$ (with a lifetime of 20 seconds) to the higher-lying $^3$D$_1$ state. Hence, we focus on the nuclear spin qubit in the metastable state, which we directly entangle with a telecom photon. A coherence time of $T^*\approx7$ sec has been demonstrated for the metastable nuclear spin qubit~\cite{Lis2023}. More broadly, it is well suited for myriad quantum computing operations~\cite{Barnes2021,Chen2022} such as mid-circuit measurements~\cite{Lis2023}, leakage error detection~\cite{Ma2023}, and Rydberg-mediated entangling gates~\cite{Ma2023,Peper2024,Muniz2024} due to its relatively strong, single-photon coupling to an S-series Rydberg manifold~\cite{Madjarov2020,Ma2023}. The relevant level structure for our work is shown in Fig.~\ref{fig:control}(a).

We initialize the clock state via a $\pi$-pulse on the $^1$S$_0$ $m_F=-1/2$ to $^3$P$_0$ $m_F=1/2$ transition with $\sigma^+$ polarization (see Appendix~\ref{app:general}). We drive the transition with a $\approx$Hz-linewidth laser and realize a Rabi frequency of $\Omega^\text{o}/2\pi\approx130$ kHz for which $\tau_\pi^\text{o}\approx4$ $\mu$s. As shown in Fig.~\ref{fig:control}(b), we can drive highly coherent Rabi oscillations on the clock transition, which is enabled by the removal of high-frequency laser noise via active feedforward~\cite{Li2022} (see Appendix~\ref{app:clock}). We observe a corrected (raw) $\pi$-pulse fidelity of \clockpiprob~(\clockpiprobraw). Our definition of fidelity and our method for correcting state preparation and measurement errors (SPAM) is described in Appendix~\ref{app:spam}. Our corrected (raw) $2\pi$-pulse fidelity of \clocktwopiprob~(\clocktwopiprobraw) suggests that our measured fidelity is limited by detuning errors rather than Rabi decoherence.

After initializing in the $m_F=1/2$ nuclear spin state of the $^3$P$_0$ clock state, we demonstrate the ability to perform fast, high-fidelity rotations of the metastable nuclear spin qubit since such operations are required for time bin remote entanglement~\cite{Bernien2013,Saha2024,Uysal2024}. We utilize stimulated Raman rotations via the $^3$D$_1$ state -- blue-detuned from the $F=3/2$, $m_F=3/2$ state by 612 MHz -- with an effective two-photon Rabi frequency of $\Omega^\text{m}/2\pi\approx1$ MHz. We operate at a magnetic field of 120 G, for which the metastable nuclear spin Zeeman splitting is $\Delta_B^\text{m}\approx137$ kHz -- much less than $\Omega^\text{m}$. We drive the Raman transition with a single monochromatic beam by operating at a ``magic" condition where the differential light shift on the qubit states exactly cancels $\Delta_B^\text{m}$~\cite{Chen2022} (see Appendix \ref{app:clockRaman}). As shown in Fig.~\ref{fig:control}(c), we can drive highly-coherent Rabi oscillations of the metastable nuclear spin qubit for which $\tau_\pi^\text{m}\approx0.6$ $\mu$s. We measure a corrected (raw) $\pi$-pulse fidelity of \ramanpiprob~(\ramanpiprobraw).

Finally, we demonstrate control of the $^3$P$_0$$\leftrightarrow$$^3$D$_1$ telecom transition. Atom-photon entanglement is based on spontaneous emission from the $^3$D$_1$ state whose lifetime is $\tau^{^3\text{D}_1}\approx330$ ns, so we must initialize the $^3$D$_1$ state much faster than 330 ns. We drive the $^3$P$_0$, $m_F=1/2$ to $^3$D$_1$, $F=3/2$, $m_F=3/2$ transition via a beam with $\sigma^+$ polarization (see Appendix~\ref{app:general}) with Rabi frequency $\Omega^\text{t}/2\pi\approx30$ MHz which has a corresponding $\pi$-pulse time of $\tau_\pi^\text{t}\approx16$ ns. As shown in Fig.~\ref{fig:control}(d), we drive highly coherent Rabi oscillations, enabled by an acousto-optic modulator with 5-ns switching time (see Appendix~\ref{app:1389}). As shown in Fig.~\ref{fig:control}(a), the $^3$D$_1$ state decays back to the $^3$P$_0$ clock state with probability $P_{^3\text{P}_0}\approx0.64$, to the $^3$P$_1$ state with probability $P_{^3\text{P}_1}\approx0.35$ (which then subsequently decays to the $^1$S$_0$ ground state in $\approx1$ $\mu$s), and to the $^3$P$_2$ state with probability $P_{^3\text{P}_2}\approx0.01$ (which is strongly anti-trapped in tweezers of this wavelength). We observe Rabi oscillations of the telecom transition by probing both the decay back to the ground state [$P_{^3\text{P}_1}$; green in Fig.~\ref{fig:control}(d)] and by swapping the population between the ground and clock states [$P_{^3\text{P}_0}$; red in Fig.~\ref{fig:control}(d)]. The sum of the two is consistent with $1-P_{^3\text{P}_2}\approx0.99$. Our measurements suggest a corrected $\pi$- ($2\pi$-)pulse fidelity of \regpiprob~(\regtwopiprob).

\section{Time bin remote entanglement}
Time bin entanglement generates the atom-photon Bell state $\ket{\psi} = (\ket{\uparrow, E} + e^{i\phi}\ket{\downarrow, L}) / \sqrt{2}$, where $\ket{E}$ and $\ket{L}$ denote single photons in the early and late bin, respectively~\cite{Bernien2013}. As shown in Fig.~\ref{fig:schematic}(c), the process begins by performing a $\pi/2$-pulse on the qubit initialized in $\ket{\uparrow}$ followed by a $\pi$-pulse on the telecom transition. The $\pi$-pulse is applied only to $\ket{\uparrow}$, which then scatters a telecom photon and returns to $\ket{\uparrow}$. Then, to make the protocol robust to photon loss and imperfect detection, we perform a second round such that both sectors of the atomic superposition have an associated photon emission. Hence, we perform a $\pi$-pulse on the qubit followed by a second $\pi$-pulse on the telecom transition. Now the qubit sector that was initially $\ket{\downarrow}$ has an associated photonic state $\ket{L}$, and we have generated the time bin-encoded atom-photon Bell state $\ket{\psi}$ (up to a $\pi$-pulse on the qubit).

The Rabi frequency of the ${}^3$P$_0 \leftrightarrow {}^3$D$_1$ telecom transition was chosen to minimize the population in the $^3$D$_1$, $F=3/2$, $m_F=1/2$ state after a $\pi$-pulse. The Zeeman splitting of the $m_F=3/2$ and $m_F=1/2$ states at 120 G is $\Delta/2\pi\approx\textcolor{black}{58}$ MHz, for which $\Delta/\Omega^\text{t}\approx2$. Hence, a $\pi$-pulse on the desired, $m_F=1/2$ to $m_F=3/2$ transition will also drive the undesired, $m_F=-1/2$ to $m_F=1/2$ transition, leading to a spin-flip error and reduce the atom-photon Bell state fidelity. Accordingly, we operate at a ``magic" Rabi frequency for which $\tau_\pi^\text{t}$ corresponds to a $2\pi$-pulse on the undesired, $m_F=-1/2$ to $m_F=1/2$ transition [see Fig.~\ref{fig:control}(a)]. This magic Rabi frequency emerges from the interplay of different Clebsch-Gordan coefficients of the desired vs undesired transitions giving rise to different Rabi frequencies ($\Omega_\uparrow=\Omega^\text{t}$ and $\Omega_\downarrow$, respectively) and the Zeeman splitting $\Delta$. The magic condition is given by $\Omega_\downarrow^\text{eff}=\sqrt{\Omega_\downarrow^2+\Delta^2}=2\Omega_\uparrow$. Appendix \ref{app:sp_em} analyzes the idealized atom-photon entanglement in the presence of $\Delta$ and $\tau^{^3\text{D}_1}$. We predict an otherwise idealized fidelity of $\approx0.98$ at the magic Rabi frequency of $\Omega^\text{t}/2\pi\approx30$ MHz, 
and $\approx 0.997$ after post-selecting on detecting $\geq 1$ emitted time-bin photons.

\section{Characterizing single photons and atom-photon entanglement}
Based on our fast, high-fidelity control of the three essential atomic transitions, we now turn to the collection and characterization of single photons and the measurement of atom-photon entanglement. We collect single photons from a single atom with a high-NA microscope objective, and image them with a 1:1 telescope onto a single SMF-28 single-mode telecom fiber [see Fig.~\ref{fig:schematic}(a); the fiber array will be discussed below]. We observe a combined collection and transmission efficiency of $\eta_\text{net}\approx0.5$\% from an atom through a single-mode fiber. For comparison, our imaging system for fluorescence detection has a $\approx4$\% atom-to-camera collection efficiency, but the use of single-mode fiber introduces another reduction factor since the dipole radiation pattern -- particularly for the stretched transition and geometry used here -- has limited overlap with a Gaussian mode. Additionally, there are other losses from polarization and wavelength filtering optics.

    To perform tomography on the photonic qubits and the atom-photon Bell pairs, we use a time-delay interferometer, as shown in Fig.~\ref{fig:schematic}(d). The length of the long arm $L_d$ is chosen to match the time delay between the two bins, $t_d$ [see Fig.~\ref{fig:schematic}(c)]. $t_d$ is primarily limited by the spontaneous emission time of the $^3$D$_1$ state. Since $\tau^{^3\text{D}_1}\approx330$ ns, we choose $t_d=2.8$ $\mu$s such that $1-\exp(-(t_d-\tau_\pi^\text{m})/\tau^{^3\text{D}_1})=0.99$ is larger than the expected Bell state fidelity ($\tau_\pi^\text{m}\approx700$ ns). 
The concomitant fiber delay length on the long arm is $L_d=c\cdot t_d=560$ m. We note that motion of the atom during this time would result in a dephasing of the emitted photon between the two bins. Hence, we must either match the time delay with the period of motion of the atom in the trap~\cite{Saha2024}, or choose a time delay that is much shorter than the trap period. We take the second approach, and we expect a reduction in the Bell state fidelity of $\approx 0.009$ (see Appendix \ref{app:motion}).

\begin{figure}[t!]
	\includegraphics[width=\linewidth]{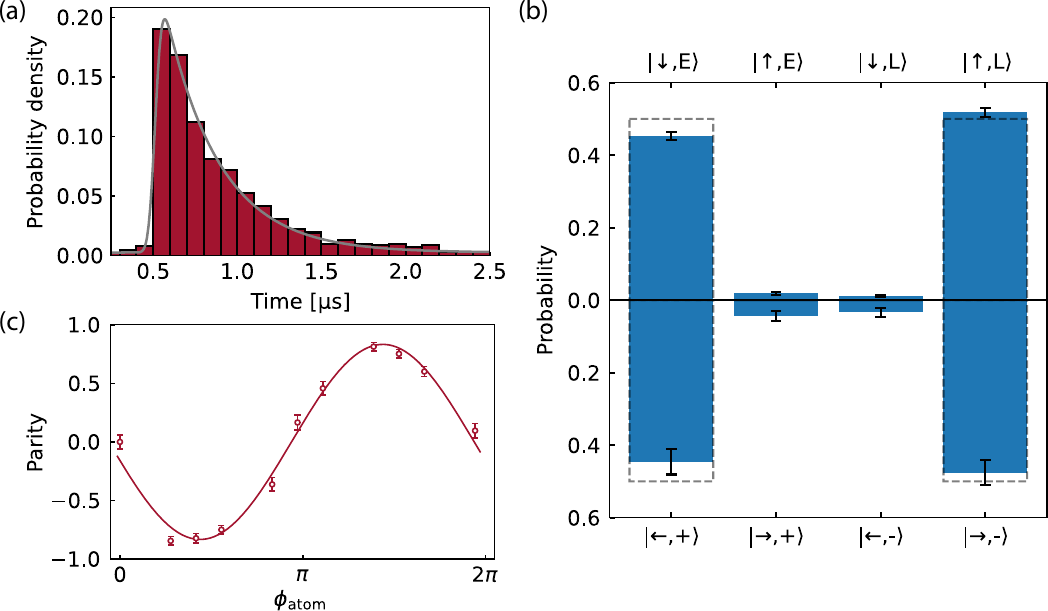}
	\caption{
        \textbf{Atom-photon entanglement}.
        (a) Arrival histogram of single photons in excellent agreement with the expected $\tau^{^3\text{D}_1}\approx330$ ns. (b) Atom-photon measurement correlation plots in the ZZ basis (top) and XX basis (bottom), dashed lines represent ideal correlations. (c) Parity contrast of the atom-photon Bell state versus the qubit readout phase.
    }
    \label{fig:photon-meas}
\end{figure}

To characterize the photons, we begin by collecting them from spontaneous emission of atoms in $\ket{\uparrow}$, sending them over a 50-m fiber link, and detecting them with a superconducting nanowire single photon detector (SNSPD). A histogram of their arrival time [see Fig.~\ref{fig:photon-meas}(a)] shows excellent agreement with the expected $\tau^{^3\text{D}_1}\approx330$ ns. To study atom-photon entanglement -- and towards the eventual goal of atom-atom entanglement -- we choose a time window of $t_\text{det}=\textcolor{black}{520}$ ns in which we count events. The probability of dark counts on the SNSPDs during $t_\text{det}$ is \textcolor{black}{$P_\text{dc}\approx3\times10^{-6}$},
and the effective signal-to-noise ratio for ZZ (XX) basis measurements are 19.6 (11.6) dB (see Appendix~\ref{app:snr}). Note that the time-delay interferometer, which adds significant losses, is only required for the XX data.

Next, we employ the time bin protocol and characterize the atom-photon Bell state in the Z basis for both the atom and the photon: we measure the atom in $\ket{\downarrow}$ or $\ket{\uparrow}$ and the photon in $\ket{E}$ or $\ket{L}$. Details of the attempt loop can be found in Appendix~\ref{app:attloop}; briefly, we attempt \textcolor{black}{128} times and the reset protocol between attempts includes \textcolor{black}{$t_\text{cool}=1.4$ ms} of cooling. The success rate of generating atom-photon entanglement is limited by a combination of $\eta_\text{net}$, $t_\text{cool}$, and the atom reload time (\textcolor{black}{$t_\text{reload}\approx1.1$ sec}) between each attempt loop. These factors limit the success rate to sub-Hz, but many improvements are possible (see the Concluding Discussion) and parallelization can boost the effective rate by a factor of $N$. We use real-time decision making to conditionally branch out of the attempt loop upon the detection of a photon in one of the two time bins. As shown in the upper plot of Fig.~\ref{fig:photon-meas}(b), we find a strong correlation between the qubit and the photon in the ZZ basis. In fact, the infidelity is dominated by readout error of the qubit (which requires a clock $\pi$-pulse back to the ground state). See Appendix~\ref{app:atom_readout} for further details.

To completely characterize the atom-photon Bell state fidelity, we must measure the atom and the photon in the XX and YY bases. For the photon, this requires the use of a time-delay interferometer (TDI) [see Fig.~\ref{fig:schematic}(d) and Appendix~\ref{app:interferometer}] as discussed above.  
To ensure that the $\ket{E}$ ($\ket{L}$) sector of the photonic state goes through the long (short) arm of the TDI, we use a Pockel's cell at the input to switch the path for each bin. As shown in Fig.~\ref{fig:photon-meas}(c),
we acquire a visibility fringe by varying the qubit measurement basis. We show the highest XX visibility condition in the lower plot of Fig.~\ref{fig:photon-meas}(b). The fringe combined with the ZZ basis measurements are sufficient to characterize the atom-photon Bell state fidelity to within a bound (see Appendix~\ref{app:Bellfidelity}). We obtain a raw fidelity of \bellfidelityraw. However, we note that this value is limited by the infidelity of transferring to the readout state ($\approx0.02$) and the infidelity of readout itself ($\approx0.03$). We thus calculate the atom measurement-corrected Bell state fidelity (see Appendix~\ref{app:spam}) to be \bellfidelity. 
Moreover, photon measurement errors contribute $\approx0.037$ to our infidelity (see Appendix~\ref{app:budget}) and can be removed with straightforward upgrades.

\begin{figure}[t!]
	\includegraphics[width=\linewidth]{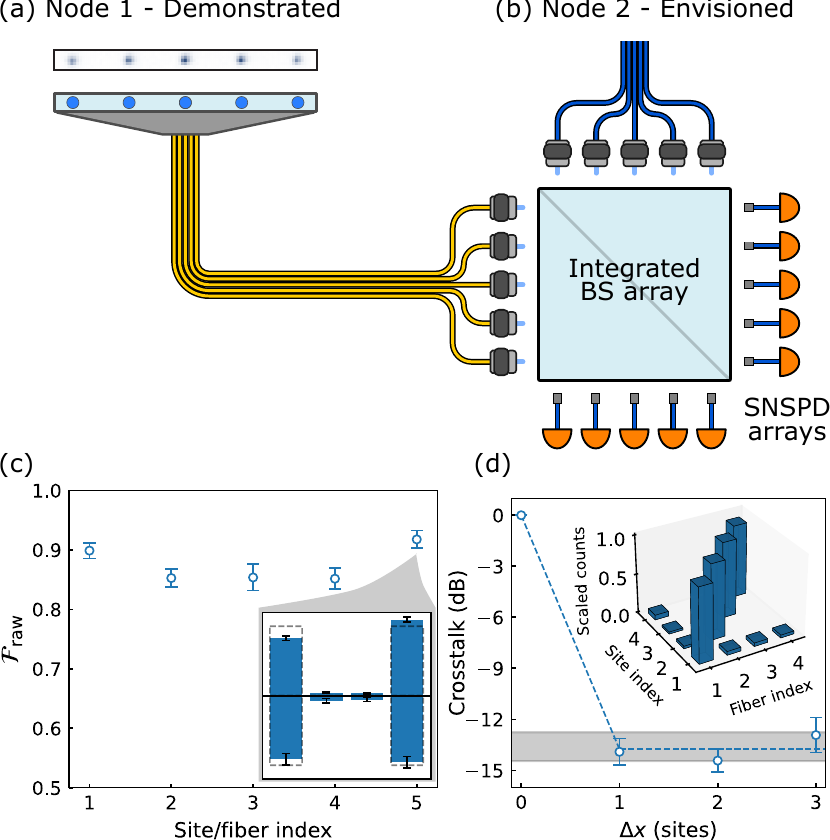}
	\caption{
        \textbf{Parallelization with a fiber array.}
        (a) and (b) We envision two atom array nodes in which the atom arrays in each are mapped to fiber arrays. The fiber arrays are directed to an integrated 50:50 beam splitter array on a photonic chip whose outputs are then directed to arrays of superconducting nanowire single photon detectors. (c) We study the atom-photon entanglement fidelity as a function of site index along the array. The inset shows one example histogram measured for one site. (d) We also study the crosstalk between channels by studying the relative photon collection efficiency from an atom on nearby sites. For $|\Delta x|\ge1$, we find an averaged crosstalk of -13.7(4) dB, limited by the -13.6(8) dB estimated noise floor (shaded region) for this measurement. The inset shows scaled photon counts collected into the fiber array post-selected on atom presence at different tweezer sites.
    }
    \label{fig:fiber-array}
\end{figure}

\section{Parallelized networking with a fiber array}
With a combined photon collection and transmission efficiency through fiber of $\eta_\text{net}$, parallelization over $N\approx1/\eta_\text{net}$ or $N\approx1/\eta_\text{net}^2$ channels would ensure a high success probability of establishing atom-photon or atom-atom entanglement in a single attempt, respectively. The attempt rate could be $\approx10^5$ s$^{-1}$, limited by atomic properties such as decay rates and qubit rotation rates. Parallelization could be realized with optical fiber arrays; fiber bundles are ubiquitous in the telecom industry. Since absolute path length (phase) stability is not required for time-bin encoding, we do not anticipate fidelity limitations associated with using an array of fibers.

In the context of remote atom-atom entanglement using two atom arrays [see Fig.~\ref{fig:fiber-array}(a)], we envision the use of a 50:50 beam splitter (BS) array [see Fig.~\ref{fig:fiber-array}(b)]. In the telecom band, it is straightforward to use silicon integrated photonics~\cite{Carolan2015,Pelucchi2022} with two fiber array inputs into the BS array and two fiber array outputs after the BS array to direct the photons to SNSPD arrays~\cite{Wollman2019,Oripov2023,Fleming2025}. The BS array could even be directly integrated with the SNSPD array. The click pattern on the SNSPD array would indicate which atomic Bell pair(s) succeeded on a given attempt, and the remaining atoms in each array that failed to generate a remote Bell pair could be used as a resource for the subsequent quantum circuit or subsequent attempts.

As a step towards this vision, we show that an array of atoms can be efficiently and independently mapped to an array of single-mode telecom fibers. We employ a lensed fiber array to better match the ``form factor" of the atom array with that of the fiber array [see Fig.~\ref{fig:fiber-array}(a)]. The core pitch of the fiber array is $d_\text{p}\approx80$ $\mu$m and the mode field diameter (MFD) is $d_\text{MFD}\approx10$ $\mu$m. To match this $d_\text{MFD}/d_\text{p}\approx13$\% ``form factor" with the atom array given the diffraction-limited spot diameter at 1389 nm of $\approx3$ $\mu$m, we choose a tweezer spacing of $\approx20$ $\mu$m, which is approximately four times as large as our nominal array spacing [see Fig.~\ref{fig:schematic}(a)]. We thus parallelize over $N=5$ channels (although we use a 20-site fiber array).

As shown in Fig.~\ref{fig:fiber-array}(c), we find a relatively uniform Bell state fidelity. This data does not include the bound, but we assume that it is again comparable to our statistical uncertainty. Crucially, we do not observe cross talk between the sites at the $-13.6(8)$ dB level corresponding to our average noise floor [Fig.~\ref{fig:fiber-array}(d)]. We characterize the crosstalk by looking at the locations of atoms in the array and the locations of detected photons. With only two SNSPD channels, we study these correlations in a pairwise fashion, as shown in the inset of Fig.~\ref{fig:fiber-array}(d). These techniques are scalable to thousands of sites. Moreover, it is possible to increase the form factor by changing the lens design on the fiber to increase the MFD. With a realistic MFD of $\approx30$ $\mu$m, the form factor matches that of typical neutral atom and trapped ion arrays. 

\section{Preserving coherence while establishing remote entanglement}

Finally, we demonstrate a technique to enable the preservation of coherence of memory qubits while establishing remote entanglement between communication qubits. In the context of distributed quantum computing, it will be essential to establish entanglement between modules in a `mid-circuit' manner in which the computations within the modules are ongoing. For long-distance quantum networking with intermediate quantum repeaters, rate optimization requires the ability to store successfully-generated Bell pairs as memory while connecting the remaining links. This need is particularly acute for multiplexed approaches~\cite{Huie2021}. Accordingly, the coherence of the data/memory qubits must be completely unaffected by the operations of the communication qubits. Figure~\ref{fig:schematic}(b) illustrates this vision.

\begin{figure}[t!]
	\includegraphics[width=\linewidth]{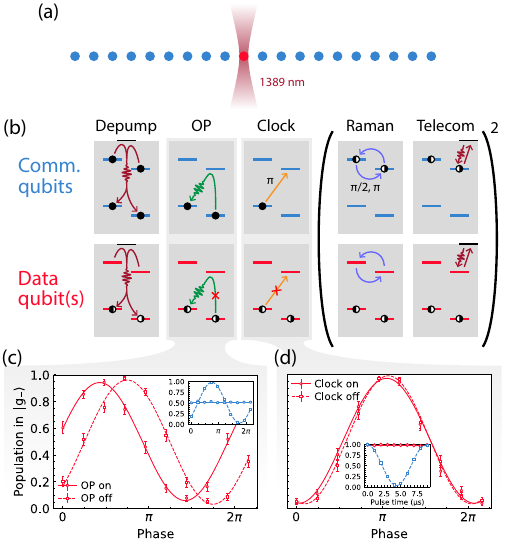}
	\caption{
        \textbf{`Mid-circuit' networking.}
        (a) A local light shift of the $^3$P$_0$ and $^3$P$_1$ states from a tightly focused 1389-nm beam ``shields" that site (data qubit; red) from the remote entanglement protocol applied to the other sites (communication qubits; blue). (b) The remote entanglement protocol contains many operations from the metastable state as well as two operations from the ground state. The local light shift causes those two ground state operations to be significantly off resonant for the data qubit that is stored in the ground nuclear spin. (c) A Ramsey fringe of the ground nuclear spin qubit for the data site (red; large) and communication sites (blue; inset) with (circle; solid line) and without (square; dashed line) the optical pumping pulse applied. (d) A Ramsey fringe of the ground nuclear spin qubit for the data qubit (red; large) with (circle; solid line) and without (square; dashed line) the clock pulse applied. The inset shows the population in the ground state for the data qubit (red) and communication qubits (blue) as the clock pulse is applied.
    }
    \label{fig:mid-circuit}
\end{figure}

Paralleling the development of techniques for `mid-circuit measurement' (MCM) in which some qubits are measured while preserving the coherence of the others, `mid-circuit networking' (MCN) can be accomplished in several ways. Four main techniques have emerged for MCM: 1) ``Shelving", in which data qubits are preserved in a portion of the atomic structure that is completely decoupled from the readout operations~\cite{Lis2023,Ma2023,Scholl2023b,Graham2023}; 2) ``Dual species" approaches in which the transition frequencies are sufficiently different between two species such that readout of one does not affect the other~\cite{Singh2022,Nakamura2024}; 3) ``Shielding" of data qubits with local light shifts that are sufficiently large to decouple them from readout operations~\cite{Norcia2023,Hu2024}; and 4) ``Zoning" techniques by which coherent transport is performed between spatially-separated readout and memory zones~\cite{Deist2022b,Bluvstein2023}. Here, we demonstrate an approach to MCN based on both ``shielding" with a local light shift at 1389 nm [see Fig.~\ref{fig:mid-circuit}(a)] and ``shelving".

Although $^{171}$Yb is well suited for the ``shelving" technique, our atom-photon entanglement scheme involves both the ground state and the metastable ``clock" state since the decay probability to the ground state is $\approx35$\% per attempt. As discussed above and shown in Fig.~\ref{fig:mid-circuit}(b) (see also Appendix~\ref{app:attloop}), each attempt involves a reset operation in the clock state (depumping to the ground state) and the ground state [optical pumping (OP) to $m_F=-1/2$]. These are followed by a $\pi$-pulse on the optical clock transition to initialize the communication qubits in $^3$P$_0$ $m_F=1/2$ (see above). Since both the OP and the clock operations directly couple to the ground state, they preclude the use of the ground nuclear spin as a data qubit. Our approach employs a light shift to shield the data qubits during the OP and clock operations. All other networking operations occur in the metastable manifold for which our data qubits are unaffected -- the core principle behind the ``shelving"-type MCM approach -- and the light shift is off.  

The shielding approach relies on a large differential light shift between the two states of interest. In our case, OP is performed on the $^1$S$_0$$\leftrightarrow$$^3$P$_1$ $F=3/2$ transition for which the polarizability ratio is $\alpha_{^3\text{P}_1}/\alpha_{^1\text{S}_0}\approx \textcolor{black}{-8}$ at 1389 nm~\cite{Tang2018}. Currently, we can only provide $\approx4$ mW of 1389-nm light to the atoms through the path with our microscope objective. Moreover, because of a focal shift between 760 nm -- the wavelength of the tweezers that defines the position of the atoms -- and 1389 nm in our microscope objective, we are unable to focus a 1389-nm spot to smaller than $\approx3$ $\mu$m waist radius. These issues prevent us from using more than one data qubit and limit our light shift on the OP transition to $\Delta_\text{LS}^\text{OP}\approx2\pi\times2$ MHz. Nevertheless, we find this shift to be sufficient to preserve Ramsey coherence of our data qubit.
As shown in Fig.~\ref{fig:mid-circuit}(c), we can look at a Ramsey fringe by varying the dark time on a scale commensurate with the Larmor precession at $\omega_g\approx2\pi\times\textcolor{black}{89}$ kHz. We use a dark time of 10 ms, during which we apply the 1389 nm shield beam to the data qubit, and conditionally apply the OP pulse for 20 $\mu$s. We find that the contrast of the Ramsey fringe of the data qubit has reduced by $\approx5(1)$\% and that the qubit has undergone a phase shift of \textcolor{black}{1.01(6)} radians. Meanwhile, the coherence of all the non-shielded atoms is completely lost.

The shielding of the optical clock pulse, by contrast, requires much lower intensity. We operate the 1389-nm beam at a detuning of $\textcolor{black}{+612}$ MHz from the $^3$P$_0$$\leftrightarrow$$^3$D$_1$ $F=3/2$ transition, for which the differential polarizability is $\alpha_{^3\text{P}_0}/\alpha_{^1\text{S}_0}\approx \textcolor{black}{-2\times10^5}$. Accordingly, only $\textcolor{black}{\approx0.01}$ $\mu$W is needed to apply a light-shift $\Delta_\text{LS}^\text{Clock}\approx2\pi\times \textcolor{black}{1}$ MHz to the data qubit's clock transition. As shown in Fig.~\ref{fig:mid-circuit}(d), this is sufficient to completely block the clock transition for the data qubit while leaving the transition for all other sites unaffected. We again show the preservation of Ramsey coherence on the data qubit with and without this clock pulse operation, finding a small reduction in coherence of $\approx0.3(3)$\%. We are confident that this technique can be readily extended to many data qubits with sufficient power (or by using 1539-nm light, close to the $^3$P$_1$$\leftrightarrow$$^3$D$_1$ transition, for shielding from OP and cooling instead of 1389 nm), and that a stronger shield would enable the preservation of coherence of the data qubits while undergoing tens or hundreds of attempts needed for heralded remote entanglement.

\section{Concluding discussion}

We have demonstrated parallelizable, high fidelity atom-photon entanglement in the telecom band in an array of ytterbium-171 atoms. There are many straightforward improvements that could increase both the rate and the fidelity. While still using a microscope objective to collect photons, the collection efficiency could be improved with better mode matching to the fiber and with aberration correction. Additionally, the number and rate of attempts per loop could potentially be increased if the tweezers could be quickly switched off during the telecom pulses to reduce heating and loss. It is also possible to use a tweezer array reservoir~\cite{Endres2016} or continuous reloading~\cite{Norcia2024,Gyger2024} to further mitigate this cost. 

The error budget for our atom-photon Bell state is discussed in Appendix~\ref{app:budget}. Our atom measurement-corrected fidelity $\mathcal{F}_\text{MD}=$ \bellfidelity$\rm{ }$ is limited by mundane effects such as detector dark counts ($\approx0.021$) and finite visibility of the interferometer fringe ($\approx0.016$). Both would disappear by improving the collection efficiency with an optical cavity and by directly studying atom-atom entanglement. The remaining infidelity ($\approx0.013$) is dominated by atomic motion between the time bins ($0.009$) and decay during the Raman pulse ($0.0035$), which could both be improved by Purcell-enhancing the emission rate with an optical cavity.

The use of an optical cavity~\cite{Periwal2021,Deist2022b,Hartung2024,Hu2024,Peters2024,Grinkemeyer2024} or an array of optical cavities~\cite{Trupke2007,Derntl2014,Menon2024,Shadmany2024} would substantially improve the rate. The collection efficiency could be increased from approximately $0.5$\% to $50$\%, giving an approximately $10^2$ ($10^4$) fold boost to the atom-photon (atom-atom) entanglement rate. Moreover, the Purcell enhancement of the desired decay path would drastically improve the effective branching fraction. Appendix~\ref{app:cavity} describes these advantages in detail. Additionally, the improved success rate would have the secondary benefit that the effects of heating and loss are mitigated (see Appendix \ref{app:general} for atomic heating and loss throughout the entanglement attempt loop). Our parallelized scheme based on a fiber array is fully compatible with both nanophotonic~\cite{Menon2024} and free-space high-NA~\cite{Shadmany2024} cavity arrays, for which we anticipate that the atom-atom entanglement rate could approach $\approx N \times 10^5$ s$^{-1}$ which quickly surpasses the clock rate of neutral atom processors~\cite{Graham2022,Bluvstein2023}.  

Finally, we envision many new scientific directions that are uniquely enabled by our work. First, our work is a major step towards quantum networking of atomic processors for distributed~\cite{Jiang2007,Monroe2014} or blind~\cite{Wootters1982,Fitzsimons2017} quantum computing. Our demonstration of `mid-circuit networking' enables efficient scheduling of algorithms that involve both intra- and inter-module two-qubit gates~\cite{Jiang2007,Monroe2014}. Additionally, our use of ytterbium opens the door to establishing a quantum network of optical atomic clocks~\cite{Komar2014,Nichol2022} which could be used for quantum-secured dissemination of a time standard~\cite{Komar2014} or for distributed quantum sensing of spatially-extended phenomena ranging from gravity~\cite{Pfister2016,Borregaard2024,Covey2025} to dark matter~\cite{Wcislo2018,Kennedy2020}.\\

\textit{Acknowledgments}.---We acknowledge Hannes Bernien, Jeff Thompson, Mark Saffman, Elizabeth Goldschmidt, and Brian DeMarco for stimulating discussions. We thank Hannes Bernien and Adam Kaufman for critical reading of the manuscript. We also thank Chris Anderson for generously sharing his SNSPD system. We acknowledge funding from the NSF QLCI for Hybrid Quantum Architectures and Networks (NSF award 2016136); the NSF PHY Division (NSF awards 2112663 and 2339487); the NSF Quantum Interconnects Challenge for Transformational Advances in Quantum Systems (NSF award 2137642); the ONR Young Investigator Program (ONR award N00014-22-1-2311); the AFOSR Young Investigator Program (AFOSR award FA9550-23-1-0059); and the U.S. Department of Energy, Office of Science, National Quantum Information Science Research Centers.

\setcounter{section}{0}



\appendix
\renewcommand\appendixname{APPENDIX}
\renewcommand\thesection{\Alph{section}}
\renewcommand\thesubsection{\arabic{subsection}}



\setcounter{figure}{0}
\renewcommand{\thefigure}{S\arabic{figure}}

\section{Details of the experiment}
\label{app:general}
\subsection{Experiment layout}
\begin{figure}[t!]
	\includegraphics[width=\linewidth]{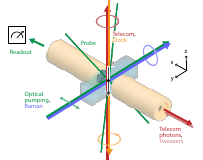}
	\caption{
        \textbf{Layout of the experiment.}
        The experiment is performed in a glass cell and the magnetic field is oriented vertically. Two microscope objectives are arranged horizontally. The first objective sends in the 760-nm optical tweezer array and collects the emitted telecom photons, and the second objective images the tweezer array for diagnostic purposes and performs fluorescence detection of the atoms. Optical pumping via $^3$P$_1$ is performed with a horizontal beam (green) that has horizontal linear polarization~\cite{Huie2023}. Raman rotations of the metastable nuclear spin qubit are performed with a horizontal beam (blue) that has near-circular polarization. The clock beam (orange) and telecom beam (red) both propagate along the B-field axis with circular polarization. They propagate in opposite directions. 
    }
    \label{fig:apparatus}
\end{figure}

The experimental apparatus is shown in Fig.~\ref{fig:apparatus} with light paths drawn for components relevant to our multi-level control of the atom array [see also Fig.~\ref{fig:control}]. Namely, we load atoms into a 20-site linear chain of optical tweezers with spacing $\approx 4.7\,\text{$\mu$m}$, depth $\approx$ 500 $\text{$\mu$K}$, and wavelength $\approx 760\,\text{nm}$. The tweezer array is oriented vertically (along z). Imaging of the atom array in the ground state is performed by collecting photons scattered through the $\rm{F}=3/2$ manifold of the $^3\rm{P}_1$ transition at 120 G with a microscope objective opposite to an identical objective that is used to form the tweezers. The collected photons are imaged onto an EMCCD. See Ref.~\cite{Huie2023} for more details. 

The clock beam that drives the atom from $\ket{g_-}=\ket{^1\rm{S}_0,\rm{ F}=1/2,\rm{ m}_{\rm{F}}=-1/2}$ to  $\ket{\uparrow}=\ket{^3\rm{P}_0,\rm{ F}=1/2,\rm{ m}_{\rm{F}}=1/2}$ is delivered parallel to the magnetic field direction with $\sigma_+$ polarization. Since the tweezer array is orientated along the direction of the magnetic field, the clock beam can be focused down to a waist of approximately 35 $\mu$m without affecting the homogeneity of the Rabi frequency across the tweezer array. The telecom pulse beam that drives the atom from the $\ket{\uparrow}$ to the $\ket{^3\rm{D}_1,\rm{ F}=3/2,\rm{ m}_{\rm{F}}=3/2}$ state propagates in the opposite direction to the clock beam, also with $\sigma_+$ polarization. The waist of the telecom pulse beam is several times larger than that of the clock beam. 

The telecom-band single photons are collected through the same objective that generates the tweezer array, which eliminates most of the common mode mechanical noise from the objective. A dichroic mirror (Thorlabs\textsuperscript{\textregistered} DMSP1020B) is used to separate the 1389 nm photons from the 760 nm tweezer light. Nevertheless, we measure a significant amount of noise photons entering the collection path, likely generated through a frequency down-conversion process inside the objective. Additional cascaded narrow-band filters (CHROMA\textsuperscript{\textregistered} RS1390/30) were therefore placed after the dichroic mirror. \textcolor{black}{In the future, we plan to obviate this problem by quickly blinking off the tweezers during the photon collection window of several hundreds of nanoseconds}. The single photon collection path is a 4-f system that images the atom onto the fiber core with a magnification factor of 5. A picomotor-actuated mirror (Newport\textsuperscript{\textregistered} 8807) is used to adjust the alignment in the Fourier plane.

The Raman beam driving the metastable nuclear transition between $\ket{\uparrow}$ and $\ket{\downarrow}=\ket{^3\rm{P}_0,\rm{ F}=1/2,\rm{ m}_{\rm{F}}=-1/2}$ co-propagates with the optical pumping beam. The Raman beam has a diameter of approximately 1 mm, much larger than the length of the tweezer array (approximately 100 $\mu$m). A polarizing beam splitter (PBS) and a quarter wave plate (QWP) placed right before the glass cell adjusts the polarization of the Raman beam to a ``magic" angle that maximizes the Rabi frequency and $\pi$-pulse fidelity (see Appendix~\ref{app:clockRaman} for further details).

\subsection{Photon detection}
\label{app:collection}

The telecom-band photons are collected using the same objective lens (Special Optics\textsuperscript{\textregistered} customized coating including 556, 760 and 1389 nm) that generates the tweezer array. Due to a chromatic shift of approximately 50 $\mu$m between the focus of a 760 nm beam and a 1389 nm beam, the 1389-nm photons collected through the objective are weakly divergent, and the 4-f imaging system had to be tuned out of focus to compensate for this shift. Even with the focal shift corrected, the chromatic shift still introduces significant aberrations to the spatial mode of the single photons, reducing the collection efficiency because of non-optimal coupling to a single-mode fiber.

The magnification of the single-photon collection system is approximately 5 to best match a single-mode fiber (SMF-28) with a 10 $\mu$m mode field diameter (MFD), which we use to characterize atom-photon entanglement shown in Fig~\ref{fig:photon-meas}. The customized fiber array (Keystone photonics\textsuperscript{\textregistered}) has a quoted MFD of 30 $\mu$m at 1550-nm, and a spacing of 82 $\mu$m per manufacturer specification. However, we observe an MFD of only 10 $\mu$m, similar to that of a single-mode SMF-28 fiber. The smaller than quoted MFD requires the use of a much larger tweezer spacing to match the form factor of the fiber array, limiting the number of parallelized networking tweezer sites to five in this work.

We use superconducting nanowire single-photon detector (SNSPD, Quantum Opus\textsuperscript{\textregistered} Opus One SN156, D10124) with two channels that work between 1300 and 1600 nm for single-photon detection. We optimize the bias currents for the two channels such that they operate close to the saturation point of their quantum efficiency without significantly raising the dark count rate. Typically, we measure dark count rates of approximately 11 Hz for both channels, which we use as a reference to adjust the bias currents sporadically when needed.

\subsection{Estimation of photon collection rate and signal-to-noise ratio}
\label{app:snr}
\begin{figure}[t!]
	\includegraphics[width=0.9\linewidth]{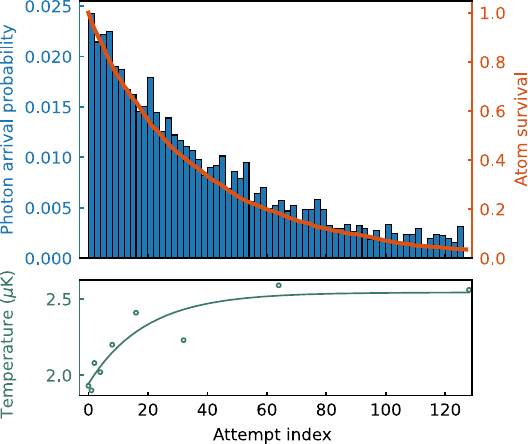}
	\caption{
        \textbf{Measured photon arrival possibility, atom survival simulation, and atomic temperature versus number of attempts.} The blue histogram shows the summarized photon arrival versus attempt index of all the X-basis measurements. The orange curve shows the Monte Carlo simulation of the atom survival versus attempt index under same condition, in agreement with the photon arrival probability. The green data shows atomic temperature after a number of entanglement attemps measured through release-recapture, solid line shows an exponential fit.
        }
    \label{fig:reg_survival}
\end{figure}

To calculate our single-photon collection efficiency, we estimate the number of photons emitted from the atom per attempt loop. For each excitation that drives the atom from $^3$P$_0$ to $^3$D$_1$, there is a 64$\%$ chance that the atom decays back to $^3$P$_0$ to emit a 1389-nm single photon, and 1$\%$ chance that the atom decays to the non-trappable $^3$P$_2$ state and is subsequently ejected. In the case where the atom decays to $^3$P$_0$, there is an approximately 3$\%$ chance that the reset pulses pump the atom to $^3$P$_2$. Given these loss mechanisms per attempt loop, we expect an exponentially decaying atom survival rate as a function of the number of attempt loops. We then ran Monte Carlo simulations of atom survival and single-photon emission in our typical experimental sequence. We ignore loss mechanisms from heating in the trap and finite vacuum lifetimes since our cooling routines keep the atom temperature well below the trap depth (see Fig. \ref{fig:reg_survival} for atomic temperature vs. number of attempts), and the entanglement sequence is much shorter than our tweezer lifetime (approximately 10s). We note that there are several different estimations for this branching ratio from $^3$D$_1$ state~\cite{Cho2012,Porsev1999,Scazza2015,Bettermann2022} and we choose the one that we believe is the most accurate. Using branching ratios of 0.35, 0.64 and 0.01~\cite{Bettermann2022} 
from $^3$D$_1$ to $^3$P$_1$, $^3$P$_0$ and $^3$P$_2$, respectively, we simulated that an average of 23 single photons at 1389 nm are emitted after 128 attempts, with an average atom survival rate of 0.28. We plot the measured photon arrival probabilities alongside the simulated atom survival rate against the number of attempts in Fig.~\ref{fig:reg_survival}, which shows good agreement.

Given the expected number of photons to be emitted per experimental sequence, we estimate the photon collection efficiency to be approximately 0.06$\%$ for XX measurements and 0.3$\%$ for ZZ measurements. We note that these numbers represent the overall efficiency from the photon collection path (75$\%$), time-delay interferometer (20$\%$, XX measurements only), quantum efficiency (70$\%$) of the SNSPD, and also the truncated time window for photon triggering (520 ns). Taking the finite triggering window and atom survival rate into account, we also estimate the number of dark counts collected given the measured dark rates of the SNSPDs, and calculated the signal-to-noise ratio to be approximately 11.6 dB (19.6 dB) for XX (ZZ) measurements.

\label{app:attloop}
\begin{figure*}[t!]
	\includegraphics[width=0.8\linewidth]{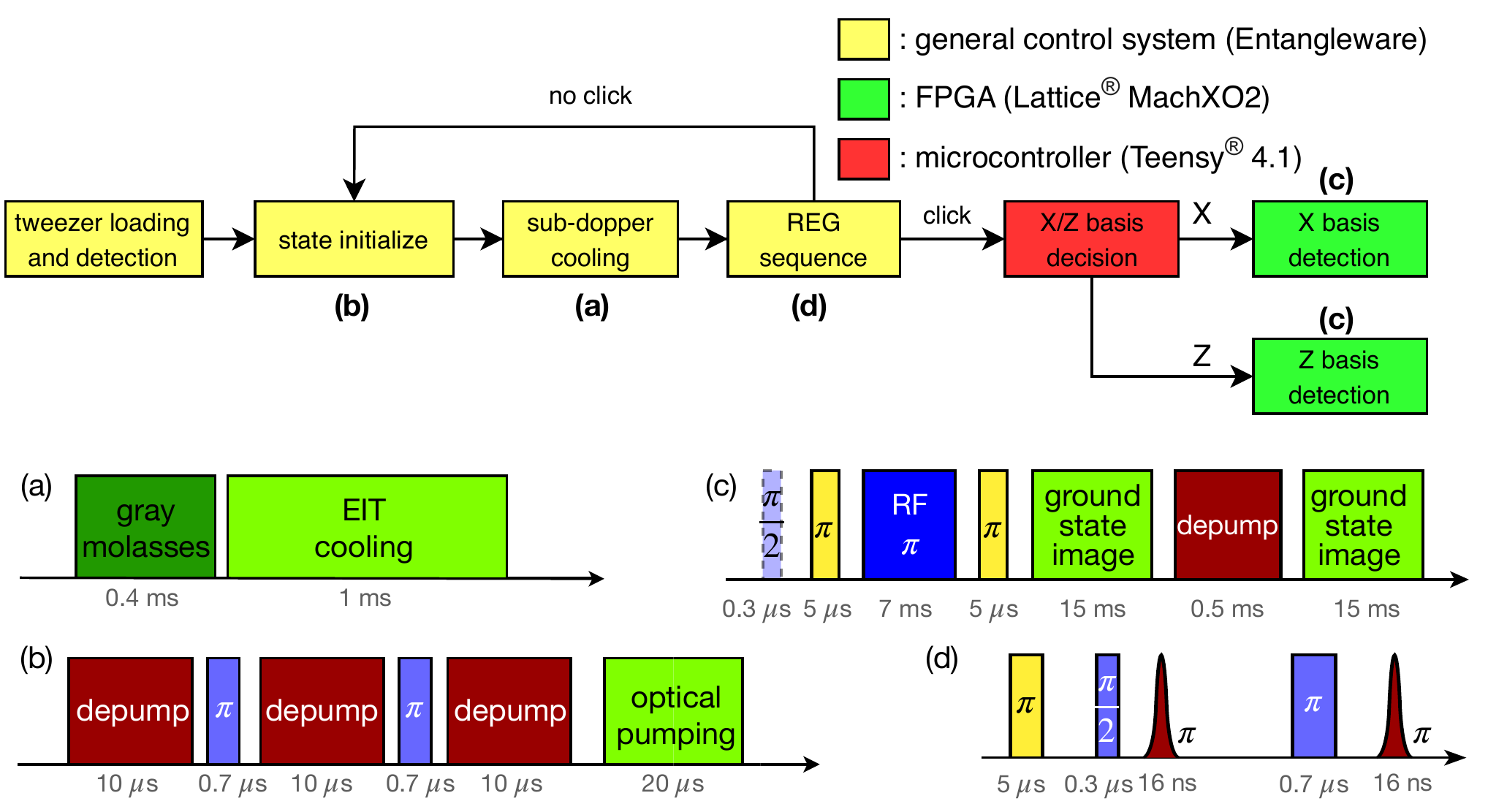}
	\caption{
        \textbf{Experiment sequence for the attempt loop:} (a) Sub-dopper cooling. (b) State initialize to $\left|g_-\right>$. (c) Atomic X(Z)-basis measurement with (without) a Raman $\pi/2$ pulse before atomic state readout. (d) Atom-photon entanglement generation, a clock pulse transfer the atom from $\left|g_-\right>$ to $\left|\uparrow\right>$ at the start.
        }
    \label{fig:attempt_loop}
\end{figure*}

\subsection{Timing and hardware: attempt loop}
\label{app:timing}

As mentioned in the previous section, because of the relatively low photon collection efficiency and limited atom loading efficiency in a single tweezer, the normal single-shot measurement is not suitable for the data collection of this experiment. According to the Monte Carlo simulation, the 1/e lifetime is about 40 attempts, and we perform 128 attempts in the experiment.

Each attempt loop can be divided into three parts: atomic state initialization, sub-Doppler cooling, and entanglement generation. The state initialization depumps the $\ket{\uparrow}$ state atom back to either $\ket{g_-}$ or $\ket{g_+}$ state through the F=3/2 and F=1/2 manifold of the $^3\textrm{P}_1$ state via $^3\textrm{D}_1$. For the atom in the $\ket{\downarrow}$ state, the Raman $\pi$ pulse is applied after the first and followed by the second depump pulse. A second Raman $\pi$ pulse and a third depump pulse are added to remove  the residual population of the atom in $\ket{\downarrow}$ state caused by the limited fidelity of the Raman pulse.

Following the state initialization, the atom in the ground state is cooled by the in-loop cooling sequence, which comprises gray-molasses cooling for 0.4 ms and EIT cooling for 1 ms. A detailed description can be found in Appendix \ref{app:gm-eit}. Without these cooling pulses, the atom temperature rises to approximately 17 $\mu$K after 128 attempts, which results in poor clock pulse fidelity that significantly reduces the entanglement generation rate and the atomic state detection fidelity.

The remote entanglement generation (REG) sequence starts from a clock $\pi$ pulse that transfers the atom from $\ket{g_-}$ to $\ket{\uparrow}$. The Raman $\pi/2$ pulse is applied 5 $\mu$s before the first telecom excitation pulse, to prevent photon leakage into the time-delay interferometer during the first photon detection window. A detailed description of the interferometer can be found in Appendix \ref{app:interferometer}. The Raman $\pi$ pulse is applied about 1.2 $\mu$s after the first telecom excitation pulse, which is long enough to avoid spin-flip errors caused by spontaneous photon emission during the Raman pulse. We estimate the spin-flip error to be approximately 0.35\% for 1.2 $\mu$s delay time and 0.7 $\mu$s Raman $\pi$ pulse time. The spacing between the first and second telecom excitation pulse is adjusted to match the time delay of the interferometer. We note that the delay between the two telecom excitation pulses has to be chosen to balance between reducing spin-flip errors that decrease with a longer delay time, and reducing dephasing errors from atomic motion that increase with a longer delay time. Motional effects are further described in Appendix \ref{app:motion}.

A microcontroller (Teensy\textsuperscript{\textregistered} 4.1) is connected to the single-photon detector to tag the arrival time of the photons relative to the trigger signal sent from the main experimental control system (\textcolor{black}{NI\textsuperscript{\textregistered} FPGA-based digital I/O module controlled via} Entangleware\textsuperscript{\textregistered}) that marks the detection window. After the successful detection of a photon within the predetermined window time (see the following subsection), the microcontroller makes a decision to run either a Z or an X measurement based on the specific time that a photon arrives. The real-time decision is made within 100 ns after the detection window ends, and a trigger signal is sent from the microcontroller to the homemade FPGA control system based on a FPGA evaluation board (Lattice Semiconductor\textsuperscript{\textregistered} LCMXO2-7000HE-B-EVN). After receiving the trigger signal, the homemade control system takes over \textcolor{black}{a subset of channels from} the main experimental control system and executes atomic state detection in the X or Z basis based on the command from microcontroller. Control is given back to the main experimental control system after the execution of the state detection sub-routine. 

For atomic state measurements in the Z basis, a clock $\pi$ pulse transfers atomic population from $\ket{\uparrow}$ to $\ket{g_-}$ immediately after the entanglement generation sequence, and atoms in $\ket{\downarrow}$ will remain in the $^3\textrm{P}_0$ state, which is dark to the ground state imaging pulses. Due to finite atomic temperature, the tweezer depth is adiabatically lowered from 500 to 50 $\mu$K to improve the clock $\pi$ pulse fidelity. To further improve atomic state readout fidelity, we apply an RF pulse to transfer the atom from $\ket{g_-}$ to $\ket{g_+}$ (see Ref.~\cite{Huie2023}) and a second clock $\pi$ pulse to transfer residual population in $\ket{\downarrow}$ to $\ket{g_-}$ that may remain due to limited clock pulse fidelity. To distinguish atom loss from residual population in $\ket{\downarrow}$, a 500 $\mu$s depump pulse transfers the $\ket{\downarrow}$ state back to the ground state for a final round of readout. The X-basis measurement differs from the Z-basis measurement only by a Raman $\pi/2$ pulse at the beginning of the measurement sequence that rotates the projection axis into X.

\subsection{Single photon detection and analysis window}
As discussed in the previous subsection, we wait 1.2 $\mu$s after the first telecom excitation pulse before we apply the Raman rotation pulse to minimize spin-flip error. We thus program a 1.0 $\mu$s window immediately after the telecom excitation pulse during which the microcontroller may trigger the homemade control system to take over the main control system for atomic state detection if a single photon is detected. The arrival times of the single photons are tagged, and we post-select on such detections with a narrower time window within the aforementioned 1.0 $\mu$s experimental detection window as a part of data analysis. The exact position and width of the analysis window are optimized for best parity visibility (see Appendix \ref{app:Bellfidelity} for a detailed discussion on atom-photon Bell state fidelity) by reducing leak-through light from the telecom excitation/Raman pulses with a delayed window start, and by maximizing signal-to-noise ratio with a narrower window width while retaining a number of collection events sufficient for statistical significance. Such analysis is performed separately for XX and ZZ atom-photon measurements due to differences in both experimental sequence and setup, but otherwise applied consistently with the same analysis window for all data measured under the same experimental procedure.

\section{Gray molasses and EIT cooling}
\label{app:gm-eit}

\begin{figure}[t!]
	\includegraphics[width=\linewidth]{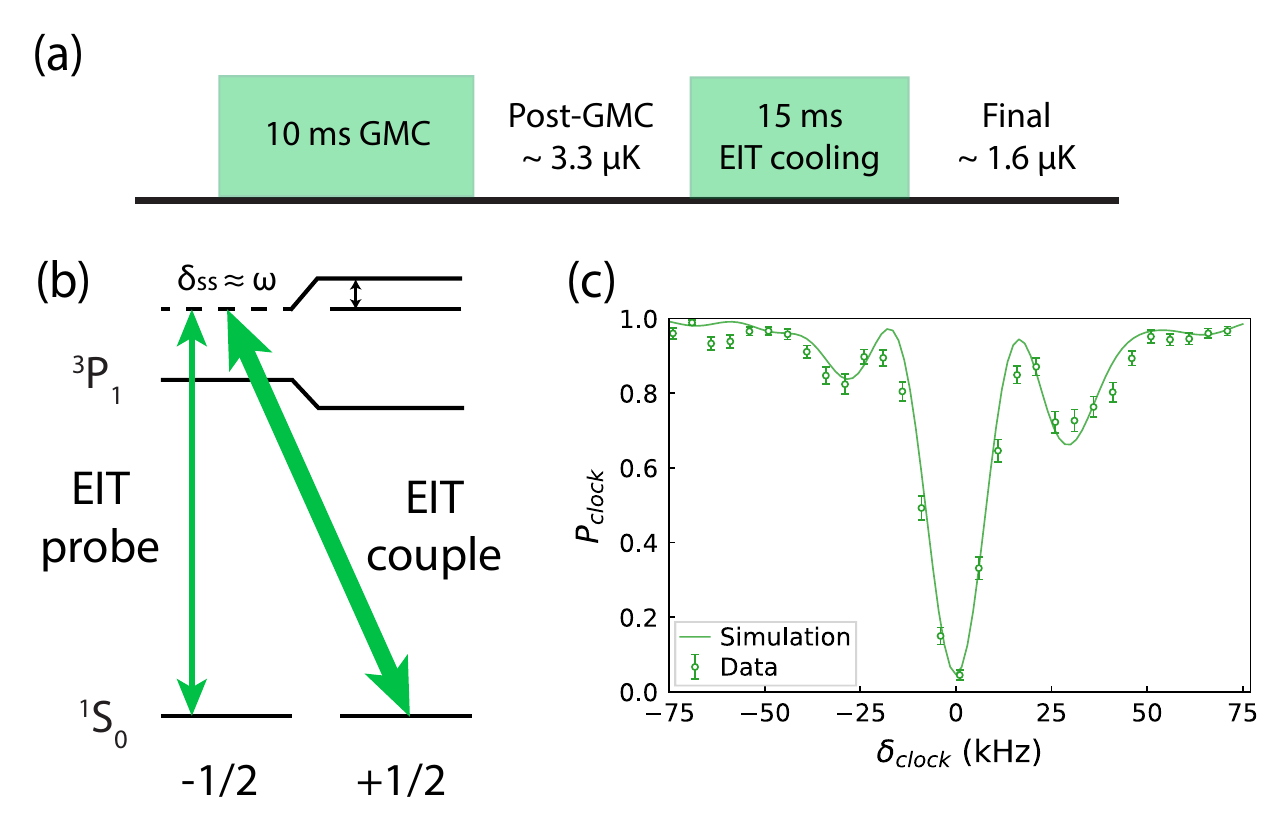}
	\caption{
        \textbf{Gray molasses and EIT cooling.} (a) The sub-Doppler cooling sequence after atom loading. After GM and EIT cooling, the atom temperature is reduced to $\approx 3.3~\mu$K and $\approx 1.6~\mu$K respectively. All atom temperatures are measured using release and recapture. (b) The schematic of implementing EIT cooling. A strong, blue-detuned coupling beam creates a dressed state while a weak, blue-detuned probe beam drives the enhanced red-sideband transition. (c) Sideband-resolved clock frequency spectrum after GM and EIT cooling. The experiment data fits nicely to the simulation curve with $\bar{n}=0.5$.
    }
    \label{fig:GM-EIT}
\end{figure}

After single-atom loading, we perform gray molasses cooling (GMC) using the green MOT beam blue-detuned from ${}^3$P$_1$ states. The measured atom temperature is $\approx3.3$ $\mu$K after applying GMC for 10 ms. After GMC, we then perform electromagnetically induced transparency (EIT) cooling to further reduce the atom temperature~\cite{Lis2023}. EIT cooling is an efficient broad-band ground-state cooling technique that takes advantage of the Fano-like profile of the EIT absorption spectrum to enhance the red-sideband transition while suppressing carrier and blue-sideband transitions \cite{Morigi2000}. Our EIT cooling is performed with OP and probe beams as the EIT coupling and probe beams, respectively, with both beams blue-detuned from ${}^3$P$_1$ states (see Fig.~\ref{fig:GM-EIT}(b)). 
After 15 ms of EIT cooling we measure the atom temperature by release-and-recapture (RNR), giving a final atom temperature of $\approx 1.6~\mu$K which agrees with sideband spectroscopy on the clock transition in Fig. \ref{fig:GM-EIT}(c) which gives an average motional quanta of $\bar{n} \approx 0.5$.

Additionally, we perform short GMC and EIT cooling pulses within each REG loop to reduce the atom temperature after absorbing and emitting 1389 photons in order to achieve higher atom survival and better Bell state fidelity (see section~\ref{app:motion}). With a 0.4-ms GMC pulse and a 1-ms EIT pulse we cool the atom to an event-averaged temperature of $\approx 2.3$ $\mu$K at the beginning of each REG loop (see Fig.~\ref{fig:reg_survival}).

\section{Clock laser system with phase and intensity noise cancellation}
\label{app:clock}
\begin{figure*}[t!]
	\includegraphics[width=0.9\linewidth]{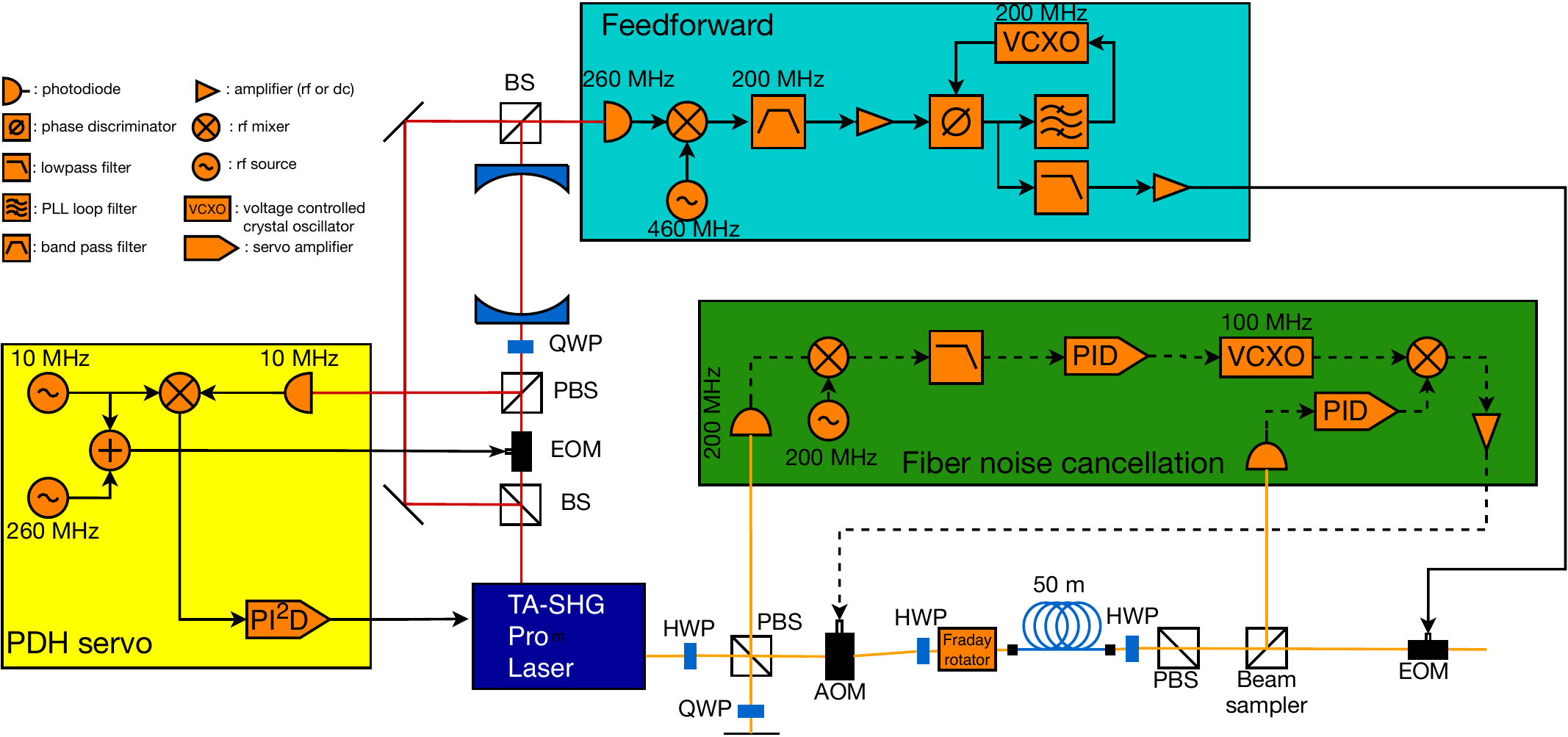}
	\caption{
        \textbf{The SHG clock laser with feed-forward noise cancellation} A Toptica\textsuperscript{\textregistered} TA-SHG laser is offset locked to a high finesse optical cavity (Stable Laser Systems\textsuperscript{\textregistered} notched cavity, $\mathcal{F}\approx 4\times 10^5$) at 1156 nm. The phase noise bump from the frequency doubled 578 nm laser is removed after a 50 m fiber with fiber noise cancellation.
    }
    \label{fig:clocklaser}
\end{figure*}

We adapt phase noise cancellation techniques to our frequency-doubled clock laser. The setup is similar to that described in our previous work~\cite{Li2022}, with several modifications described below and in Fig. \ref{fig:clocklaser}.

To achieve a noise suppression better than 20 dB between 0 and 1 MHz, we have upgraded the design of the phase measurement circuit in Ref. \cite{Li2022} and the feedforward bandwidth is estimated to have increased from 4.5 MHz to about 40 MHz (small signal bandwidth of the phase detector circuit). The upgraded schematic and PCB design can be found in~\cite{FFcircuit}.

Since our laser is offset-locked to an optical cavity, there is no additional acoustic-optic modulator (AOM) needed to get the heterodyne beat signal between the cavity transmission and reflection. For a variable offset frequency, the voltage controlled crystal oscillator (VCXO) can be replaced with a low noise RF source with frequency modulation input. In our clock laser setup, approximately 70 nW of cavity transmission beam beats with a 200 $\mu$W output from the seeding laser on a homemade amplified InGaAs photodiode with 1 GHz bandwidth. The beat signal is further amplified to $-$30 dBm and down-converted in frequency from 260 to 200 MHz using a mixer (Mini-circuits\textsuperscript{\textregistered} ZMF-2+) with a local oscillator (LO) working at 460 MHz. The selection of LO frequency (60 MHz or 460 MHz) is based on the sign of the laser locking frequency offset from the cavity mode.

Aside from the laser setup in Ref. \cite{Li2022}, the 50-m delay fiber is also used for delivering clock laser light to the experiment setup about 10 m away. Fiber noise cancellation is performed separately using fiber tip reflection~\cite{Ma1994}, and a homemade PID servo gives a closed-loop feedback bandwidth of approximately 1 kHz. The intensity noise of the clock laser is stabilized before feedforward by modulating the RF signal amplitude sent to the fiber noise-cancellation AOM, similar to our previous setup~\cite{Li2022}. We do not observe a significant intensity noise contribution from the feedforward electro-optic modulator (EOM).

\begin{figure}[t!]
	\includegraphics[width=0.8\linewidth]{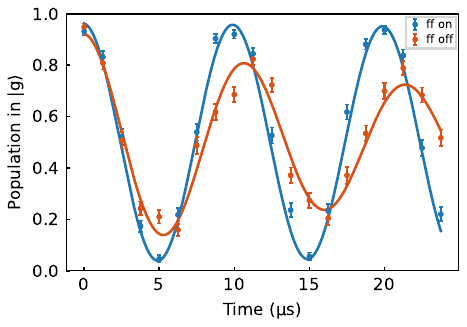}
	\caption{
        \textbf{The clock Rabi oscillation with feed-forward} The blue (orange) data points show the clock Rabi oscillations from $\ket{g_-}$ to $\ket{\uparrow}$ with (without) laser phase noise feed-forward enabled.
    }
    \label{fig:clockrabi}
\end{figure}

Since the cavity transmission light is not frequency-doubled, the feedforward setup is first tested with an infrared cavity transmission. The phase noise cancellation can be understood as tuning the delay time of the laser beam wavefront. Thus, the feedforward system gain should stay the EOM also has a linear dependence of phase modulation gain on laser wavelength. After coarse adjustment using the infrared cavity transmission, we use the atomic Rabi oscillation to perform fine adjustment of both the gain of the EOM amplifier and the electrical delay relative to the SHG laser; the tuning method can be found in~\cite{Li2022}. As shown in Fig~\ref{fig:clockrabi}, we find a significant improvement in the clock pulse fidelity after enabling the phase noise feedforward. We note that the oscillation contrast was not optimized for these curves.

\section{The 1389-nm laser system}
\label{app:1389}

The 1389 nm laser is sourced from a Toptica\textsuperscript{\textregistered} DL-pro that is offset-locked to a reference cavity using the PDH technique. Although the telecom excitation pulse is fast ($\Omega\approx30$ MHz) and shares common mode phase noise with the Raman pulse, we require laser noise feedforward to remove differential phase noise in the emitted photons between the early and late time bin, which decreases the XX-basis measurement contrast with the interferometer (see Appendix~\ref{app:interferometer} for a detailed description of the interferometer). The locking cavity has a finesse of approximately $5\times10^4$, and approximately 10 $\mu$W of transmission light is beat with the laser output for feedforward phase stabilization to be applied after a 10-m delay fiber. We note that this optical delay for feedforward is significantly shorter than the clock laser setup mainly due to group delay difference of the RF band-pass filter after the photodiode. 

\subsection{Remote entanglement generation pulse generation}

We estimate that a Rabi frequency of approximately 30 MHz maximizes atom-photon entanglement fidelity at a B-field of 120-G. A detailed analysis can be found in Appendix~\ref{app:fidelity}. We thus require a $\pi$ pulse time of approximately 15 ns for a square pulse, which is typically difficult to achieve with standard AOMs. At the same time, we wish to minimize leakage through the pulsing AOM on the $\mu$s scale that the REG loop takes place in. Since the two time scales differ by two orders of magnitude, the required extinction level of the telecom excitation pulse is above $10\log_{10}(300^2)\approx50$ dB. 

To achieve both fast modulation and good extinction, we use a fast AOM with 5.3-ns rise time and a modulation bandwidth of approximately 100 MHz (Brimrose\textsuperscript{\textregistered} GPM-400-100-1389). However, the extinction ratio of this model is only approximately 30 dB ($1000:1$). To achieve a better extinction, we double-pass the light through this AOM with a Faraday rotator (Thorlabs\textsuperscript{\textregistered} I1390R5) placed in front of the AOM. We use a Faraday rotator instead of a quarter waveplate because this model only works for horizontally polarized light. The efficiency of the double-pass setup is about 12$\%$ with approximately 2.5-mW of 1389-nm light delivered to the atoms. Due to the double-pass configuration and the limited rise time of the RF switch (5 ns), the total telecom excitation $\pi$-pulse duration is approximately 25-ns, with a 16-ns full width at half maximum (FWHM) and a Rabi frequency of approximately 30 MHz.

\subsection{Stabilized Raman pulses generation}

Since the double-passed AOM for the 1389-nm laser has relatively low efficiency, the 1389 nm laser power is divided with time division multiplexing (TDM) between the Raman and the telecom excitation pulse systems by making use of the 0th order output of the excitation pulse AOM. The 0th order light first passes through an intensity-servo AOM that stabilizes the Raman pulse power delivered to the experiment.

The intensity-stabilized Raman light is first sent to a pulsing AOM with the output coupled to a short polarization-maintaining (PM) fiber that leads to the final set of optics before the atoms. A polarizing beam splitter (PBS) is placed after the short fiber to ensure the purity of the pulse polarization. A slow drift of Raman pulse power after the PBS can be observed due to limited extinction in the PM fiber. Due to the direct proportionality between Raman pulse power and Rabi frequency and the typical length of the experiment (tens of hours), the slow power drift introduces non-negligible pulse errors. To further stabilize the pulse power, we briefly turn the AOM on to send a short pre-pulse at the beginning of every experimental sequence. A fast homemade peak detector with a sample-and-hold circuit measures this pre-pulse and adjusts the Raman pulse intensity servo set point accordingly. With this pulse amplitude feedback enabled, we achieve a pulse intensity stability better than 0.5$\%$.

\section{Raman gates of the metastable nuclear spin}
\label{app:clockRaman}

We perform fast, high-fidelity metastable nuclear spin Raman Rabi oscillations via a single 1389 nm laser beam perpendicular to the magnetic field. Unlike the previous demonstration in \cite{Lis2023}, we intend to operate under a magnetic field of 120G for higher REG and readout fidelity, which can potentially introduce a significant detuning error during Rabi oscillation. Following our previous analysis in \cite{Chen2022}, we look for a ``magic'' condition under which the differential Stark shift between the two nuclear spin states cancels the nuclear Zeeman splitting, thereby eliminating the detuning error during Raman rotations.

\begin{figure}[t!]
	\includegraphics[width=\linewidth]{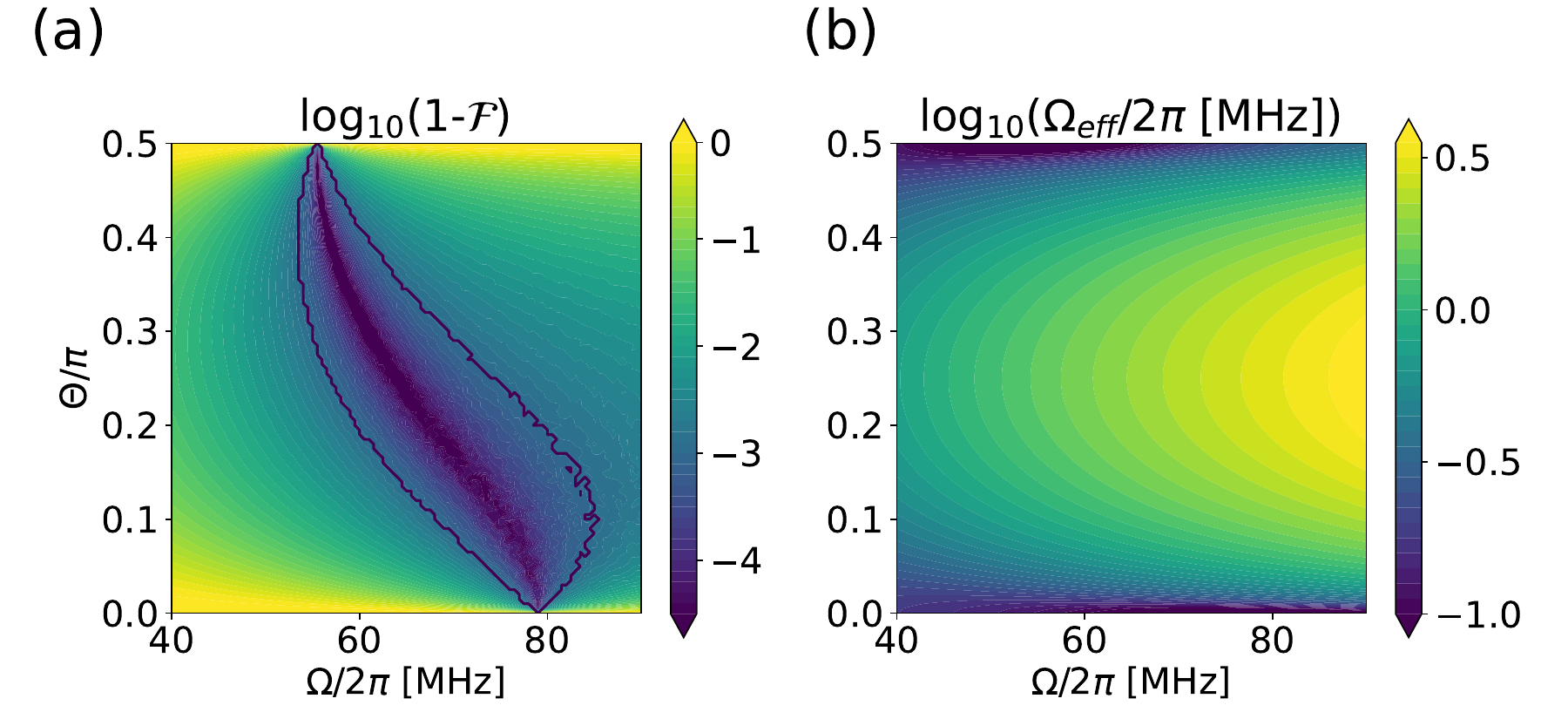}
	\caption{
        \textbf{Magic conditions for metastable Raman transition.} (a) Simulated Raman $\pi$ pulse fidelity under different single-photon Rabi frequencies and phase offsets between horizontal and vertical components. Here we set $B=120$ G, $\Delta(3/2)=+612$ MHz, and consider spontaneous emission during Raman transition. Within the contour we expect the $\pi$-pulse fidelity to exceed 99.9\%.
        (b) The corresponding two-photon Raman Rabi frequency.
    }
    \label{fig:magic_angle}
\end{figure}

We study the magic condition when driving Raman transitions via the $^3$D$_1$, $F=3/2$ intermediate states. To minimize destructive interference between the two possible Raman paths, we fix the phase difference between the horizontal and vertical components of the Raman beam to $\pi/2$ \cite{Jenkins2022} by rotating the polarization of the Raman light with a quarter-wave plate. Under a magnetic field of 120 G and with a blue-detuning from the $m_F=3/2$ substate by 612 MHz, we vary the single-photon Rabi frequency $\Omega$ and the ratio between the amplitude of the horizontal and vertical components $\Omega_H/\Omega_V\equiv\tan\Theta$ (perpendicular and parallel to the B-field, respectively) and found a magic condition that give $< 10^{-3}$ infidelity for $\pi$-rotations, as shown in Fig. \ref{fig:magic_angle}(a). The corresponding two-photon Rabi frequency $\Omega_{\textrm{eff}}$ under different parameter settings is shown in Fig.~\ref{fig:magic_angle}(b).

\section{State preparation and measurement (SPAM)}
\label{app:spam}

\subsection{Atomic state measurement}
\label{app:atom_readout}
Measuring the atomic state encoded in metastable nuclear spin states requires transferring the population in one metastable spin state to the ground state with high fidelity. After transferring, a ground-state readout is performed to detect whether the atom is in the ground state (bright) or in the metastable states (dark). One intuitive approach is to drive a clock $\pi$-pulse between $\ket{\uparrow}$ and $\ket{g_{-}}$ states to shelve the population in $\ket{\uparrow}$ down to the ground state, as shown in Fig.~\ref{fig:control}. However, due to heating during the attempt loop (see Fig.~\ref{fig:reg_survival}), that leads to a decrease in the shelving fidelity, which would bias the atom readout to $\ket{\downarrow}$ as the remaining population in $\ket{\uparrow}$ and the population in $\ket{\downarrow}$ cannot be distinguished. 

To resolve the issue, we perform a two-round de-shelving protocol from $\ket{\uparrow}$ to suppress the remaining population in $\ket{\uparrow}$ to second order. After applying the clock $\pi$-pulse, we apply a ground state RF $\pi$-pulse to transfer the population in $\ket{g_{-}}$ to $\ket{g_{+}}$ (see Ref.~\cite{Huie2023}). The RF $\pi$-pulse is insensitive to temperature, and with a fidelity exceeding 99\%~\cite{Huie2023}. Then we apply another clock $\pi$-pulse to bring the remaining population in $\ket{\uparrow}$ down to $\ket{g_{-}}$, and finally perform ground-state readout. Using this readout sequence, the ZZ-basis measurement in Fig.~\ref{fig:photon-meas}(c) shows no obvious bias to $\ket{\downarrow}$.

\subsection{Graph-based SPAM correction}
We refer to Ref.~\cite{Huie2023} for details on the use of directed, acyclic graphs (DAGs) to correct errors in the preparation and measurement of atomic states. Generically, the DAGs describe the processes accessible to the atoms under an applied experimental sequence, where each node represents a possible state of the physical system and the edges between nodes are weighted by transition probabilities. By traversing the graph, it is then possible to generate models that predict relevant measured probabilities as functions of ``true'' probabilities and other fixed parameters. A fully constrained system of these models is then inverted numerically to obtain corrected quantities. To evaluate uncertainties, sets of sample values of all constant parameters are randomly drawn from appropriate beta distributions. The values reported here are the result of averaging over these Monte Carlo draws, with uncertainties calculated as the standard deviations of the resulting distributions over corrected values.

Specifically, we are interested in the fidelities $\mathcal{F}_\text{clock-$\pi$}$, $\mathcal{F}_\text{Raman-$\pi$}$, and $\mathcal{F}_\text{reg-$\pi$}$ of driving $\pi$-pulses on the $\ket{g_-} \leftrightarrow \ket{\uparrow}$ (clock), $\ket{\uparrow} \leftrightarrow \ket{\downarrow}$ (Raman), and $\ket{\uparrow} \leftrightarrow {}^3\t{D}_1 ~ F = 3/2 ~ \ket{m_F = +3/2}$ (REG) transitions, respectively [see Fig.~\ref{fig:control}]. We are also interested in their counterparts for a $2 \pi$-pulse, $\mathcal{F}_\text{clock-$2\pi$}$, $\mathcal{F}_\text{Raman-$2\pi$}$, and $\mathcal{F}_\text{reg-$2\pi$}$. We construct DAGs for the experimental sequences implementing these rotations, each of which is based on a three-image archetype wherein the $\ket{g_-}$ state is nondestructively imaged three times with various operations to map onto $\ket{g_-}$ as necessary. The first and second images measure the population in the appropriate initial and final atomic states; both states are then collapsed into $\ket{g_-}$ before the third image, the outcome of which is post-selected on to filter out instances where atoms may be missing from the tweezer. We define the above pulse fidelities in terms of the probability of measuring bright outcomes for the first two images ($B_0 \, B_1$), conditioned on a bright outcome for the third image ($B_2$), i.e.
\begin{align}
    \label{eq:pifidelity}
    \mathcal{F}_\text{X-$\pi$}
        &= 1 - \Pr(B_0 \, B_1 | B_2)
    \\
    \label{eq:2pifidelity}
    \mathcal{F}_\text{X-$2\pi$}
        &= \Pr(B_0 \, B_1 | B_2)
,\end{align}
with the understanding that the initial state--$\ket{g_-}$ for the clock transition and $\ket{\uparrow}$ for the Raman and REG transitions--correspond to bright outcomes in the first two images. Additionally, we renormalize these probabilities measured from the REG transition sequence based on estimated branching ratios for the scattering process taking the excited state down to $\ket{g_-}$.

The models derived from these sequence DAGs are functions of the desired ``true'' fidelities of interest, as well as several other parameters such as the tweezer loading fraction, fluorescence imaging discrimination fidelity, post-imaging survival and spin-flip probabilities, etc. whose values we take from previous work \cite{Huie2023}. However, these models also depend on the probability of repumping from $\ket{\uparrow}$ back down to the ${}^1S_0$ manifold for detection $\eta_\text{repump}$ and the probability of tweezer-induced atom loss $\varepsilon_\text{depump}$ .

\begin{figure*}[t!]
    \centering
    \includegraphics[width=\linewidth]{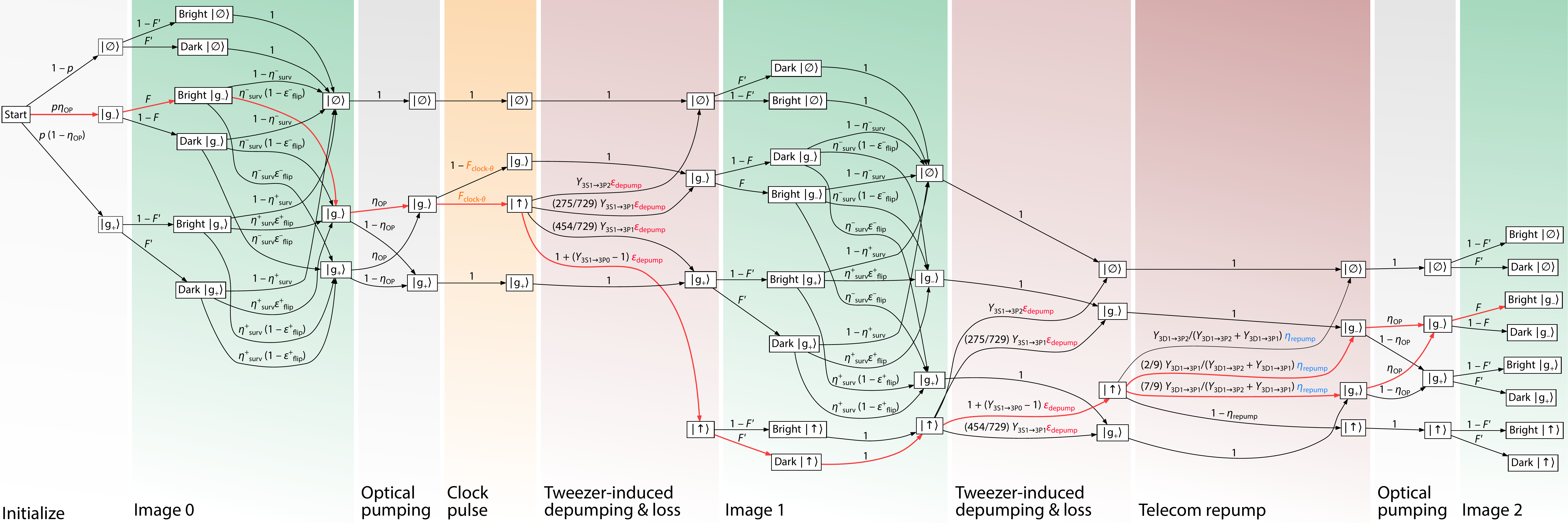}
    \caption{
        \textbf{Sequence DAG for clock transition sequences.}
        Each node represents a possible state of the system and each edge gives the probability of transition to another. $\ket{\varnothing}$ is used as a shorthand for ``empty tweezer or lost atom'' and $\mathcal{F}_\text{clock-$\theta$}$ stands in for either of the clock pulse fidelities $\mathcal{F}_\text{clock-$\pi$}$ or $\mathcal{F}_\text{clock-$2\pi$}$. States are grouped into columns to show distinct physical processes acting on the entire state space, as indicated by labels at the bottom of the figure. SPAM-corrected quantities $\mathcal{F}_\text{clock-$\theta$}$, $\varepsilon_\text{depump}$, and $\eta_\text{repump}$ in the complete system of equations (involving other DAGs not shown) are highlighted in orange, red, and blue, respectively, with all others held fixed at values shown in Table~\ref{tab:spam-consts}. Paths where no SPAM error occurs are shown in red.
    }
    \label{fig:clock-spam}
\end{figure*}

\begin{table}[h]
    \centering
    \caption{
        \textbf{SPAM model parameters.}
        Fixed constants in the $\pi$- and $2\pi$- pulse fidelity SPAM correction models for the clock, Raman, and REG experimental sequences. Values are all taken from previous work \cite{Huie2023} with the exception of the bright and dark state discrimination fidelities $F$ and $F'$, which were re-derived from more recent data; and the branching fractions, which are fixed to results from a mixture of theory and experiment~\cite{Porsev1999,Bettermann2022}.
    }
    \label{tab:spam-consts}
    \begin{tabular}{lrl}
        \hline\hline
        \textbf{Symbol} & \textbf{Value} & \textbf{Description} \\
        \hline
        $p$ & 0.603(5) & Loading fraction \\
        $\eta_\t{OP}$ & 0.98(1) & Ground state optical pumping efficiency \\
        $F$ & 0.97(1) & Bright state discrimination fidelity \\
        $F'$ & 0.996(1) & Dark state discrimination fidelity \\
        $\varepsilon^-_\t{flip}$ & 0.013(2) & Imaging spin-flip probability from $\ket{g_-}$ \\
        $\varepsilon^+_\t{flip}$ & 0.010(6) & Imaging spin-flip probability from $\ket{g_+}$ \\
        $\eta^-_\t{surv}$ & 0.99(1) & Bright state survival probability \\
        $\eta^+_\t{surv}$ & 0.9960(1) & Dark state survival probability \\
        $Y_{{}^3\t{S}_1 \rightarrow {}^3\t{P}_1}$ & 0.367 & ${}^3\t{S}_1 \rightarrow {}^3\t{P}_1$ branching fraction \\
        $Y_{{}^3\t{S}_1 \rightarrow {}^3\t{P}_2}$ & 0.503 & ${}^3\t{S}_1 \rightarrow {}^3\t{P}_2$ branching fraction \\
        $Y_{{}^3\t{D}_1 \rightarrow {}^3\t{P}_1}$ & 0.35 & ${}^3\t{D}_1 \rightarrow {}^3\t{P}_1$ branching fraction \\
        $Y_{{}^3\t{D}_1 \rightarrow {}^3\t{P}_2}$ & 0.01 & ${}^3\t{D}_1 \rightarrow {}^3\t{P}_2$ branching fraction \\
        --- & 275/729 & $\ket{\uparrow} \rightarrow \ket{g_-}$ branch. frac. via ${}^3\t{S}_1 \rightarrow {}^3\t{P}_1$ \\
        --- & 454/729 & $\ket{\uparrow} \rightarrow \ket{g_+}$ branch. frac. via ${}^3\t{S}_1 \rightarrow {}^3\t{P}_1$ \\
        --- & 2/9 & $\ket{\uparrow} \rightarrow \ket{g_-}$ branch. frac. via ${}^3\t{D}_1 \rightarrow {}^3\t{P}_1$ \\
        --- & 7/9 & $\ket{\uparrow} \rightarrow \ket{g_+}$ branch. frac. via ${}^3\t{D}_1 \rightarrow {}^3\t{P}_1$ \\
        \hline\hline
    \end{tabular}
\end{table}

We therefore add two more experimental sequences (and models thereof) to the system in order to fully constrain all degrees of freedom. These sequences focused on a long delay time inserted after initializing atoms in $\ket{\uparrow}$. This was performed with and without telecom light in order to measure the state's lifetime, in the sense of the atom remaining in $\ket{\uparrow}$ before being repumped down to the ground state via decay through ${}^3P_1$ in the former case, and lost from the tweezer via off-resonant scatter to the nearby ${}^3\t{S}_1$ manifold in the latter, assuming a $\pi$-linearly polarized tweezer with respect to the available ${}^3\t{P}_0 \leftrightarrow {}^3\t{S}_1$ transitions. These two sequences introduce additional models to constrain the probabilities $\eta_\text{repump}$ and $\varepsilon_\text{depump}$, respectively, of these events occurring, which fully constrains the system of models used for SPAM correction. These processes additionally depend on the rates of spontaneous decay from the ${}^3\t{D}_1$~\cite{Bettermann2022} and ${}^3\t{S}_1$~\cite{Porsev1999} excited states to the ${}^3\t{P}_J$ manifold, whose values we fix to those provided by a mix of experiment and theory. See Fig.~\ref{fig:clock-spam} and Table~\ref{tab:spam-consts} for an example sequence DAG for the clock transition and a description of associated parameters.

We report SPAM-corrected values of the above fidelities $\mathcal{F}_\text{clock-$\pi$} = \clockpiprob$, $\mathcal{F}_\text{clock-$2\pi$} = \clocktwopiprob$, $\mathcal{F}_\text{raman-$\pi$} = \ramanpiprob$, $\mathcal{F}_\text{raman-$2\pi$} = \ramantwopiprob$, $\mathcal{F}_\text{REG-$\pi$} = \regpiprob$, $\mathcal{F}_\text{REG-$2\pi$} = \regtwopiprob$, $\eta_\text{repump} = \repumpprob$, and $\varepsilon_\text{depump} = \depumpprob$.

\begin{figure}[t!]
    \includegraphics[width=\linewidth]{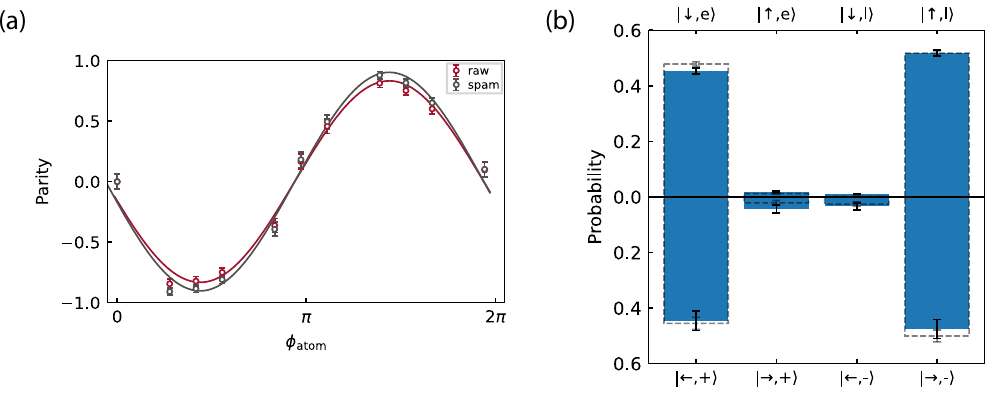}
	\caption{
        \textbf{SPAM-correction comparison}
        (a) Comparison between raw/spam-corrected values for Fig. \ref{fig:photon-meas}(c). Red/gray solid lines are fits to raw/spam-corrected data points, respectively. (b) Comparison between raw/spam-corrected values for Fig. \ref{fig:photon-meas}(b). Solid/dashed bars are raw/spam-corrected correlation probabilities, respectively.
    }
    \label{fig:fidel-spam}
\end{figure}

For SPAM correction relevant to atom-photon Bell state measurement, we extend to appropriate models for the combined measurement of both the atomic state and photon time bin. We apply correction to the atom-photon correlation probabilities (i.e. Fig. \ref{fig:photon-meas}(b)) using pulse fidelities fixed to the corrected values above. For XX measurements, we apply this correction at each of the atomic phase set point, and then refit the resulting parity curve to obtain a new amplitude that we use to calculate the SPAM-corrected fidelity. Fig. \ref{fig:fidel-spam} shows comparisons between raw and spam corrected values for Fig. \ref{fig:photon-meas}(b) and (c). See Appendix \ref{app:Bellfidelity} for a detailed discussion on Bell state tomography.

\section{Time-delay interferometer}
\label{app:interferometer}
\begin{figure*}
    \centering
    \includegraphics[width=0.8\linewidth]{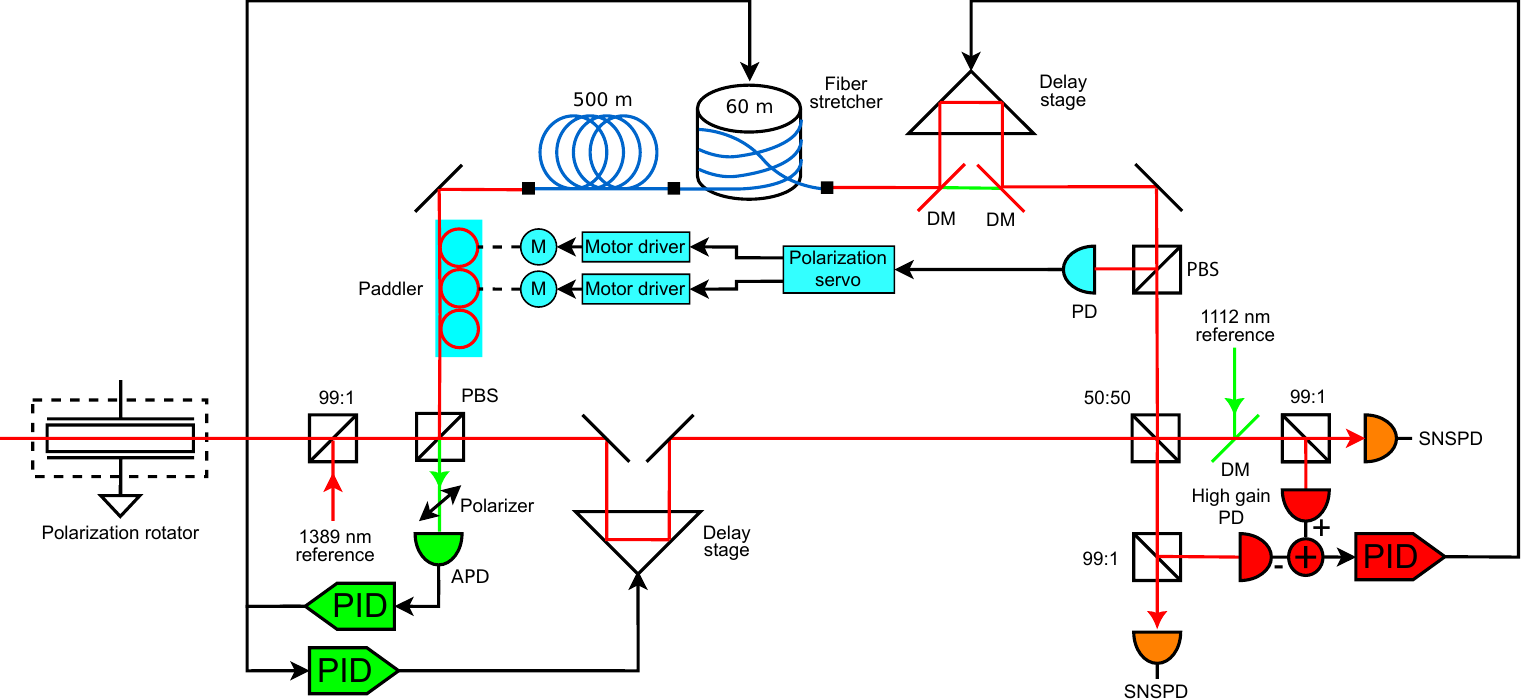}
    \caption{
        \textbf{Time-delay interferometer (TDI).} An amplitude-modulating EOM directs single photons into either the free-space or the fiber-delay arm of the interferometer. The relative phase of the interferometer is locked continuously to a 1112-nm reference, and intermittently to a 1389-nm reference (see text for detail). A motor-driven fiber paddler maintains polarization stability in the fiber.
    }
    \label{fig:inteferometer}
\end{figure*}
\begin{figure}[t!]
    \centering
    \includegraphics[width=\linewidth]{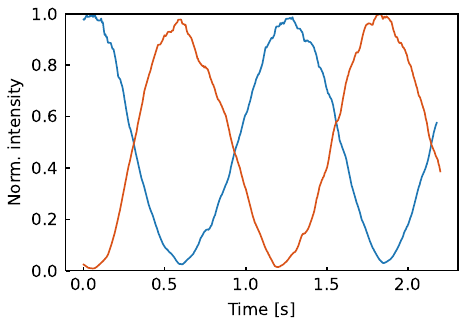}
    \caption{
        \textbf{TDI visibility.} Pulse modulated 1389-nm CW light were sent through the same single-photon collection fiber into the short/long arm of the interferometer. A ramping voltage applied to the fast piezo actuator results in two interference fringes after the 50:50 BS, and are measured by two regular PDs in place of the SNSPDs, shown in red/blue. From the two fringes we extract an averaged visibility of 96.7(2)\%.
    }
    \label{fig:visibility}
\end{figure}
Tomography of the time-bin encoded photonic qubits is performed with a time-delay interferometer (Fig. \ref{fig:inteferometer}), where a 560-m fiber (SMF-28) erases the which-bin information of the arriving photons.

In order to maximize the signal-to-noise ratio for XX basis measurements, we use an amplitude-modulating EOM (Thorlabs\textsuperscript{\textregistered} 	
EO-AM-NR-C3) to vary the polarization of the single photons, such that photons emitted after the first (second) excitation pulse are directed towards the fiber (free-space) arm of the interferometer [see Fig. \ref{fig:schematic}(d)]. Fast modulation of the EOM voltage is driven by a homemade EOM amplifier using a high voltage operational amplifier (APEX\textsuperscript{\textregistered} PA194). 

The 500-m delay fiber is placed inside an insulated box for improved passive temperature stability. Two locking circuits are used for active stabilization against fluctuations in relative path length in the interferometer. First, a 1112-nm reference light (PDH-locked to the same cavity that 1389 nm laser is locked to) of approximately 50 nW is injected with a dichroic mirror (DM). The interference fringe is measured on a homemade InGaAs avalanche photodiode (APD) with a bandwidth of approximately 100 kHz, and subsequently fed into a homemade PID-circuit to stabilize the phase of the interferometer using a piezoelectric fiber stretcher (Paulssion, Inc\textsuperscript{\textregistered} PZ2 high efficiency fiber stretcher) with a homemade low-noise piezo driver that achieves a closed-loop bandwidth of about 10 kHz. The 1112-nm locking beam power is chosen to suppress Raman scattering in the delay fiber to a level significantly below the inherent dark count rate of the SNSPD, which allows the interferometer to be locked to the 1112-nm reference continuously during the REG loop. An additional free-space delay stage (Thorlabs\textsuperscript{\textregistered} VCFL35) with homemade driver and PID servo is employed together with the fiber stretcher to remove the long-term drift of the fiber path, which can otherwise exceed the tuning range of the fiber stretcher (4.4 mm). For context, a coefficient of thermal expansion of $\approx10^{-5}$/$^\circ$C means that the length of a 600-m fiber will change by 6 mm per $^\circ$C.

A slow phase drift of the 1389-nm light can be observed when the interferometer is locked to the 1112-nm reference due to the difference in index of refraction between 1389-nm and 1112-nm inside the delay fiber.
To actively compensate for this differential phase drift, we employ another homemade servo with a sample-and-hold circuit involving the injection of a 1389-nm reference light (PDH-locked to the same cavity as the 1112-nm reference) and a free-space delay stage that shifts only the 1389-nm path (see Fig. \ref{fig:inteferometer}) with a piezo actuator (Thorlabs\textsuperscript{\textregistered} APF705) driven by a homemade low-noise piezo driver. To minimize loss on the single-photon path, three beam samplers with less than 1$\%$ reflection are used to combine the reference light into the path and sample the beat signals of the two detector paths. A balanced detector is used to maintain the locking point stability against power fluctuation of the reference light. The injection light power is about 100 $\mu$W, and less than 10 nW enters the homemade high-gain differential photodiode with 100 Hz bandwidth. The seemingly narrow bandwidth of the photodiode is acceptable because of an even slower closed-loop feedback bandwidth (1 Hz). The 1389-nm reference light is blocked intermittently by shutters, and the servo is briefly paused when collecting single photons from atoms. Additionally, a fast piezo actuator (Thorlabs\textsuperscript{\textregistered} AE0505D08F) is stacked on the APF705, and is triggered in sequence to vary the measurement phase of the single photon without affecting the locking routine outside of the photon detection window.

Polarization stability is achieved by two radio control (rc) servomotors that are attached to the paddles of the manual fiber polarization controller (Thorlabs\textsuperscript{\textregistered} FPC030). When the optical power from the PBS reflection rises above a threshold (around 2$\%$), the polarization servo is switched on to minimize the PBS reflection through lock-in detection. The status of the polarization and phase servo is recorded synchronously with the experiment, and is analyzed to identify and discard experimental shots when the servo misbehaves due to significant environmental interference, such as temperature fluctuation or mechanical noise. Incidentally, these records also tell us if or when either of our 1112-nm or 1389-nm lasers are unlocked or unstable.

With all the locking routines discussed in the previous paragraphs enabled, we send classical 1389-nm pulses with 400-ns pulse widths through the single-photon collection fiber into the interferometer to measure its visibility. A ramping voltage is applied on the fast piezo actuator to obtain an oscillating fringe that we measure with two classical PDs in place of the SNSPDs after the 50:50 BS. Fig. \ref{fig:visibility} shows the normalized intensity measured with the two PDs, from which we extract an averaged visibility of 96.7(2)\%.

\section{Atom-photon Bell state fidelity characterization}
\label{app:Bellfidelity}

We characterize the fidelity of the final atom-photon Bell state comparing with a maximally-entangled Bell state. The state fidelity is characterized by joint measurement of the atomic state and the photon arrival time and TDI output port, in both ZZ- and XX-basis.

Given the atom-photon state density matrix $\rho$, the Bell state fidelity is defined as
\begin{align}
    \mathcal{F} = \frac{1}{2}\left(\rho_{\uparrow E, \uparrow E} + \rho_{\downarrow L, \downarrow L} + 2|\rho_{\uparrow E, \downarrow L}|\right).
\end{align}
The first two terms, $\rho_{\uparrow E, \uparrow E}$ and $\rho_{\downarrow L, \downarrow L}$, are from measuring the atom-photon state in the ZZ basis, given by the probability of the photon arriving in the early (late) bin and the atom readout giving a $\uparrow$ ($\downarrow$) state. As the ZZ correlation between the atom and photon states is classical, one can bypass the interferometer and directly use an SNSPD to measure the arrival time of collected photons. We use this method for ZZ-basis measurement to remove photon loss from the interferometer, which in turn improves the rate at which data can be taken.

The last term, $|\rho_{\uparrow E, \downarrow L}|$, is obtained by measuring in the XX-basis and can be extracted by applying $\pi/2$-rotations on both the photon and atom qubits before performing ZZ basis measurements. The measurement result (parity fringe) can be simplified to
\begin{align}
    P = 2\Re\left(e^{i(\phi_a + \phi_p)}\rho_{\uparrow E, \downarrow L} + e^{i(\phi_a - \phi_p)}\rho_{\uparrow L, \downarrow E}\right)
    \label{eqn:parity}
\end{align}
where $\phi_{a}$ and $\phi_p$ are the phases of the $\pi/2$-rotations on atom and photon qubits. To extract $|\rho_{\uparrow E, \downarrow L}|$, the ``parity scan'' method is often implemented by varying $\phi_a$ and $\phi_p$ together, i.e., $\phi_{a,p} \rightarrow \phi_{a,p} + \Delta \phi$. Scanning $\Delta \phi$ from 0 to $\pi$ gives a sinusoidal oscillation of $P$, the amplitude of which is equal to $|\rho_{\uparrow E, \downarrow L}|$. Parity scans are commonly used when characterizing two-qubit gate fidelities when the phases of the two $\pi/2$-rotations can be precisely controlled.

We control $\phi_a$ by varying the time delay between the final telecom excitation $\pi$-pulse and the Raman $\pi/2$-pulse for the parity scan. Since we use a light shift to eliminate the qubit splitting during the metastable state Raman transitions, the time delay introduces a $\sigma_Z$-rotation at the qubit splitting frequency (Larmor precession). $\phi_p$ can be controlled by changing the phase offset between the two arms of the interferometer. In our case, for experimental simplicity, we instead fix $\phi_p$ and only scan $\phi_a$ from 0 to $2\pi$. Without loss of generality, we take $\phi_p=0$, and up to a global phase offset, take $\rho_{\uparrow E, \downarrow L}\rightarrow |\rho_{\uparrow E, \downarrow L}|$ along with $\rho_{\uparrow L, \downarrow E} \rightarrow |\rho_{\uparrow L, \downarrow E}| e^{i \theta}$.
Then Eq.~\ref{eqn:parity} becomes
\begin{align}
    P_0 = 2\cos\phi_a|\rho_{\uparrow E, \downarrow L}| + 2\cos(\phi_a - \theta)|\rho_{\uparrow L, \downarrow E}|.
    \label{eqn:parity_0}
\end{align}
Depending on the phase offset $\theta$, the amplitude $A$ of the measured parity fringe curve is bounded by 
\begin{align}
    2|\rho_{\uparrow E, \downarrow L}| - 2|\rho_{\uparrow L, \downarrow E}| \leq A \leq 2|\rho_{\uparrow E, \downarrow L}| + 2|\rho_{\uparrow L, \downarrow E}|.
\end{align}

Finally, noticing that $|\rho_{\uparrow L, \downarrow E}| \leq \sqrt{\rho_{\uparrow L, \uparrow L} \cdot \rho_{\downarrow E, \downarrow E}}$, we can upper- and lower-bound $|\rho_{\uparrow L, \downarrow E}|$ using the $ZZ$-measurement results:
\begin{align}
    \frac{A}{2} - \sqrt{\rho_{\uparrow L, \uparrow L} \cdot \rho_{\downarrow E, \downarrow E}} \leq |\rho_{\uparrow L, \downarrow E}| \leq \frac{A}{2} + \sqrt{\rho_{\uparrow L, \uparrow L} \cdot \rho_{\downarrow E, \downarrow E}}
    \label{eqn:XX_bound}
\end{align}
Substituting Eq.~\ref{eqn:XX_bound} into Eq.~\ref{eqn:parity} we can bound the atom-photon Bell state fidelity. As our ZZ measurement gives values of $\rho_{\uparrow L, \uparrow L}$ and $\rho_{\downarrow E, \downarrow E}$ close to zero , the upper- and lower-bounds are reasonably tight -- comparable with our statistical uncertainties. We quote our fidelity upper- and lower-bound using the format $\frac{1}{2}\left(\rho_{\uparrow E, \uparrow E} + \rho_{\downarrow L, \downarrow L} + A\right) \pm \sqrt{\rho_{\uparrow L, \uparrow L} \cdot \rho_{\downarrow E, \downarrow E}}_{\text{bound}}$ with additional statistical uncertainties.

\section{Fidelity estimation of entanglement generation and tomography}
\label{app:fidelity}

\subsection{Spontaneous emission and off-resonant telecom excitation}
\label{app:sp_em}

The atom-photon entanglement fidelity is ultimately limited by spontaneous emission during the telecom $\pi$-pulse and the residual off-resonant excitation from the $\ket{\downarrow}$ state to the $\ket{{}^3D_1, F=3/2, m_F=1/2}$ state. To suppress spontaneous emission, we note that when the telecom Rabi frequency is comparable to or larger than the transition linewidth, the contribution to infidelity from spontaneous emission during a telecom $\pi$-pulse is inversely proportional to the telecom Rabi frequency. This indicates that a larger Rabi frequency with a shorter pulse can reduce the spontaneous emission during a $\pi$-pulse.

We want the off-resonant excitation from $\ket{\downarrow}$ to $\ket{{}^3D_1, F=3/2, m_F=1/2}$ to be minimized at the $\pi$-time of the stretched $\ket{\uparrow}\leftrightarrow\ket{{}^3D_1, F=3/2, m_F=3/2}$ transition, i.e., the off-resonant transition should evolve through an integer multiple of $2\pi$. This requires the magnetic field $B$ and Rabi frequency $\Omega$ on the stretched transition to satisfy

\begin{align}
    \sqrt{(\Omega/\sqrt{3})^2 + (g\mu_B B)^2} \cdot \frac{\pi}{\Omega} = 2m \pi,~m = 1, 2, \cdots
\end{align}

\begin{figure}[t!]
    \centering
    \includegraphics[width=\linewidth]{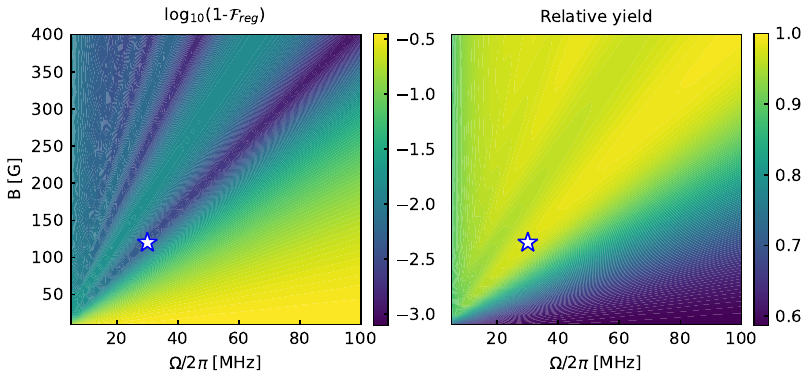}
    \caption{\textbf{Atom-photon entanglement infidelity and relative yield under different telecom Rabi frequencies and magnetic fields after post-selecting on receiving $\geq$1 1389-nm photons.} We find that using higher Rabi frequency and operating under higher B-field strength result in better atom-photon entanglement fidelities. Under a fixed $B$-field strength, there exist several local minima of atom-photon entanglement infidelity where the off-resonant transition from $\ket{\uparrow}$ goes through multiples of $2\pi$. The star indicates the settings used our experiment.}
    \label{fig:REG_infidelity}
\end{figure}

Here $\sqrt{3}$ comes from a Clebsch-Gordan coefficient, $g=1/3$ is the Landé $g$-factor for $^3\text{D}_1$ manifold and $\mu_B$ is the Bohr magneton. We further assume post-selection on receiving $\geq$1 REG photons in the collection window, in agreement with our experiment in practice. Figure \ref{fig:REG_infidelity} shows the simulated minimum entanglement infidelity and relative yield after post-selection under different settings of $\Omega$ and $B$. The multiple regions surrounding local minima in the fidelity plot correspond to the settings where the residual off-resonant transition is mitigated, and within one local minimum region the infidelity decreases as $\Omega$ and $B$ increase. We choose to operate at $\Omega/2\pi=30$ MHz and $B=120$ G, giving a maximum REG fidelity of 99.7\% with post-selection. The remaining infidelity after post-selection is from double excitation within and after the REG attempt loop.

\subsection{Effects from atom motion between time bins}
\label{app:motion}

One specific challenge for time-bin entanglement with atomic qubits, compared with solid-state spin qubits, is the recoil on motional states from photon absorption and emission. Let $\Delta \vec{k}$ be the change in atom momentum after REG photon absorption and emission (here we set $\hbar = 1$), $H_m = \omega a^{\dagger}a$ be the Hamiltonian of the harmonic trap potential, and $\rho_{m}$ be the initial motional state density matrix of the atom. Assuming the early bin and late bin are separated by time $\Delta t$ (including photon collection time $t_d$ and Raman $\pi$-time $\tau^{m}_{\pi}$), the atom-photon-motion state after the REG sequence is

\begin{align}
    \frac{1}{2} & \left( \ket{\uparrow, E} \bra{\uparrow, E} e^{-i H_m \Delta t} e^{i \Delta \vec{k} \cdot \vec{x}} \rho_{m} e^{-i \Delta \vec{k} \cdot \vec{x}} e^{i H_m \Delta t} \right. \nonumber\\
    & + \ket{\uparrow, E} \bra{\downarrow, L} e^{-i H_m \Delta t} e^{i \Delta \vec{k} \cdot \vec{x}} \rho_{m} e^{i H_m \Delta t} e^{-i \Delta \vec{k} \cdot \vec{x}} \nonumber\\
    & + \ket{\downarrow, L} \bra{\uparrow, E} e^{i \Delta \vec{k} \cdot \vec{x}} e^{-i H_m \Delta t} \rho_{m} e^{-i \Delta \vec{k} \cdot \vec{x}} e^{i H_m \Delta t} \nonumber\\
     & \left. + \ket{\downarrow, L} \bra{\downarrow, L} e^{i \Delta \vec{k} \cdot \vec{x}} e^{-i H_m \Delta t} \rho_{m} e^{i H_m \Delta t} e^{-i \Delta \vec{k} \cdot \vec{x}}\right),
     \label{eqn:dm_motion}
\end{align}
with $\omega$ being the trap frequency. The motional states can be entangled with atom and photon qubits and reduce the atom-photon entanglement fidelity. For fixed $\Delta t$ and $\omega$, higher temperature leads to increased dephasing in the atom-photon entangled state. The analytical expression of the infidelity can be found in Ref.~\cite{Saha2024}.

\begin{figure}[t!]
	\includegraphics[width=\linewidth]{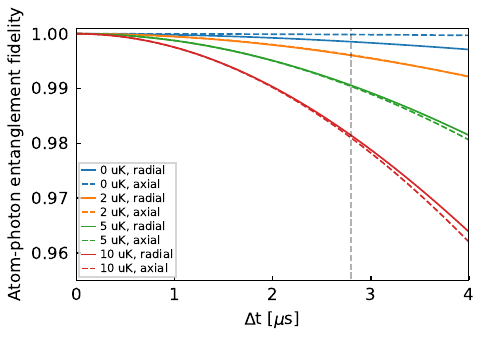}
	\caption{
        \textbf{Atom-photon entanglement fidelity degradation from radial and axial atom-motion entanglement as a function of time-bin separation.} As our $\Delta\vec{k}$ has components along both radial and axial directions, we calculate effects on the overall atom-photon Bell state fidelity from both axial and radial thermal states for different temperatures. The grey dashed line indicates $\Delta t = 2.8~\mu$s which we use in our experimental sequence. We expect a $\approx 0.9\%$ fidelity reduction from both radial and axial motion combined.
    }
    \label{fig:atom-motion}
\end{figure}

One way to eliminate this effect is to set $\Delta t$ to be an integer multiple of the trap period, as used in Ref. \cite{Saha2024}. For tweezers, however, the trap period is much longer than  the typical REG sequence length, and the inhomogeneity across the array leads to decoherence over the array as the system size increases. Instead, we work in the regime where $\Delta t$ is much smaller than the trap period, ``freezing'' the atom movement between the early and late time bin. In this regime, the difference between axial and radial motion is negligible at finite temperature. Figure \ref{fig:atom-motion} shows simulated atom-photon Bell state fidelity as a function of $\Delta t$ for various axial and radial temperatures. We choose to operate at $\omega_{\text{radial}}/2\pi = 32$ kHz, $\omega_{\text{axial}}/2\pi = 3$ kHz and $\Delta t = 2.8$ $\mu$s. Given our expected atom temperature of $T\approx 2.33$ $\mu$K during our attempt loop, we anticipate the upper limit of our atom-photon Bell state fidelity to be $\approx$ 99.1\%.  

\section{Bell state fidelity error budget}
\label{app:budget}

In Table~\ref{table:errorbudget} we summarize the major sources of error contributing to measured Bell state infidelity. The largest source of error is SNSPD dark counts considering the SNSPD dark count rate and our collection efficiency (see Appendix~\ref{app:snr}), leading to an error contribution of 2.1(5)\%.
Imperfect interferometer visibility is expected to reduce fidelity by 1.6(1)\% (see Appendix~\ref{app:interferometer}). 
We expect 0.9\% error to be contributed by the atomic motion between the two time bins at our expected average temperature (see Appendix~\ref{app:motion}).
Spontaneous emission during the Raman $\pi$ pulse cause the spin-flip error of 0.35\% (see Appendix~\ref{app:timing}). Double-excitation during 1389 photon generation is expected to contribute 0.3\% (see Appendix~\ref{app:sp_em}). 

This total error should be compared with the infidelity our atom measurement-corrected Bell state fidelity, $1-\big($\bellfidelity$\big)\approx0.05$. We see that the two values are in good agreement.

\begin{table} [h]
    \centering
\caption{Error budget for entanglement generation and measurement.}
\label{table:errorbudget}
   \begin{tabular}{l|r} \hline 
         & \textbf{Fidelity}\\ 
        \textbf{Source of error (see Appendix $X$)} & \textbf{Error\hspace{2mm} }\\ \hline 
         \hline        
          SNSPD dark counts (\ref{app:snr})        & $0.021(5)$\phantom{}\\ 
         \hline 
         Interferometer visibility (\ref{app:interferometer}) & $ 0.016(1) $\phantom{}\\ \hline
         Atom motion between time bins (\ref{app:motion})    & $0.009$\phantom{}\\ 
          \hline 
       Spin-flip during Raman (\ref{app:timing})               & $0.0035$\phantom{}\\ 
         \hline 
       REG double-excitation (\ref{app:sp_em})               & $0.003$\phantom{}\\ 
         \hline
         \textbf{TOTAL}                         & \textbf{0.053(5)}\phantom{} \\ \hline
    \end{tabular}

\end{table}

\section{Expected performance with an optical cavity}
\label{app:cavity}

Here we 
suppose a hypothetical optical cavity to enhance photon collection and estimate the corresponding improvement to the average Bell state generation rate
following the methods developed in~\cite{Young2022, LiThompson2024}.

\subsection{Cavity-enhanced collection efficiency and branching ratio}

Coupling an atomic transition to an optical cavity can improve overall photon collection efficiency by causing the atom to preferentially emit photons not only into the cavity mode, but along the specific decay pathways resonant with that mode as well
~\cite{Young2022}.

The collection efficiency $\eta$ of a two-level atom in a resonant cavity is given by 
\begin{align*}
    \eta & =  \left(\frac{C}{C+1}\right) \left(\frac{\kappa}{\kappa+\Gamma}\right) \left(\frac{T}{T+L} \right),
\end{align*}
where  $T$ is the transmission of the output mirror of a one-sided cavity, 
$\kappa$ the photonic mode decay rate of the cavity,
$L$ the total cavity loss, 
and
\begin{align*}
    C &= \frac{4g^2}{\kappa \Gamma} = \frac{6}{\pi^3} \frac{\lambda^2}{w^2} F(T,L)
\end{align*}
is the single-atom cooperativity with finesse $F(T,L)$.

One can optimize the collection $\eta$ over mirror properties $(T,L)$. 
In our case of interest, considering the telecom transition $\lambda=1389\text{~nm}$ with $\Gamma^{^3\text{D}_1}=1/\tau^{^3\text{D}_1}$, 
and a near-concentric Fabry-Perot cavity with curvature $R_c=8\text{~mm}$, $g_\text{cav}=1-L_\text{cav}/R_c=-0.9$ (yielding a Gaussian mode waist $w_0\approx 27.8~\mu\text{m}$), 
$T_\text{in}=20\text{~ppm}$ and assumed loss of $L=100\text{~ppm}$, 
we find $\eta \approx 0.51$ can be achieved for $T\approx 1070\text{~ppm}$, with a corresponding $C\approx 2.6$.

While the above collection considers a single decay channel (as relevant for stretched transitions), for more realistic systems with multiple possible decay channels the emission
along the desired transition is correspondingly reduced.
In our case we have a branching ratio (in free space) of $f=\{0.64:0.35:0.01 \}$, with relative probability $f_0=0.64$ of emission into the desired path.
However, by coupling the telecom transition to the cavity mode, we expect the modified branching rato to be 
$f'=\{ \frac{0.64(1+C)}{(0.64(1+C))+0.36},
\frac{0.35}{(0.64(1+C))+0.36},
\frac{0.01}{(0.64(1+C))+0.36}\}$.
Then considering the above optimized $T\approx 1070\text{~ppm}$,
we find $f_0' \approx 0.86$,
and a corresponding modified collection efficiency of $\eta'=f_0'\eta \approx 0.44$.

\subsection{Expected success probability and rate of remote atom-atom entanglement generation rate}

With the cavity-enhanced collection efficiency, we can estimate the expected success probability of remote atom-atom entanglement between two identical experimental setups, given by
\begin{align*}
    P_{aa}=\frac12 (\eta' \eta_\text{det})^2,
\end{align*}
where $ \eta_\text{det}$ characterizes the efficiency of the detection path.
Given a SNSPD detection efficiency around $\eta_\text{det} \approx 0.8$ and ignoring the fiber loss for short to intermediate networks at telecom wavelengths, $P_{aa} \approx 0.063$ can be reached.

To estimate the resulting remote atom-atom entanglement generation rate, 
note that the expected binomially distributed process of a successful Bell-state measurement will give on average 
$N P_{aa}$ out of $N$ attempts.
Thus in our case we expect $N^*=1/P_{aa}\approx 17$ attempts per success.

For a single atom coupled to the cavity mode, the required experimental time to generate entanglement can be parameterized as
\begin{align*}
    T_{aa} & = N_\text{cycle}(n_\text{a} (T_\text{init}+T_\text{REG}) + T_\text{cool} ),
\end{align*}
where $n_\text{a}$ attempts of state initialization and remote entanglement generation [of duration $(T_\text{init}+T_\text{REG})$] are made before a cooling cycle [of duration $T_\text{cool} $], 
with the requirement  $(N_\text{cycle} n_\text{a}) = N^*$ to successfully generate entanglement on average.
As $T_\text{cool}\gg T_\text{init,REG}$ for typical systems, 
even with optimized cooling time of $T_\text{cool}\approx 0.1\text{~ms}$ this sets a ceiling for the entanglement generation rate at $T_{aa}^{-1}\lesssim 10^3-10^4\text{/s}$.

However, by combining an array of atoms with local control and leveraging the high collection efficiency afforded by the cavity, it becomes possible to exceed $T_{aa}^{-1}\gtrsim 10^4\text{/s}$. 
In such a case, the large initial time cost (to cool and/or move an array of atoms in parallel) can be compensated by the sufficiently large number of atoms made available to (sequentially) undergo REG attempts.
To estimate the average Bell pair generation rate, 
we adopt the method developed in~\cite{LiThompson2024} which envisions an array of 
a few hundred atoms coupled to the cavity,
with strong local light control to far-detune all but a single atom undergoing a REG attempt.
Then the average Bell pair generation rate after $m$ rounds of array-entangling attempts is given by
\begin{align*}
    R_\text{bp}&=\frac{\sum_{i=1}^m N_i P_{aa}}{T_\text{move}+m (T_\text{init}+T_\text{depump})+
    T_\text{REG}\sum_{i=1}^m N_i},
\end{align*}
where $N_i=N_{i-1}(1-P_{aa})$ with $N_1=N_a$.

 The scheme in ~\cite{LiThompson2024} utilizes a cooling/moving time of $T_\text{move}\approx 100~\mu\text{s}$ and a reset time (to depump $^3P_0$ to $^1S_0$ and optically pump in $^1S_0$) of  $T_\text{depump}\approx 6.1~\mu\text{s}$, as well as an array size of $N_a=204$ [chosen to exceed $\approx (T_\text{move}+T_\text{init})/T_\text{REG} \approx 100$ to offset the large initial time cost, while maximally filling the cavity mode volume to achieve interaction strength $g\gtrsim 0.9 g_\text{max}$].
Thus combining the numbers used in this work for atom-photon entanglement,
\begin{align*}
    T_\text{init} &= t_\pi^\text{clock}+t_{\pi/2}^\text{Raman}
    \approx 5.3~\mu\text{s},\\
    T_\text{REG} & = 2(t_\pi^\text{REG}+t_\text{wait})+  t_{\pi}^\text{Raman}\\
    & \approx 1.752~\mu\text{s},
\end{align*}
where we have set $t_\text{wait}=6.454/\kappa \approx 0.55~\mu\text{s}$ 
to collect $\gtrsim 0.99$ of leaked cavity mode,
we find a rate of $2.6\times 10^4\text{/s}$ after $m=1$ round, 
reaching a maximum of $\approx3.1\times 10^4\text{/s}$ after 5 rounds.

\bibliography{library}

\begin{thebibliography}{81}%
\makeatletter
\providecommand \@ifxundefined [1]{%
 \@ifx{#1\undefined}
}%
\providecommand \@ifnum [1]{%
 \ifnum #1\expandafter \@firstoftwo
 \else \expandafter \@secondoftwo
 \fi
}%
\providecommand \@ifx [1]{%
 \ifx #1\expandafter \@firstoftwo
 \else \expandafter \@secondoftwo
 \fi
}%
\providecommand \natexlab [1]{#1}%
\providecommand \enquote  [1]{``#1''}%
\providecommand \bibnamefont  [1]{#1}%
\providecommand \bibfnamefont [1]{#1}%
\providecommand \citenamefont [1]{#1}%
\providecommand \href@noop [0]{\@secondoftwo}%
\providecommand \href [0]{\begingroup \@sanitize@url \@href}%
\providecommand \@href[1]{\@@startlink{#1}\@@href}%
\providecommand \@@href[1]{\endgroup#1\@@endlink}%
\providecommand \@sanitize@url [0]{\catcode `\\12\catcode `\$12\catcode `\&12\catcode `\#12\catcode `\^12\catcode `\_12\catcode `\%12\relax}%
\providecommand \@@startlink[1]{}%
\providecommand \@@endlink[0]{}%
\providecommand \url  [0]{\begingroup\@sanitize@url \@url }%
\providecommand \@url [1]{\endgroup\@href {#1}{\urlprefix }}%
\providecommand \urlprefix  [0]{URL }%
\providecommand \Eprint [0]{\href }%
\providecommand \doibase [0]{https://doi.org/}%
\providecommand \selectlanguage [0]{\@gobble}%
\providecommand \bibinfo  [0]{\@secondoftwo}%
\providecommand \bibfield  [0]{\@secondoftwo}%
\providecommand \translation [1]{[#1]}%
\providecommand \BibitemOpen [0]{}%
\providecommand \bibitemStop [0]{}%
\providecommand \bibitemNoStop [0]{.\EOS\space}%
\providecommand \EOS [0]{\spacefactor3000\relax}%
\providecommand \BibitemShut  [1]{\csname bibitem#1\endcsname}%
\let\auto@bib@innerbib\@empty
\bibitem [{\citenamefont {Gisin}\ \emph {et~al.}(2002)\citenamefont {Gisin}, \citenamefont {Ribordy}, \citenamefont {Tittel},\ and\ \citenamefont {Zbinden}}]{Gisin2002}%
  \BibitemOpen
  \bibfield  {author} {\bibinfo {author} {\bibfnamefont {N.}~\bibnamefont {Gisin}}, \bibinfo {author} {\bibfnamefont {G.}~\bibnamefont {Ribordy}}, \bibinfo {author} {\bibfnamefont {W.}~\bibnamefont {Tittel}},\ and\ \bibinfo {author} {\bibfnamefont {H.}~\bibnamefont {Zbinden}},\ }\bibfield  {title} {\bibinfo {title} {{Quantum cryptography}},\ }\href {https://doi.org/10.1103/RevModPhys.74.145} {\bibfield  {journal} {\bibinfo  {journal} {Rev. Mod. Phys.}\ }\textbf {\bibinfo {volume} {74}},\ \bibinfo {pages} {145} (\bibinfo {year} {2002})}\BibitemShut {NoStop}%
\bibitem [{\citenamefont {Pirandola}\ \emph {et~al.}(2020{\natexlab{a}})\citenamefont {Pirandola}, \citenamefont {Andersen}, \citenamefont {Banchi}, \citenamefont {Berta}, \citenamefont {Bunandar}, \citenamefont {Colbeck}, \citenamefont {Englund}, \citenamefont {Gehring}, \citenamefont {Lupo}, \citenamefont {Ottaviani}, \citenamefont {Pereira}, \citenamefont {Razavi}, \citenamefont {{Shamsul Shaari}}, \citenamefont {Tomamichel}, \citenamefont {Usenko}, \citenamefont {Vallone}, \citenamefont {Villoresi},\ and\ \citenamefont {Wallden}}]{Pirandola2019}%
  \BibitemOpen
  \bibfield  {author} {\bibinfo {author} {\bibfnamefont {S.}~\bibnamefont {Pirandola}}, \bibinfo {author} {\bibfnamefont {U.~L.}\ \bibnamefont {Andersen}}, \bibinfo {author} {\bibfnamefont {L.}~\bibnamefont {Banchi}}, \bibinfo {author} {\bibfnamefont {M.}~\bibnamefont {Berta}}, \bibinfo {author} {\bibfnamefont {D.}~\bibnamefont {Bunandar}}, \bibinfo {author} {\bibfnamefont {R.}~\bibnamefont {Colbeck}}, \bibinfo {author} {\bibfnamefont {D.}~\bibnamefont {Englund}}, \bibinfo {author} {\bibfnamefont {T.}~\bibnamefont {Gehring}}, \bibinfo {author} {\bibfnamefont {C.}~\bibnamefont {Lupo}}, \bibinfo {author} {\bibfnamefont {C.}~\bibnamefont {Ottaviani}}, \bibinfo {author} {\bibfnamefont {J.~L.}\ \bibnamefont {Pereira}}, \bibinfo {author} {\bibfnamefont {M.}~\bibnamefont {Razavi}}, \bibinfo {author} {\bibfnamefont {J.}~\bibnamefont {{Shamsul Shaari}}}, \bibinfo {author} {\bibfnamefont {M.}~\bibnamefont {Tomamichel}}, \bibinfo {author} {\bibfnamefont {V.~C.}\ \bibnamefont {Usenko}}, \bibinfo {author} {\bibfnamefont
  {G.}~\bibnamefont {Vallone}}, \bibinfo {author} {\bibfnamefont {P.}~\bibnamefont {Villoresi}},\ and\ \bibinfo {author} {\bibfnamefont {P.}~\bibnamefont {Wallden}},\ }\bibfield  {title} {\bibinfo {title} {{Advances in quantum cryptography}},\ }\href {https://doi.org/10.1364/AOP.361502} {\bibfield  {journal} {\bibinfo  {journal} {Adv. Opt. Photonics}\ }\textbf {\bibinfo {volume} {12}},\ \bibinfo {pages} {1012} (\bibinfo {year} {2020}{\natexlab{a}})}\BibitemShut {NoStop}%
\bibitem [{\citenamefont {Pirandola}\ \emph {et~al.}(2020{\natexlab{b}})\citenamefont {Pirandola}, \citenamefont {Andersen}, \citenamefont {Banchi}, \citenamefont {Berta}, \citenamefont {Bunandar}, \citenamefont {Colbeck}, \citenamefont {Englund}, \citenamefont {Gehring}, \citenamefont {Lupo}, \citenamefont {Ottaviani}, \citenamefont {Pereira}, \citenamefont {Razavi}, \citenamefont {{Shamsul Shaari}}, \citenamefont {Tomamichel}, \citenamefont {Usenko}, \citenamefont {Vallone}, \citenamefont {Villoresi},\ and\ \citenamefont {Wallden}}]{Pirandola2020}%
  \BibitemOpen
  \bibfield  {author} {\bibinfo {author} {\bibfnamefont {S.}~\bibnamefont {Pirandola}}, \bibinfo {author} {\bibfnamefont {U.~L.}\ \bibnamefont {Andersen}}, \bibinfo {author} {\bibfnamefont {L.}~\bibnamefont {Banchi}}, \bibinfo {author} {\bibfnamefont {M.}~\bibnamefont {Berta}}, \bibinfo {author} {\bibfnamefont {D.}~\bibnamefont {Bunandar}}, \bibinfo {author} {\bibfnamefont {R.}~\bibnamefont {Colbeck}}, \bibinfo {author} {\bibfnamefont {D.}~\bibnamefont {Englund}}, \bibinfo {author} {\bibfnamefont {T.}~\bibnamefont {Gehring}}, \bibinfo {author} {\bibfnamefont {C.}~\bibnamefont {Lupo}}, \bibinfo {author} {\bibfnamefont {C.}~\bibnamefont {Ottaviani}}, \bibinfo {author} {\bibfnamefont {J.~L.}\ \bibnamefont {Pereira}}, \bibinfo {author} {\bibfnamefont {M.}~\bibnamefont {Razavi}}, \bibinfo {author} {\bibfnamefont {J.}~\bibnamefont {{Shamsul Shaari}}}, \bibinfo {author} {\bibfnamefont {M.}~\bibnamefont {Tomamichel}}, \bibinfo {author} {\bibfnamefont {V.~C.}\ \bibnamefont {Usenko}}, \bibinfo {author} {\bibfnamefont
  {G.}~\bibnamefont {Vallone}}, \bibinfo {author} {\bibfnamefont {P.}~\bibnamefont {Villoresi}},\ and\ \bibinfo {author} {\bibfnamefont {P.}~\bibnamefont {Wallden}},\ }\bibfield  {title} {\bibinfo {title} {{Advances in quantum cryptography}},\ }\href {https://doi.org/10.1364/AOP.361502} {\bibfield  {journal} {\bibinfo  {journal} {Adv. Opt. Photonics}\ }\textbf {\bibinfo {volume} {12}},\ \bibinfo {pages} {1012} (\bibinfo {year} {2020}{\natexlab{b}})}\BibitemShut {NoStop}%
\bibitem [{\citenamefont {Gottesman}\ \emph {et~al.}(2012)\citenamefont {Gottesman}, \citenamefont {Jennewein},\ and\ \citenamefont {Croke}}]{Gottesman2012}%
  \BibitemOpen
  \bibfield  {author} {\bibinfo {author} {\bibfnamefont {D.}~\bibnamefont {Gottesman}}, \bibinfo {author} {\bibfnamefont {T.}~\bibnamefont {Jennewein}},\ and\ \bibinfo {author} {\bibfnamefont {S.}~\bibnamefont {Croke}},\ }\bibfield  {title} {\bibinfo {title} {{Longer-Baseline Telescopes Using Quantum Repeaters}},\ }\href {https://doi.org/10.1103/PhysRevLett.109.070503} {\bibfield  {journal} {\bibinfo  {journal} {Phys. Rev. Lett.}\ }\textbf {\bibinfo {volume} {109}},\ \bibinfo {pages} {070503} (\bibinfo {year} {2012})}\BibitemShut {NoStop}%
\bibitem [{\citenamefont {Malia}\ \emph {et~al.}(2022)\citenamefont {Malia}, \citenamefont {Wu}, \citenamefont {Mart{\'{i}}nez-Rinc{\'{o}}n},\ and\ \citenamefont {Kasevich}}]{Malia2022}%
  \BibitemOpen
  \bibfield  {author} {\bibinfo {author} {\bibfnamefont {B.~K.}\ \bibnamefont {Malia}}, \bibinfo {author} {\bibfnamefont {Y.}~\bibnamefont {Wu}}, \bibinfo {author} {\bibfnamefont {J.}~\bibnamefont {Mart{\'{i}}nez-Rinc{\'{o}}n}},\ and\ \bibinfo {author} {\bibfnamefont {M.~A.}\ \bibnamefont {Kasevich}},\ }\bibfield  {title} {\bibinfo {title} {{Distributed quantum sensing with mode-entangled spin-squeezed atomic states}},\ }\href {https://doi.org/10.1038/s41586-022-05363-z} {\bibfield  {journal} {\bibinfo  {journal} {Nature}\ }\textbf {\bibinfo {volume} {612}},\ \bibinfo {pages} {661} (\bibinfo {year} {2022})}\BibitemShut {NoStop}%
\bibitem [{\citenamefont {K{\'{o}}m{\'{a}}r}\ \emph {et~al.}(2014)\citenamefont {K{\'{o}}m{\'{a}}r}, \citenamefont {Kessler}, \citenamefont {Bishof}, \citenamefont {Jiang}, \citenamefont {S{\o}rensen}, \citenamefont {Ye},\ and\ \citenamefont {Lukin}}]{Komar2014}%
  \BibitemOpen
  \bibfield  {author} {\bibinfo {author} {\bibfnamefont {P.}~\bibnamefont {K{\'{o}}m{\'{a}}r}}, \bibinfo {author} {\bibfnamefont {E.~M.}\ \bibnamefont {Kessler}}, \bibinfo {author} {\bibfnamefont {M.}~\bibnamefont {Bishof}}, \bibinfo {author} {\bibfnamefont {L.}~\bibnamefont {Jiang}}, \bibinfo {author} {\bibfnamefont {A.~S.}\ \bibnamefont {S{\o}rensen}}, \bibinfo {author} {\bibfnamefont {J.}~\bibnamefont {Ye}},\ and\ \bibinfo {author} {\bibfnamefont {M.~D.}\ \bibnamefont {Lukin}},\ }\bibfield  {title} {\bibinfo {title} {{A quantum network of clocks}},\ }\href {https://doi.org/10.1038/nphys3000} {\bibfield  {journal} {\bibinfo  {journal} {Nat. Phys.}\ }\textbf {\bibinfo {volume} {10}},\ \bibinfo {pages} {582} (\bibinfo {year} {2014})}\BibitemShut {NoStop}%
\bibitem [{\citenamefont {Nichol}\ \emph {et~al.}(2022)\citenamefont {Nichol}, \citenamefont {Srinivas}, \citenamefont {Nadlinger}, \citenamefont {Drmota}, \citenamefont {Main}, \citenamefont {Araneda}, \citenamefont {Ballance},\ and\ \citenamefont {Lucas}}]{Nichol2022}%
  \BibitemOpen
  \bibfield  {author} {\bibinfo {author} {\bibfnamefont {B.~C.}\ \bibnamefont {Nichol}}, \bibinfo {author} {\bibfnamefont {R.}~\bibnamefont {Srinivas}}, \bibinfo {author} {\bibfnamefont {D.~P.}\ \bibnamefont {Nadlinger}}, \bibinfo {author} {\bibfnamefont {P.}~\bibnamefont {Drmota}}, \bibinfo {author} {\bibfnamefont {D.}~\bibnamefont {Main}}, \bibinfo {author} {\bibfnamefont {G.}~\bibnamefont {Araneda}}, \bibinfo {author} {\bibfnamefont {C.~J.}\ \bibnamefont {Ballance}},\ and\ \bibinfo {author} {\bibfnamefont {D.~M.}\ \bibnamefont {Lucas}},\ }\bibfield  {title} {\bibinfo {title} {{An elementary quantum network of entangled optical atomic clocks}},\ }\href {https://doi.org/10.1038/s41586-022-05088-z} {\bibfield  {journal} {\bibinfo  {journal} {Nature}\ }\textbf {\bibinfo {volume} {609}},\ \bibinfo {pages} {689} (\bibinfo {year} {2022})}\BibitemShut {NoStop}%
\bibitem [{\citenamefont {Wootters}\ and\ \citenamefont {Zurek}(1982)}]{Wootters1982}%
  \BibitemOpen
  \bibfield  {author} {\bibinfo {author} {\bibfnamefont {W.~K.}\ \bibnamefont {Wootters}}\ and\ \bibinfo {author} {\bibfnamefont {W.~H.}\ \bibnamefont {Zurek}},\ }\bibfield  {title} {\bibinfo {title} {{A single quantum cannot be cloned}},\ }\href {https://doi.org/10.1038/299802a0} {\bibfield  {journal} {\bibinfo  {journal} {Nature}\ }\textbf {\bibinfo {volume} {299}},\ \bibinfo {pages} {802} (\bibinfo {year} {1982})}\BibitemShut {NoStop}%
\bibitem [{\citenamefont {Barz}\ \emph {et~al.}(2012)\citenamefont {Barz}, \citenamefont {Kashefi}, \citenamefont {Broadbent}, \citenamefont {Fitzsimons}, \citenamefont {Zeilinger},\ and\ \citenamefont {Walther}}]{Barz2012}%
  \BibitemOpen
  \bibfield  {author} {\bibinfo {author} {\bibfnamefont {S.}~\bibnamefont {Barz}}, \bibinfo {author} {\bibfnamefont {E.}~\bibnamefont {Kashefi}}, \bibinfo {author} {\bibfnamefont {A.}~\bibnamefont {Broadbent}}, \bibinfo {author} {\bibfnamefont {J.~F.}\ \bibnamefont {Fitzsimons}}, \bibinfo {author} {\bibfnamefont {A.}~\bibnamefont {Zeilinger}},\ and\ \bibinfo {author} {\bibfnamefont {P.}~\bibnamefont {Walther}},\ }\bibfield  {title} {\bibinfo {title} {{Demonstration of Blind Quantum Computing}},\ }\href {https://doi.org/10.1126/science.1214707} {\bibfield  {journal} {\bibinfo  {journal} {Science}\ }\textbf {\bibinfo {volume} {335}},\ \bibinfo {pages} {303} (\bibinfo {year} {2012})}\BibitemShut {NoStop}%
\bibitem [{\citenamefont {Fitzsimons}(2017)}]{Fitzsimons2017}%
  \BibitemOpen
  \bibfield  {author} {\bibinfo {author} {\bibfnamefont {J.~F.}\ \bibnamefont {Fitzsimons}},\ }\bibfield  {title} {\bibinfo {title} {{Private quantum computation: an introduction to blind quantum computing and related protocols}},\ }\href {https://doi.org/10.1038/s41534-017-0025-3} {\bibfield  {journal} {\bibinfo  {journal} {npj Quantum Inf.}\ }\textbf {\bibinfo {volume} {3}},\ \bibinfo {pages} {23} (\bibinfo {year} {2017})}\BibitemShut {NoStop}%
\bibitem [{\citenamefont {Jiang}\ \emph {et~al.}(2007)\citenamefont {Jiang}, \citenamefont {Taylor}, \citenamefont {S{\o}rensen},\ and\ \citenamefont {Lukin}}]{Jiang2007}%
  \BibitemOpen
  \bibfield  {author} {\bibinfo {author} {\bibfnamefont {L.}~\bibnamefont {Jiang}}, \bibinfo {author} {\bibfnamefont {J.~M.}\ \bibnamefont {Taylor}}, \bibinfo {author} {\bibfnamefont {A.~S.}\ \bibnamefont {S{\o}rensen}},\ and\ \bibinfo {author} {\bibfnamefont {M.~D.}\ \bibnamefont {Lukin}},\ }\bibfield  {title} {\bibinfo {title} {{Distributed quantum computation based on small quantum registers}},\ }\href {https://doi.org/10.1103/PhysRevA.76.062323} {\bibfield  {journal} {\bibinfo  {journal} {Phys. Rev. A}\ }\textbf {\bibinfo {volume} {76}},\ \bibinfo {pages} {062323} (\bibinfo {year} {2007})}\BibitemShut {NoStop}%
\bibitem [{\citenamefont {Monroe}\ \emph {et~al.}(2014)\citenamefont {Monroe}, \citenamefont {Raussendorf}, \citenamefont {Ruthven}, \citenamefont {Brown}, \citenamefont {Maunz}, \citenamefont {Duan},\ and\ \citenamefont {Kim}}]{Monroe2014}%
  \BibitemOpen
  \bibfield  {author} {\bibinfo {author} {\bibfnamefont {C.}~\bibnamefont {Monroe}}, \bibinfo {author} {\bibfnamefont {R.}~\bibnamefont {Raussendorf}}, \bibinfo {author} {\bibfnamefont {A.}~\bibnamefont {Ruthven}}, \bibinfo {author} {\bibfnamefont {K.~R.}\ \bibnamefont {Brown}}, \bibinfo {author} {\bibfnamefont {P.}~\bibnamefont {Maunz}}, \bibinfo {author} {\bibfnamefont {L.-M.}\ \bibnamefont {Duan}},\ and\ \bibinfo {author} {\bibfnamefont {J.}~\bibnamefont {Kim}},\ }\bibfield  {title} {\bibinfo {title} {{Large-scale modular quantum-computer architecture with atomic memory and photonic interconnects}},\ }\href {https://doi.org/10.1103/PhysRevA.89.022317} {\bibfield  {journal} {\bibinfo  {journal} {Phys. Rev. A}\ }\textbf {\bibinfo {volume} {89}},\ \bibinfo {pages} {022317} (\bibinfo {year} {2014})}\BibitemShut {NoStop}%
\bibitem [{\citenamefont {Stephenson}\ \emph {et~al.}(2020)\citenamefont {Stephenson}, \citenamefont {Nadlinger}, \citenamefont {Nichol}, \citenamefont {An}, \citenamefont {Drmota}, \citenamefont {Ballance}, \citenamefont {Thirumalai}, \citenamefont {Goodwin}, \citenamefont {Lucas},\ and\ \citenamefont {Ballance}}]{Stephenson2020}%
  \BibitemOpen
  \bibfield  {author} {\bibinfo {author} {\bibfnamefont {L.~J.}\ \bibnamefont {Stephenson}}, \bibinfo {author} {\bibfnamefont {D.~P.}\ \bibnamefont {Nadlinger}}, \bibinfo {author} {\bibfnamefont {B.~C.}\ \bibnamefont {Nichol}}, \bibinfo {author} {\bibfnamefont {S.}~\bibnamefont {An}}, \bibinfo {author} {\bibfnamefont {P.}~\bibnamefont {Drmota}}, \bibinfo {author} {\bibfnamefont {T.~G.}\ \bibnamefont {Ballance}}, \bibinfo {author} {\bibfnamefont {K.}~\bibnamefont {Thirumalai}}, \bibinfo {author} {\bibfnamefont {J.~F.}\ \bibnamefont {Goodwin}}, \bibinfo {author} {\bibfnamefont {D.~M.}\ \bibnamefont {Lucas}},\ and\ \bibinfo {author} {\bibfnamefont {C.~J.}\ \bibnamefont {Ballance}},\ }\bibfield  {title} {\bibinfo {title} {{High-Rate, High-Fidelity Entanglement of Qubits Across an Elementary Quantum Network}},\ }\href {https://doi.org/10.1103/PhysRevLett.124.110501} {\bibfield  {journal} {\bibinfo  {journal} {Phys. Rev. Lett.}\ }\textbf {\bibinfo {volume} {124}},\ \bibinfo {pages} {110501} (\bibinfo {year}
  {2020})}\BibitemShut {NoStop}%
\bibitem [{\citenamefont {Saha}\ \emph {et~al.}(2024)\citenamefont {Saha}, \citenamefont {Shalaev}, \citenamefont {O'Reilly}, \citenamefont {Goetting}, \citenamefont {Toh}, \citenamefont {Kalakuntla}, \citenamefont {Yu},\ and\ \citenamefont {Monroe}}]{Saha2024}%
  \BibitemOpen
  \bibfield  {author} {\bibinfo {author} {\bibfnamefont {S.}~\bibnamefont {Saha}}, \bibinfo {author} {\bibfnamefont {M.}~\bibnamefont {Shalaev}}, \bibinfo {author} {\bibfnamefont {J.}~\bibnamefont {O'Reilly}}, \bibinfo {author} {\bibfnamefont {I.}~\bibnamefont {Goetting}}, \bibinfo {author} {\bibfnamefont {G.}~\bibnamefont {Toh}}, \bibinfo {author} {\bibfnamefont {A.}~\bibnamefont {Kalakuntla}}, \bibinfo {author} {\bibfnamefont {Y.}~\bibnamefont {Yu}},\ and\ \bibinfo {author} {\bibfnamefont {C.}~\bibnamefont {Monroe}},\ }\bibfield  {title} {\bibinfo {title} {{High-fidelity remote entanglement of trapped atoms mediated by time-bin photons}},\ }\href {http://arxiv.org/abs/2406.01761} {\bibfield  {journal} {\bibinfo  {journal} {arXiv Prepr.}\ }\textbf {\bibinfo {volume} {2406.01761}} (\bibinfo {year} {2024})},\ \Eprint {https://arxiv.org/abs/2406.01761} {arXiv:2406.01761} \BibitemShut {NoStop}%
\bibitem [{\citenamefont {Hensen}\ \emph {et~al.}(2015)\citenamefont {Hensen}, \citenamefont {Bernien}, \citenamefont {Dr{\'{e}}au}, \citenamefont {Reiserer}, \citenamefont {Kalb}, \citenamefont {Blok}, \citenamefont {Ruitenberg}, \citenamefont {Vermeulen}, \citenamefont {Schouten}, \citenamefont {Abell{\'{a}}n}, \citenamefont {Amaya}, \citenamefont {Pruneri}, \citenamefont {Mitchell}, \citenamefont {Markham}, \citenamefont {Twitchen}, \citenamefont {Elkouss}, \citenamefont {Wehner}, \citenamefont {Taminiau},\ and\ \citenamefont {Hanson}}]{Hensen2015}%
  \BibitemOpen
  \bibfield  {author} {\bibinfo {author} {\bibfnamefont {B.}~\bibnamefont {Hensen}}, \bibinfo {author} {\bibfnamefont {H.}~\bibnamefont {Bernien}}, \bibinfo {author} {\bibfnamefont {A.~E.}\ \bibnamefont {Dr{\'{e}}au}}, \bibinfo {author} {\bibfnamefont {A.}~\bibnamefont {Reiserer}}, \bibinfo {author} {\bibfnamefont {N.}~\bibnamefont {Kalb}}, \bibinfo {author} {\bibfnamefont {M.~S.}\ \bibnamefont {Blok}}, \bibinfo {author} {\bibfnamefont {J.}~\bibnamefont {Ruitenberg}}, \bibinfo {author} {\bibfnamefont {R.~F.~L.}\ \bibnamefont {Vermeulen}}, \bibinfo {author} {\bibfnamefont {R.~N.}\ \bibnamefont {Schouten}}, \bibinfo {author} {\bibfnamefont {C.}~\bibnamefont {Abell{\'{a}}n}}, \bibinfo {author} {\bibfnamefont {W.}~\bibnamefont {Amaya}}, \bibinfo {author} {\bibfnamefont {V.}~\bibnamefont {Pruneri}}, \bibinfo {author} {\bibfnamefont {M.~W.}\ \bibnamefont {Mitchell}}, \bibinfo {author} {\bibfnamefont {M.}~\bibnamefont {Markham}}, \bibinfo {author} {\bibfnamefont {D.~J.}\ \bibnamefont {Twitchen}}, \bibinfo {author}
  {\bibfnamefont {D.}~\bibnamefont {Elkouss}}, \bibinfo {author} {\bibfnamefont {S.}~\bibnamefont {Wehner}}, \bibinfo {author} {\bibfnamefont {T.~H.}\ \bibnamefont {Taminiau}},\ and\ \bibinfo {author} {\bibfnamefont {R.}~\bibnamefont {Hanson}},\ }\bibfield  {title} {\bibinfo {title} {{Loophole-free Bell inequality violation using electron spins separated by 1.3 kilometres}},\ }\href {https://doi.org/10.1038/nature15759} {\bibfield  {journal} {\bibinfo  {journal} {Nature}\ }\textbf {\bibinfo {volume} {526}},\ \bibinfo {pages} {682} (\bibinfo {year} {2015})}\BibitemShut {NoStop}%
\bibitem [{\citenamefont {van Leent}\ \emph {et~al.}(2022)\citenamefont {van Leent}, \citenamefont {Bock}, \citenamefont {Fertig}, \citenamefont {Garthoff}, \citenamefont {Eppelt}, \citenamefont {Zhou}, \citenamefont {Malik}, \citenamefont {Seubert}, \citenamefont {Bauer}, \citenamefont {Rosenfeld}, \citenamefont {Zhang}, \citenamefont {Becher},\ and\ \citenamefont {Weinfurter}}]{vanLeent2022}%
  \BibitemOpen
  \bibfield  {author} {\bibinfo {author} {\bibfnamefont {T.}~\bibnamefont {van Leent}}, \bibinfo {author} {\bibfnamefont {M.}~\bibnamefont {Bock}}, \bibinfo {author} {\bibfnamefont {F.}~\bibnamefont {Fertig}}, \bibinfo {author} {\bibfnamefont {R.}~\bibnamefont {Garthoff}}, \bibinfo {author} {\bibfnamefont {S.}~\bibnamefont {Eppelt}}, \bibinfo {author} {\bibfnamefont {Y.}~\bibnamefont {Zhou}}, \bibinfo {author} {\bibfnamefont {P.}~\bibnamefont {Malik}}, \bibinfo {author} {\bibfnamefont {M.}~\bibnamefont {Seubert}}, \bibinfo {author} {\bibfnamefont {T.}~\bibnamefont {Bauer}}, \bibinfo {author} {\bibfnamefont {W.}~\bibnamefont {Rosenfeld}}, \bibinfo {author} {\bibfnamefont {W.}~\bibnamefont {Zhang}}, \bibinfo {author} {\bibfnamefont {C.}~\bibnamefont {Becher}},\ and\ \bibinfo {author} {\bibfnamefont {H.}~\bibnamefont {Weinfurter}},\ }\bibfield  {title} {\bibinfo {title} {{Entangling single atoms over 33 km telecom fibre}},\ }\href {https://doi.org/10.1038/s41586-022-04764-4} {\bibfield  {journal} {\bibinfo
  {journal} {Nature}\ }\textbf {\bibinfo {volume} {607}},\ \bibinfo {pages} {69} (\bibinfo {year} {2022})}\BibitemShut {NoStop}%
\bibitem [{\citenamefont {Krutyanskiy}\ \emph {et~al.}(2023)\citenamefont {Krutyanskiy}, \citenamefont {Canteri}, \citenamefont {Meraner}, \citenamefont {Bate}, \citenamefont {Krcmarsky}, \citenamefont {Schupp}, \citenamefont {Sangouard},\ and\ \citenamefont {Lanyon}}]{Krutyanskiy2023}%
  \BibitemOpen
  \bibfield  {author} {\bibinfo {author} {\bibfnamefont {V.}~\bibnamefont {Krutyanskiy}}, \bibinfo {author} {\bibfnamefont {M.}~\bibnamefont {Canteri}}, \bibinfo {author} {\bibfnamefont {M.}~\bibnamefont {Meraner}}, \bibinfo {author} {\bibfnamefont {J.}~\bibnamefont {Bate}}, \bibinfo {author} {\bibfnamefont {V.}~\bibnamefont {Krcmarsky}}, \bibinfo {author} {\bibfnamefont {J.}~\bibnamefont {Schupp}}, \bibinfo {author} {\bibfnamefont {N.}~\bibnamefont {Sangouard}},\ and\ \bibinfo {author} {\bibfnamefont {B.~P.}\ \bibnamefont {Lanyon}},\ }\bibfield  {title} {\bibinfo {title} {{Telecom-Wavelength Quantum Repeater Node Based on a Trapped-Ion Processor}},\ }\href {https://doi.org/10.1103/PhysRevLett.130.213601} {\bibfield  {journal} {\bibinfo  {journal} {Phys. Rev. Lett.}\ }\textbf {\bibinfo {volume} {130}},\ \bibinfo {pages} {213601} (\bibinfo {year} {2023})}\BibitemShut {NoStop}%
\bibitem [{\citenamefont {Uysal}\ \emph {et~al.}(2024)\citenamefont {Uysal}, \citenamefont {Dusanowski}, \citenamefont {Xu}, \citenamefont {Horvath}, \citenamefont {Ourari}, \citenamefont {Cava}, \citenamefont {de~Leon},\ and\ \citenamefont {Thompson}}]{Uysal2024}%
  \BibitemOpen
  \bibfield  {author} {\bibinfo {author} {\bibfnamefont {M.~T.}\ \bibnamefont {Uysal}}, \bibinfo {author} {\bibfnamefont {{\L}.}~\bibnamefont {Dusanowski}}, \bibinfo {author} {\bibfnamefont {H.}~\bibnamefont {Xu}}, \bibinfo {author} {\bibfnamefont {S.~P.}\ \bibnamefont {Horvath}}, \bibinfo {author} {\bibfnamefont {S.}~\bibnamefont {Ourari}}, \bibinfo {author} {\bibfnamefont {R.~J.}\ \bibnamefont {Cava}}, \bibinfo {author} {\bibfnamefont {N.~P.}\ \bibnamefont {de~Leon}},\ and\ \bibinfo {author} {\bibfnamefont {J.~D.}\ \bibnamefont {Thompson}},\ }\bibfield  {title} {\bibinfo {title} {{Spin-photon entanglement of a single Er$^{3+}$ ion in the telecom band}},\ }\href {http://arxiv.org/abs/2406.06515} {\bibfield  {journal} {\bibinfo  {journal} {arXiv Prepr.}\ }\textbf {\bibinfo {volume} {2406.06515}} (\bibinfo {year} {2024})}\BibitemShut {NoStop}%
\bibitem [{\citenamefont {Covey}\ \emph {et~al.}(2023)\citenamefont {Covey}, \citenamefont {Weinfurter},\ and\ \citenamefont {Bernien}}]{Covey2023}%
  \BibitemOpen
  \bibfield  {author} {\bibinfo {author} {\bibfnamefont {J.~P.}\ \bibnamefont {Covey}}, \bibinfo {author} {\bibfnamefont {H.}~\bibnamefont {Weinfurter}},\ and\ \bibinfo {author} {\bibfnamefont {H.}~\bibnamefont {Bernien}},\ }\bibfield  {title} {\bibinfo {title} {{Quantum networks with neutral atom processing nodes}},\ }\href {https://doi.org/10.1038/s41534-023-00759-9} {\bibfield  {journal} {\bibinfo  {journal} {npj Quantum Inf.}\ }\textbf {\bibinfo {volume} {9}},\ \bibinfo {pages} {90} (\bibinfo {year} {2023})}\BibitemShut {NoStop}%
\bibitem [{\citenamefont {Moehring}\ \emph {et~al.}(2007)\citenamefont {Moehring}, \citenamefont {Maunz}, \citenamefont {Olmschenk}, \citenamefont {Younge}, \citenamefont {Matsukevich}, \citenamefont {Duan},\ and\ \citenamefont {Monroe}}]{Moehring2007}%
  \BibitemOpen
  \bibfield  {author} {\bibinfo {author} {\bibfnamefont {D.~L.}\ \bibnamefont {Moehring}}, \bibinfo {author} {\bibfnamefont {P.}~\bibnamefont {Maunz}}, \bibinfo {author} {\bibfnamefont {S.}~\bibnamefont {Olmschenk}}, \bibinfo {author} {\bibfnamefont {K.~C.}\ \bibnamefont {Younge}}, \bibinfo {author} {\bibfnamefont {D.~N.}\ \bibnamefont {Matsukevich}}, \bibinfo {author} {\bibfnamefont {L.-M.}\ \bibnamefont {Duan}},\ and\ \bibinfo {author} {\bibfnamefont {C.}~\bibnamefont {Monroe}},\ }\bibfield  {title} {\bibinfo {title} {{Entanglement of single-atom quantum bits at a distance}},\ }\href {https://doi.org/10.1038/nature06118} {\bibfield  {journal} {\bibinfo  {journal} {Nature}\ }\textbf {\bibinfo {volume} {449}},\ \bibinfo {pages} {68} (\bibinfo {year} {2007})}\BibitemShut {NoStop}%
\bibitem [{\citenamefont {Bernien}\ \emph {et~al.}(2013)\citenamefont {Bernien}, \citenamefont {Hensen}, \citenamefont {Pfaff}, \citenamefont {Koolstra}, \citenamefont {Blok}, \citenamefont {Robledo}, \citenamefont {Taminiau}, \citenamefont {Markham}, \citenamefont {Twitchen}, \citenamefont {Childress},\ and\ \citenamefont {Hanson}}]{Bernien2013}%
  \BibitemOpen
  \bibfield  {author} {\bibinfo {author} {\bibfnamefont {H.}~\bibnamefont {Bernien}}, \bibinfo {author} {\bibfnamefont {B.}~\bibnamefont {Hensen}}, \bibinfo {author} {\bibfnamefont {W.}~\bibnamefont {Pfaff}}, \bibinfo {author} {\bibfnamefont {G.}~\bibnamefont {Koolstra}}, \bibinfo {author} {\bibfnamefont {M.~S.}\ \bibnamefont {Blok}}, \bibinfo {author} {\bibfnamefont {L.}~\bibnamefont {Robledo}}, \bibinfo {author} {\bibfnamefont {T.~H.}\ \bibnamefont {Taminiau}}, \bibinfo {author} {\bibfnamefont {M.}~\bibnamefont {Markham}}, \bibinfo {author} {\bibfnamefont {D.~J.}\ \bibnamefont {Twitchen}}, \bibinfo {author} {\bibfnamefont {L.}~\bibnamefont {Childress}},\ and\ \bibinfo {author} {\bibfnamefont {R.}~\bibnamefont {Hanson}},\ }\bibfield  {title} {\bibinfo {title} {{Heralded entanglement between solid-state qubits separated by three metres}},\ }\href {https://doi.org/10.1038/nature12016} {\bibfield  {journal} {\bibinfo  {journal} {Nature}\ }\textbf {\bibinfo {volume} {497}},\ \bibinfo {pages} {86} (\bibinfo
  {year} {2013})}\BibitemShut {NoStop}%
\bibitem [{\citenamefont {Bhaskar}\ \emph {et~al.}(2020)\citenamefont {Bhaskar}, \citenamefont {Riedinger}, \citenamefont {Machielse}, \citenamefont {Levonian}, \citenamefont {Nguyen}, \citenamefont {Knall}, \citenamefont {Park}, \citenamefont {Englund}, \citenamefont {Lon{\v{c}}ar}, \citenamefont {Sukachev},\ and\ \citenamefont {Lukin}}]{Bhaskar2020}%
  \BibitemOpen
  \bibfield  {author} {\bibinfo {author} {\bibfnamefont {M.~K.}\ \bibnamefont {Bhaskar}}, \bibinfo {author} {\bibfnamefont {R.}~\bibnamefont {Riedinger}}, \bibinfo {author} {\bibfnamefont {B.}~\bibnamefont {Machielse}}, \bibinfo {author} {\bibfnamefont {D.~S.}\ \bibnamefont {Levonian}}, \bibinfo {author} {\bibfnamefont {C.~T.}\ \bibnamefont {Nguyen}}, \bibinfo {author} {\bibfnamefont {E.~N.}\ \bibnamefont {Knall}}, \bibinfo {author} {\bibfnamefont {H.}~\bibnamefont {Park}}, \bibinfo {author} {\bibfnamefont {D.}~\bibnamefont {Englund}}, \bibinfo {author} {\bibfnamefont {M.}~\bibnamefont {Lon{\v{c}}ar}}, \bibinfo {author} {\bibfnamefont {D.~D.}\ \bibnamefont {Sukachev}},\ and\ \bibinfo {author} {\bibfnamefont {M.~D.}\ \bibnamefont {Lukin}},\ }\bibfield  {title} {\bibinfo {title} {{Experimental demonstration of memory-enhanced quantum communication}},\ }\href {https://doi.org/10.1038/s41586-020-2103-5} {\bibfield  {journal} {\bibinfo  {journal} {Nature}\ }\textbf {\bibinfo {volume} {580}},\ \bibinfo {pages} {60}
  (\bibinfo {year} {2020})}\BibitemShut {NoStop}%
\bibitem [{\citenamefont {Stolk}\ \emph {et~al.}(2022)\citenamefont {Stolk}, \citenamefont {van~der Enden}, \citenamefont {Roehsner}, \citenamefont {Teepe}, \citenamefont {Faes}, \citenamefont {Bradley}, \citenamefont {Cadot}, \citenamefont {van Rantwijk}, \citenamefont {te~Raa}, \citenamefont {Hagen}, \citenamefont {Verlaan}, \citenamefont {Biemond}, \citenamefont {Khorev}, \citenamefont {Vollmer}, \citenamefont {Markham}, \citenamefont {Edmonds}, \citenamefont {Morits}, \citenamefont {Taminiau}, \citenamefont {van Zwet},\ and\ \citenamefont {Hanson}}]{Stolk2022}%
  \BibitemOpen
  \bibfield  {author} {\bibinfo {author} {\bibfnamefont {A.}~\bibnamefont {Stolk}}, \bibinfo {author} {\bibfnamefont {K.}~\bibnamefont {van~der Enden}}, \bibinfo {author} {\bibfnamefont {M.-C.}\ \bibnamefont {Roehsner}}, \bibinfo {author} {\bibfnamefont {A.}~\bibnamefont {Teepe}}, \bibinfo {author} {\bibfnamefont {S.}~\bibnamefont {Faes}}, \bibinfo {author} {\bibfnamefont {C.}~\bibnamefont {Bradley}}, \bibinfo {author} {\bibfnamefont {S.}~\bibnamefont {Cadot}}, \bibinfo {author} {\bibfnamefont {J.}~\bibnamefont {van Rantwijk}}, \bibinfo {author} {\bibfnamefont {I.}~\bibnamefont {te~Raa}}, \bibinfo {author} {\bibfnamefont {R.}~\bibnamefont {Hagen}}, \bibinfo {author} {\bibfnamefont {A.}~\bibnamefont {Verlaan}}, \bibinfo {author} {\bibfnamefont {J.}~\bibnamefont {Biemond}}, \bibinfo {author} {\bibfnamefont {A.}~\bibnamefont {Khorev}}, \bibinfo {author} {\bibfnamefont {R.}~\bibnamefont {Vollmer}}, \bibinfo {author} {\bibfnamefont {M.}~\bibnamefont {Markham}}, \bibinfo {author} {\bibfnamefont {A.}~\bibnamefont
  {Edmonds}}, \bibinfo {author} {\bibfnamefont {J.}~\bibnamefont {Morits}}, \bibinfo {author} {\bibfnamefont {T.}~\bibnamefont {Taminiau}}, \bibinfo {author} {\bibfnamefont {E.}~\bibnamefont {van Zwet}},\ and\ \bibinfo {author} {\bibfnamefont {R.}~\bibnamefont {Hanson}},\ }\bibfield  {title} {\bibinfo {title} {{Telecom-Band Quantum Interference of Frequency-Converted Photons from Remote Detuned NV Centers}},\ }\href {https://doi.org/10.1103/PRXQuantum.3.020359} {\bibfield  {journal} {\bibinfo  {journal} {PRX Quantum}\ }\textbf {\bibinfo {volume} {3}},\ \bibinfo {pages} {020359} (\bibinfo {year} {2022})}\BibitemShut {NoStop}%
\bibitem [{\citenamefont {Bersin}\ \emph {et~al.}(2024)\citenamefont {Bersin}, \citenamefont {Sutula}, \citenamefont {Huan}, \citenamefont {Suleymanzade}, \citenamefont {Assumpcao}, \citenamefont {Wei}, \citenamefont {Stas}, \citenamefont {Knaut}, \citenamefont {Knall}, \citenamefont {Langrock}, \citenamefont {Sinclair}, \citenamefont {Murphy}, \citenamefont {Riedinger}, \citenamefont {Yeh}, \citenamefont {Xin}, \citenamefont {Bandyopadhyay}, \citenamefont {Sukachev}, \citenamefont {Machielse}, \citenamefont {Levonian}, \citenamefont {Bhaskar}, \citenamefont {Hamilton}, \citenamefont {Park}, \citenamefont {Lon{\v{c}}ar}, \citenamefont {Fejer}, \citenamefont {Dixon}, \citenamefont {Englund},\ and\ \citenamefont {Lukin}}]{Bersin2024}%
  \BibitemOpen
  \bibfield  {author} {\bibinfo {author} {\bibfnamefont {E.}~\bibnamefont {Bersin}}, \bibinfo {author} {\bibfnamefont {M.}~\bibnamefont {Sutula}}, \bibinfo {author} {\bibfnamefont {Y.~Q.}\ \bibnamefont {Huan}}, \bibinfo {author} {\bibfnamefont {A.}~\bibnamefont {Suleymanzade}}, \bibinfo {author} {\bibfnamefont {D.~R.}\ \bibnamefont {Assumpcao}}, \bibinfo {author} {\bibfnamefont {Y.-C.}\ \bibnamefont {Wei}}, \bibinfo {author} {\bibfnamefont {P.-J.}\ \bibnamefont {Stas}}, \bibinfo {author} {\bibfnamefont {C.~M.}\ \bibnamefont {Knaut}}, \bibinfo {author} {\bibfnamefont {E.~N.}\ \bibnamefont {Knall}}, \bibinfo {author} {\bibfnamefont {C.}~\bibnamefont {Langrock}}, \bibinfo {author} {\bibfnamefont {N.}~\bibnamefont {Sinclair}}, \bibinfo {author} {\bibfnamefont {R.}~\bibnamefont {Murphy}}, \bibinfo {author} {\bibfnamefont {R.}~\bibnamefont {Riedinger}}, \bibinfo {author} {\bibfnamefont {M.}~\bibnamefont {Yeh}}, \bibinfo {author} {\bibfnamefont {C.}~\bibnamefont {Xin}}, \bibinfo {author} {\bibfnamefont
  {S.}~\bibnamefont {Bandyopadhyay}}, \bibinfo {author} {\bibfnamefont {D.~D.}\ \bibnamefont {Sukachev}}, \bibinfo {author} {\bibfnamefont {B.}~\bibnamefont {Machielse}}, \bibinfo {author} {\bibfnamefont {D.~S.}\ \bibnamefont {Levonian}}, \bibinfo {author} {\bibfnamefont {M.~K.}\ \bibnamefont {Bhaskar}}, \bibinfo {author} {\bibfnamefont {S.}~\bibnamefont {Hamilton}}, \bibinfo {author} {\bibfnamefont {H.}~\bibnamefont {Park}}, \bibinfo {author} {\bibfnamefont {M.}~\bibnamefont {Lon{\v{c}}ar}}, \bibinfo {author} {\bibfnamefont {M.~M.}\ \bibnamefont {Fejer}}, \bibinfo {author} {\bibfnamefont {P.~B.}\ \bibnamefont {Dixon}}, \bibinfo {author} {\bibfnamefont {D.~R.}\ \bibnamefont {Englund}},\ and\ \bibinfo {author} {\bibfnamefont {M.~D.}\ \bibnamefont {Lukin}},\ }\bibfield  {title} {\bibinfo {title} {{Telecom Networking with a Diamond Quantum Memory}},\ }\href {https://doi.org/10.1103/PRXQuantum.5.010303} {\bibfield  {journal} {\bibinfo  {journal} {PRX Quantum}\ }\textbf {\bibinfo {volume} {5}},\ \bibinfo {pages}
  {010303} (\bibinfo {year} {2024})}\BibitemShut {NoStop}%
\bibitem [{\citenamefont {Zhong}\ \emph {et~al.}(2017)\citenamefont {Zhong}, \citenamefont {Kindem}, \citenamefont {Bartholomew}, \citenamefont {Rochman}, \citenamefont {Craiciu}, \citenamefont {Miyazono}, \citenamefont {Bettinelli}, \citenamefont {Cavalli}, \citenamefont {Verma}, \citenamefont {Nam}, \citenamefont {Marsili}, \citenamefont {Shaw}, \citenamefont {Beyer},\ and\ \citenamefont {Faraon}}]{Zhong2017}%
  \BibitemOpen
  \bibfield  {author} {\bibinfo {author} {\bibfnamefont {T.}~\bibnamefont {Zhong}}, \bibinfo {author} {\bibfnamefont {J.~M.}\ \bibnamefont {Kindem}}, \bibinfo {author} {\bibfnamefont {J.~G.}\ \bibnamefont {Bartholomew}}, \bibinfo {author} {\bibfnamefont {J.}~\bibnamefont {Rochman}}, \bibinfo {author} {\bibfnamefont {I.}~\bibnamefont {Craiciu}}, \bibinfo {author} {\bibfnamefont {E.}~\bibnamefont {Miyazono}}, \bibinfo {author} {\bibfnamefont {M.}~\bibnamefont {Bettinelli}}, \bibinfo {author} {\bibfnamefont {E.}~\bibnamefont {Cavalli}}, \bibinfo {author} {\bibfnamefont {V.}~\bibnamefont {Verma}}, \bibinfo {author} {\bibfnamefont {S.~W.}\ \bibnamefont {Nam}}, \bibinfo {author} {\bibfnamefont {F.}~\bibnamefont {Marsili}}, \bibinfo {author} {\bibfnamefont {M.~D.}\ \bibnamefont {Shaw}}, \bibinfo {author} {\bibfnamefont {A.~D.}\ \bibnamefont {Beyer}},\ and\ \bibinfo {author} {\bibfnamefont {A.}~\bibnamefont {Faraon}},\ }\bibfield  {title} {\bibinfo {title} {{Nanophotonic rare-earth quantum memory with optically
  controlled retrieval}},\ }\href {https://doi.org/10.1126/science.aan5959} {\bibfield  {journal} {\bibinfo  {journal} {Science}\ }\textbf {\bibinfo {volume} {357}},\ \bibinfo {pages} {1392} (\bibinfo {year} {2017})}\BibitemShut {NoStop}%
\bibitem [{\citenamefont {Dibos}\ \emph {et~al.}(2018)\citenamefont {Dibos}, \citenamefont {Raha}, \citenamefont {Phenicie},\ and\ \citenamefont {Thompson}}]{Dibos2018}%
  \BibitemOpen
  \bibfield  {author} {\bibinfo {author} {\bibfnamefont {A.~M.}\ \bibnamefont {Dibos}}, \bibinfo {author} {\bibfnamefont {M.}~\bibnamefont {Raha}}, \bibinfo {author} {\bibfnamefont {C.~M.}\ \bibnamefont {Phenicie}},\ and\ \bibinfo {author} {\bibfnamefont {J.~D.}\ \bibnamefont {Thompson}},\ }\bibfield  {title} {\bibinfo {title} {{Atomic Source of Single Photons in the Telecom Band}},\ }\href {https://doi.org/10.1103/PhysRevLett.120.243601} {\bibfield  {journal} {\bibinfo  {journal} {Phys. Rev. Lett.}\ }\textbf {\bibinfo {volume} {120}},\ \bibinfo {pages} {243601} (\bibinfo {year} {2018})}\BibitemShut {NoStop}%
\bibitem [{\citenamefont {Kindem}\ \emph {et~al.}(2020)\citenamefont {Kindem}, \citenamefont {Ruskuc}, \citenamefont {Bartholomew}, \citenamefont {Rochman}, \citenamefont {Huan},\ and\ \citenamefont {Faraon}}]{Kindem2020}%
  \BibitemOpen
  \bibfield  {author} {\bibinfo {author} {\bibfnamefont {J.~M.}\ \bibnamefont {Kindem}}, \bibinfo {author} {\bibfnamefont {A.}~\bibnamefont {Ruskuc}}, \bibinfo {author} {\bibfnamefont {J.~G.}\ \bibnamefont {Bartholomew}}, \bibinfo {author} {\bibfnamefont {J.}~\bibnamefont {Rochman}}, \bibinfo {author} {\bibfnamefont {Y.~Q.}\ \bibnamefont {Huan}},\ and\ \bibinfo {author} {\bibfnamefont {A.}~\bibnamefont {Faraon}},\ }\bibfield  {title} {\bibinfo {title} {{Control and single-shot readout of an ion embedded in a nanophotonic cavity}},\ }\href {https://doi.org/10.1038/s41586-020-2160-9} {\bibfield  {journal} {\bibinfo  {journal} {Nature}\ }\textbf {\bibinfo {volume} {580}},\ \bibinfo {pages} {201} (\bibinfo {year} {2020})}\BibitemShut {NoStop}%
\bibitem [{\citenamefont {Hucul}\ \emph {et~al.}(2015)\citenamefont {Hucul}, \citenamefont {Inlek}, \citenamefont {Vittorini}, \citenamefont {Crocker}, \citenamefont {Debnath}, \citenamefont {Clark},\ and\ \citenamefont {Monroe}}]{Hucul2015}%
  \BibitemOpen
  \bibfield  {author} {\bibinfo {author} {\bibfnamefont {D.}~\bibnamefont {Hucul}}, \bibinfo {author} {\bibfnamefont {I.~V.}\ \bibnamefont {Inlek}}, \bibinfo {author} {\bibfnamefont {G.}~\bibnamefont {Vittorini}}, \bibinfo {author} {\bibfnamefont {C.}~\bibnamefont {Crocker}}, \bibinfo {author} {\bibfnamefont {S.}~\bibnamefont {Debnath}}, \bibinfo {author} {\bibfnamefont {S.~M.}\ \bibnamefont {Clark}},\ and\ \bibinfo {author} {\bibfnamefont {C.}~\bibnamefont {Monroe}},\ }\bibfield  {title} {\bibinfo {title} {{Modular entanglement of atomic qubits using photons and phonons}},\ }\href {https://doi.org/10.1038/nphys3150} {\bibfield  {journal} {\bibinfo  {journal} {Nat. Phys.}\ }\textbf {\bibinfo {volume} {11}},\ \bibinfo {pages} {37} (\bibinfo {year} {2015})}\BibitemShut {NoStop}%
\bibitem [{\citenamefont {Huie}\ \emph {et~al.}(2021)\citenamefont {Huie}, \citenamefont {Menon}, \citenamefont {Bernien},\ and\ \citenamefont {Covey}}]{Huie2021}%
  \BibitemOpen
  \bibfield  {author} {\bibinfo {author} {\bibfnamefont {W.}~\bibnamefont {Huie}}, \bibinfo {author} {\bibfnamefont {S.~G.}\ \bibnamefont {Menon}}, \bibinfo {author} {\bibfnamefont {H.}~\bibnamefont {Bernien}},\ and\ \bibinfo {author} {\bibfnamefont {J.~P.}\ \bibnamefont {Covey}},\ }\bibfield  {title} {\bibinfo {title} {{Multiplexed telecommunication-band quantum networking with atom arrays in optical cavities}},\ }\href {https://doi.org/10.1103/PhysRevResearch.3.043154} {\bibfield  {journal} {\bibinfo  {journal} {Phys. Rev. Res.}\ }\textbf {\bibinfo {volume} {3}},\ \bibinfo {pages} {043154} (\bibinfo {year} {2021})}\BibitemShut {NoStop}%
\bibitem [{\citenamefont {Li}\ and\ \citenamefont {Thompson}(2024)}]{LiThompson2024}%
  \BibitemOpen
  \bibfield  {author} {\bibinfo {author} {\bibfnamefont {Y.}~\bibnamefont {Li}}\ and\ \bibinfo {author} {\bibfnamefont {J.~D.}\ \bibnamefont {Thompson}},\ }\bibfield  {title} {\bibinfo {title} {{High-Rate and High-Fidelity Modular Interconnects between Neutral Atom Quantum Processors}},\ }\href {https://doi.org/10.1103/PRXQuantum.5.020363} {\bibfield  {journal} {\bibinfo  {journal} {PRX Quantum}\ }\textbf {\bibinfo {volume} {5}},\ \bibinfo {pages} {020363} (\bibinfo {year} {2024})}\BibitemShut {NoStop}%
\bibitem [{\citenamefont {Canteri}\ \emph {et~al.}(2024)\citenamefont {Canteri}, \citenamefont {Koong}, \citenamefont {Bate}, \citenamefont {Winkler}, \citenamefont {Krutyanskiy},\ and\ \citenamefont {Lanyon}}]{Canteri2024}%
  \BibitemOpen
  \bibfield  {author} {\bibinfo {author} {\bibfnamefont {M.}~\bibnamefont {Canteri}}, \bibinfo {author} {\bibfnamefont {Z.~X.}\ \bibnamefont {Koong}}, \bibinfo {author} {\bibfnamefont {J.}~\bibnamefont {Bate}}, \bibinfo {author} {\bibfnamefont {A.}~\bibnamefont {Winkler}}, \bibinfo {author} {\bibfnamefont {V.}~\bibnamefont {Krutyanskiy}},\ and\ \bibinfo {author} {\bibfnamefont {B.~P.}\ \bibnamefont {Lanyon}},\ }\bibfield  {title} {\bibinfo {title} {{A photon-interfaced ten qubit quantum network node}},\ }\href {http://arxiv.org/abs/2406.09480} {\bibfield  {journal} {\bibinfo  {journal} {arXiv Prepr.}\ }\textbf {\bibinfo {volume} {2406.09480}} (\bibinfo {year} {2024})}\BibitemShut {NoStop}%
\bibitem [{\citenamefont {Hartung}\ \emph {et~al.}(2024)\citenamefont {Hartung}, \citenamefont {Seubert}, \citenamefont {Welte}, \citenamefont {Distante},\ and\ \citenamefont {Rempe}}]{Hartung2024}%
  \BibitemOpen
  \bibfield  {author} {\bibinfo {author} {\bibfnamefont {L.}~\bibnamefont {Hartung}}, \bibinfo {author} {\bibfnamefont {M.}~\bibnamefont {Seubert}}, \bibinfo {author} {\bibfnamefont {S.}~\bibnamefont {Welte}}, \bibinfo {author} {\bibfnamefont {E.}~\bibnamefont {Distante}},\ and\ \bibinfo {author} {\bibfnamefont {G.}~\bibnamefont {Rempe}},\ }\bibfield  {title} {\bibinfo {title} {{A quantum-network register assembled with optical tweezers in an optical cavity}},\ }\href {https://doi.org/10.1126/science.ado6471} {\bibfield  {journal} {\bibinfo  {journal} {Science}\ }\textbf {\bibinfo {volume} {385}},\ \bibinfo {pages} {179} (\bibinfo {year} {2024})}\BibitemShut {NoStop}%
\bibitem [{\citenamefont {Sunami}\ \emph {et~al.}(2024)\citenamefont {Sunami}, \citenamefont {Tamiya}, \citenamefont {Inoue}, \citenamefont {Yamasaki},\ and\ \citenamefont {Goban}}]{Sunami2024}%
  \BibitemOpen
  \bibfield  {author} {\bibinfo {author} {\bibfnamefont {S.}~\bibnamefont {Sunami}}, \bibinfo {author} {\bibfnamefont {S.}~\bibnamefont {Tamiya}}, \bibinfo {author} {\bibfnamefont {R.}~\bibnamefont {Inoue}}, \bibinfo {author} {\bibfnamefont {H.}~\bibnamefont {Yamasaki}},\ and\ \bibinfo {author} {\bibfnamefont {A.}~\bibnamefont {Goban}},\ }\bibfield  {title} {\bibinfo {title} {{Scalable Networking of Neutral-Atom Qubits: Nanofiber-Based Approach for Multiprocessor Fault-Tolerant Quantum Computer}},\ }\href {http://arxiv.org/abs/2407.11111} {\bibfield  {journal} {\bibinfo  {journal} {arXiv Prepr.}\ }\textbf {\bibinfo {volume} {2407.11111}} (\bibinfo {year} {2024})}\BibitemShut {NoStop}%
\bibitem [{\citenamefont {Huie}\ \emph {et~al.}(2023)\citenamefont {Huie}, \citenamefont {Li}, \citenamefont {Chen}, \citenamefont {Hu}, \citenamefont {Jia}, \citenamefont {Sun},\ and\ \citenamefont {Covey}}]{Huie2023}%
  \BibitemOpen
  \bibfield  {author} {\bibinfo {author} {\bibfnamefont {W.}~\bibnamefont {Huie}}, \bibinfo {author} {\bibfnamefont {L.}~\bibnamefont {Li}}, \bibinfo {author} {\bibfnamefont {N.}~\bibnamefont {Chen}}, \bibinfo {author} {\bibfnamefont {X.}~\bibnamefont {Hu}}, \bibinfo {author} {\bibfnamefont {Z.}~\bibnamefont {Jia}}, \bibinfo {author} {\bibfnamefont {W.~K.~C.}\ \bibnamefont {Sun}},\ and\ \bibinfo {author} {\bibfnamefont {J.~P.}\ \bibnamefont {Covey}},\ }\bibfield  {title} {\bibinfo {title} {{Repetitive Readout and Real-Time Control of Nuclear Spin Qubits in 171Yb Atoms}},\ }\href {https://doi.org/10.1103/PRXQuantum.4.030337} {\bibfield  {journal} {\bibinfo  {journal} {PRX Quantum}\ }\textbf {\bibinfo {volume} {4}},\ \bibinfo {pages} {030337} (\bibinfo {year} {2023})}\BibitemShut {NoStop}%
\bibitem [{\citenamefont {Jenkins}\ \emph {et~al.}(2022)\citenamefont {Jenkins}, \citenamefont {Lis}, \citenamefont {Senoo}, \citenamefont {McGrew},\ and\ \citenamefont {Kaufman}}]{Jenkins2022}%
  \BibitemOpen
  \bibfield  {author} {\bibinfo {author} {\bibfnamefont {A.}~\bibnamefont {Jenkins}}, \bibinfo {author} {\bibfnamefont {J.~W.}\ \bibnamefont {Lis}}, \bibinfo {author} {\bibfnamefont {A.}~\bibnamefont {Senoo}}, \bibinfo {author} {\bibfnamefont {W.~F.}\ \bibnamefont {McGrew}},\ and\ \bibinfo {author} {\bibfnamefont {A.~M.}\ \bibnamefont {Kaufman}},\ }\bibfield  {title} {\bibinfo {title} {{Ytterbium Nuclear-Spin Qubits in an Optical Tweezer Array}},\ }\href {https://doi.org/10.1103/PhysRevX.12.021027} {\bibfield  {journal} {\bibinfo  {journal} {Phys. Rev. X}\ }\textbf {\bibinfo {volume} {12}},\ \bibinfo {pages} {021027} (\bibinfo {year} {2022})}\BibitemShut {NoStop}%
\bibitem [{\citenamefont {Morigi}\ \emph {et~al.}(2000)\citenamefont {Morigi}, \citenamefont {Eschner},\ and\ \citenamefont {Keitel}}]{Morigi2000}%
  \BibitemOpen
  \bibfield  {author} {\bibinfo {author} {\bibfnamefont {G.}~\bibnamefont {Morigi}}, \bibinfo {author} {\bibfnamefont {J.}~\bibnamefont {Eschner}},\ and\ \bibinfo {author} {\bibfnamefont {C.~H.}\ \bibnamefont {Keitel}},\ }\bibfield  {title} {\bibinfo {title} {{Ground State Laser Cooling Using Electromagnetically Induced Transparency}},\ }\href {https://doi.org/10.1103/PhysRevLett.85.4458} {\bibfield  {journal} {\bibinfo  {journal} {Phys. Rev. Lett.}\ }\textbf {\bibinfo {volume} {85}},\ \bibinfo {pages} {4458} (\bibinfo {year} {2000})}\BibitemShut {NoStop}%
\bibitem [{\citenamefont {Lis}\ \emph {et~al.}(2023)\citenamefont {Lis}, \citenamefont {Senoo}, \citenamefont {McGrew}, \citenamefont {R{\"{o}}nchen}, \citenamefont {Jenkins},\ and\ \citenamefont {Kaufman}}]{Lis2023}%
  \BibitemOpen
  \bibfield  {author} {\bibinfo {author} {\bibfnamefont {J.~W.}\ \bibnamefont {Lis}}, \bibinfo {author} {\bibfnamefont {A.}~\bibnamefont {Senoo}}, \bibinfo {author} {\bibfnamefont {W.~F.}\ \bibnamefont {McGrew}}, \bibinfo {author} {\bibfnamefont {F.}~\bibnamefont {R{\"{o}}nchen}}, \bibinfo {author} {\bibfnamefont {A.}~\bibnamefont {Jenkins}},\ and\ \bibinfo {author} {\bibfnamefont {A.~M.}\ \bibnamefont {Kaufman}},\ }\bibfield  {title} {\bibinfo {title} {{Midcircuit Operations Using the omg Architecture in Neutral Atom Arrays}},\ }\href {https://doi.org/10.1103/PhysRevX.13.041035} {\bibfield  {journal} {\bibinfo  {journal} {Phys. Rev. X}\ }\textbf {\bibinfo {volume} {13}},\ \bibinfo {pages} {041035} (\bibinfo {year} {2023})}\BibitemShut {NoStop}%
\bibitem [{\citenamefont {Barnes}\ \emph {et~al.}(2022)\citenamefont {Barnes}, \citenamefont {Battaglino}, \citenamefont {Bloom}, \citenamefont {Cassella}, \citenamefont {Coxe}, \citenamefont {Crisosto}, \citenamefont {King}, \citenamefont {Kondov}, \citenamefont {Kotru}, \citenamefont {Larsen}, \citenamefont {Lauigan}, \citenamefont {Lester}, \citenamefont {McDonald}, \citenamefont {Megidish}, \citenamefont {Narayanaswami}, \citenamefont {Nishiguchi}, \citenamefont {Notermans}, \citenamefont {Peng}, \citenamefont {Ryou}, \citenamefont {Wu},\ and\ \citenamefont {Yarwood}}]{Barnes2021}%
  \BibitemOpen
  \bibfield  {author} {\bibinfo {author} {\bibfnamefont {K.}~\bibnamefont {Barnes}}, \bibinfo {author} {\bibfnamefont {P.}~\bibnamefont {Battaglino}}, \bibinfo {author} {\bibfnamefont {B.~J.}\ \bibnamefont {Bloom}}, \bibinfo {author} {\bibfnamefont {K.}~\bibnamefont {Cassella}}, \bibinfo {author} {\bibfnamefont {R.}~\bibnamefont {Coxe}}, \bibinfo {author} {\bibfnamefont {N.}~\bibnamefont {Crisosto}}, \bibinfo {author} {\bibfnamefont {J.~P.}\ \bibnamefont {King}}, \bibinfo {author} {\bibfnamefont {S.~S.}\ \bibnamefont {Kondov}}, \bibinfo {author} {\bibfnamefont {K.}~\bibnamefont {Kotru}}, \bibinfo {author} {\bibfnamefont {S.~C.}\ \bibnamefont {Larsen}}, \bibinfo {author} {\bibfnamefont {J.}~\bibnamefont {Lauigan}}, \bibinfo {author} {\bibfnamefont {B.~J.}\ \bibnamefont {Lester}}, \bibinfo {author} {\bibfnamefont {M.}~\bibnamefont {McDonald}}, \bibinfo {author} {\bibfnamefont {E.}~\bibnamefont {Megidish}}, \bibinfo {author} {\bibfnamefont {S.}~\bibnamefont {Narayanaswami}}, \bibinfo {author} {\bibfnamefont
  {C.}~\bibnamefont {Nishiguchi}}, \bibinfo {author} {\bibfnamefont {R.}~\bibnamefont {Notermans}}, \bibinfo {author} {\bibfnamefont {L.~S.}\ \bibnamefont {Peng}}, \bibinfo {author} {\bibfnamefont {A.}~\bibnamefont {Ryou}}, \bibinfo {author} {\bibfnamefont {T.-Y.}\ \bibnamefont {Wu}},\ and\ \bibinfo {author} {\bibfnamefont {M.}~\bibnamefont {Yarwood}},\ }\bibfield  {title} {\bibinfo {title} {{Assembly and coherent control of a register of nuclear spin qubits}},\ }\href {https://doi.org/10.1038/s41467-022-29977-z} {\bibfield  {journal} {\bibinfo  {journal} {Nat. Commun.}\ }\textbf {\bibinfo {volume} {13}},\ \bibinfo {pages} {2779} (\bibinfo {year} {2022})}\BibitemShut {NoStop}%
\bibitem [{\citenamefont {Chen}\ \emph {et~al.}(2022)\citenamefont {Chen}, \citenamefont {Li}, \citenamefont {Huie}, \citenamefont {Zhao}, \citenamefont {Vetter}, \citenamefont {Greene},\ and\ \citenamefont {Covey}}]{Chen2022}%
  \BibitemOpen
  \bibfield  {author} {\bibinfo {author} {\bibfnamefont {N.}~\bibnamefont {Chen}}, \bibinfo {author} {\bibfnamefont {L.}~\bibnamefont {Li}}, \bibinfo {author} {\bibfnamefont {W.}~\bibnamefont {Huie}}, \bibinfo {author} {\bibfnamefont {M.}~\bibnamefont {Zhao}}, \bibinfo {author} {\bibfnamefont {I.}~\bibnamefont {Vetter}}, \bibinfo {author} {\bibfnamefont {C.~H.}\ \bibnamefont {Greene}},\ and\ \bibinfo {author} {\bibfnamefont {J.~P.}\ \bibnamefont {Covey}},\ }\bibfield  {title} {\bibinfo {title} {{Analyzing the Rydberg-based optical-metastable-ground architecture for $^{171}$Yb nuclear spins}},\ }\href {https://doi.org/10.1103/PhysRevA.105.052438} {\bibfield  {journal} {\bibinfo  {journal} {Phys. Rev. A}\ }\textbf {\bibinfo {volume} {105}},\ \bibinfo {pages} {052438} (\bibinfo {year} {2022})}\BibitemShut {NoStop}%
\bibitem [{\citenamefont {Ma}\ \emph {et~al.}(2023)\citenamefont {Ma}, \citenamefont {Liu}, \citenamefont {Peng}, \citenamefont {Zhang}, \citenamefont {Jandura}, \citenamefont {Claes}, \citenamefont {Burgers}, \citenamefont {Pupillo}, \citenamefont {Puri},\ and\ \citenamefont {Thompson}}]{Ma2023}%
  \BibitemOpen
  \bibfield  {author} {\bibinfo {author} {\bibfnamefont {S.}~\bibnamefont {Ma}}, \bibinfo {author} {\bibfnamefont {G.}~\bibnamefont {Liu}}, \bibinfo {author} {\bibfnamefont {P.}~\bibnamefont {Peng}}, \bibinfo {author} {\bibfnamefont {B.}~\bibnamefont {Zhang}}, \bibinfo {author} {\bibfnamefont {S.}~\bibnamefont {Jandura}}, \bibinfo {author} {\bibfnamefont {J.}~\bibnamefont {Claes}}, \bibinfo {author} {\bibfnamefont {A.~P.}\ \bibnamefont {Burgers}}, \bibinfo {author} {\bibfnamefont {G.}~\bibnamefont {Pupillo}}, \bibinfo {author} {\bibfnamefont {S.}~\bibnamefont {Puri}},\ and\ \bibinfo {author} {\bibfnamefont {J.~D.}\ \bibnamefont {Thompson}},\ }\bibfield  {title} {\bibinfo {title} {{High-fidelity gates and mid-circuit erasure conversion in an atomic qubit}},\ }\href {https://doi.org/10.1038/s41586-023-06438-1} {\bibfield  {journal} {\bibinfo  {journal} {Nature}\ }\textbf {\bibinfo {volume} {622}},\ \bibinfo {pages} {279} (\bibinfo {year} {2023})}\BibitemShut {NoStop}%
\bibitem [{\citenamefont {Peper}\ \emph {et~al.}(2024)\citenamefont {Peper}, \citenamefont {Li}, \citenamefont {Knapp}, \citenamefont {Bileska}, \citenamefont {Ma}, \citenamefont {Liu}, \citenamefont {Peng}, \citenamefont {Zhang}, \citenamefont {Horvath}, \citenamefont {Burgers},\ and\ \citenamefont {Thompson}}]{Peper2024}%
  \BibitemOpen
  \bibfield  {author} {\bibinfo {author} {\bibfnamefont {M.}~\bibnamefont {Peper}}, \bibinfo {author} {\bibfnamefont {Y.}~\bibnamefont {Li}}, \bibinfo {author} {\bibfnamefont {D.~Y.}\ \bibnamefont {Knapp}}, \bibinfo {author} {\bibfnamefont {M.}~\bibnamefont {Bileska}}, \bibinfo {author} {\bibfnamefont {S.}~\bibnamefont {Ma}}, \bibinfo {author} {\bibfnamefont {G.}~\bibnamefont {Liu}}, \bibinfo {author} {\bibfnamefont {P.}~\bibnamefont {Peng}}, \bibinfo {author} {\bibfnamefont {B.}~\bibnamefont {Zhang}}, \bibinfo {author} {\bibfnamefont {S.~P.}\ \bibnamefont {Horvath}}, \bibinfo {author} {\bibfnamefont {A.~P.}\ \bibnamefont {Burgers}},\ and\ \bibinfo {author} {\bibfnamefont {J.~D.}\ \bibnamefont {Thompson}},\ }\bibfield  {title} {\bibinfo {title} {{Spectroscopy and modeling of $^{171}$Yb Rydberg states for high-fidelity two-qubit gates}},\ }\href {http://arxiv.org/abs/2406.01482} {\bibfield  {journal} {\bibinfo  {journal} {arXiv Prepr.}\ }\textbf {\bibinfo {volume} {2406.01482}} (\bibinfo {year}
  {2024})}\BibitemShut {NoStop}%
\bibitem [{\citenamefont {Muniz}\ \emph {et~al.}(2024)\citenamefont {Muniz}, \citenamefont {Stone}, \citenamefont {Stack}, \citenamefont {Jaffe}, \citenamefont {Kindem}, \citenamefont {Wadleigh}, \citenamefont {Zalys-Geller}, \citenamefont {Zhang}, \citenamefont {Chen}, \citenamefont {Norcia}, \citenamefont {Epstein}, \citenamefont {Halperin}, \citenamefont {Hummel}, \citenamefont {Wilkason}, \citenamefont {Li}, \citenamefont {Barnes}, \citenamefont {Battaglino}, \citenamefont {Bohdanowicz}, \citenamefont {Booth}, \citenamefont {Brown}, \citenamefont {Brown}, \citenamefont {Cairncross}, \citenamefont {Cassella}, \citenamefont {Coxe}, \citenamefont {Crow}, \citenamefont {Feldkamp}, \citenamefont {Griger}, \citenamefont {Heinz}, \citenamefont {Jones}, \citenamefont {Kim}, \citenamefont {King}, \citenamefont {Kotru}, \citenamefont {Lauigan}, \citenamefont {Marjanovic}, \citenamefont {Megidish}, \citenamefont {Meredith}, \citenamefont {McDonald}, \citenamefont {Morshead}, \citenamefont {Narayanaswami},
  \citenamefont {Nishiguchi}, \citenamefont {Paule}, \citenamefont {Pawlak}, \citenamefont {Pudenz}, \citenamefont {P{\'{e}}rez}, \citenamefont {Ryou}, \citenamefont {Simon}, \citenamefont {Smull}, \citenamefont {Urbanek}, \citenamefont {van~de Veerdonk}, \citenamefont {Vendeiro}, \citenamefont {Wu}, \citenamefont {Xie},\ and\ \citenamefont {Bloom}}]{Muniz2024}%
  \BibitemOpen
  \bibfield  {author} {\bibinfo {author} {\bibfnamefont {J.~A.}\ \bibnamefont {Muniz}}, \bibinfo {author} {\bibfnamefont {M.}~\bibnamefont {Stone}}, \bibinfo {author} {\bibfnamefont {D.~T.}\ \bibnamefont {Stack}}, \bibinfo {author} {\bibfnamefont {M.}~\bibnamefont {Jaffe}}, \bibinfo {author} {\bibfnamefont {J.~M.}\ \bibnamefont {Kindem}}, \bibinfo {author} {\bibfnamefont {L.}~\bibnamefont {Wadleigh}}, \bibinfo {author} {\bibfnamefont {E.}~\bibnamefont {Zalys-Geller}}, \bibinfo {author} {\bibfnamefont {X.}~\bibnamefont {Zhang}}, \bibinfo {author} {\bibfnamefont {C.~A.}\ \bibnamefont {Chen}}, \bibinfo {author} {\bibfnamefont {M.~A.}\ \bibnamefont {Norcia}}, \bibinfo {author} {\bibfnamefont {J.}~\bibnamefont {Epstein}}, \bibinfo {author} {\bibfnamefont {E.}~\bibnamefont {Halperin}}, \bibinfo {author} {\bibfnamefont {F.}~\bibnamefont {Hummel}}, \bibinfo {author} {\bibfnamefont {T.}~\bibnamefont {Wilkason}}, \bibinfo {author} {\bibfnamefont {M.}~\bibnamefont {Li}}, \bibinfo {author} {\bibfnamefont
  {K.}~\bibnamefont {Barnes}}, \bibinfo {author} {\bibfnamefont {P.}~\bibnamefont {Battaglino}}, \bibinfo {author} {\bibfnamefont {T.~C.}\ \bibnamefont {Bohdanowicz}}, \bibinfo {author} {\bibfnamefont {G.}~\bibnamefont {Booth}}, \bibinfo {author} {\bibfnamefont {A.}~\bibnamefont {Brown}}, \bibinfo {author} {\bibfnamefont {M.~O.}\ \bibnamefont {Brown}}, \bibinfo {author} {\bibfnamefont {W.~B.}\ \bibnamefont {Cairncross}}, \bibinfo {author} {\bibfnamefont {K.}~\bibnamefont {Cassella}}, \bibinfo {author} {\bibfnamefont {R.}~\bibnamefont {Coxe}}, \bibinfo {author} {\bibfnamefont {D.}~\bibnamefont {Crow}}, \bibinfo {author} {\bibfnamefont {M.}~\bibnamefont {Feldkamp}}, \bibinfo {author} {\bibfnamefont {C.}~\bibnamefont {Griger}}, \bibinfo {author} {\bibfnamefont {A.}~\bibnamefont {Heinz}}, \bibinfo {author} {\bibfnamefont {A.~M.~W.}\ \bibnamefont {Jones}}, \bibinfo {author} {\bibfnamefont {H.}~\bibnamefont {Kim}}, \bibinfo {author} {\bibfnamefont {J.}~\bibnamefont {King}}, \bibinfo {author} {\bibfnamefont
  {K.}~\bibnamefont {Kotru}}, \bibinfo {author} {\bibfnamefont {J.}~\bibnamefont {Lauigan}}, \bibinfo {author} {\bibfnamefont {J.}~\bibnamefont {Marjanovic}}, \bibinfo {author} {\bibfnamefont {E.}~\bibnamefont {Megidish}}, \bibinfo {author} {\bibfnamefont {M.}~\bibnamefont {Meredith}}, \bibinfo {author} {\bibfnamefont {M.}~\bibnamefont {McDonald}}, \bibinfo {author} {\bibfnamefont {R.}~\bibnamefont {Morshead}}, \bibinfo {author} {\bibfnamefont {S.}~\bibnamefont {Narayanaswami}}, \bibinfo {author} {\bibfnamefont {C.}~\bibnamefont {Nishiguchi}}, \bibinfo {author} {\bibfnamefont {T.}~\bibnamefont {Paule}}, \bibinfo {author} {\bibfnamefont {K.~A.}\ \bibnamefont {Pawlak}}, \bibinfo {author} {\bibfnamefont {K.~L.}\ \bibnamefont {Pudenz}}, \bibinfo {author} {\bibfnamefont {D.~R.}\ \bibnamefont {P{\'{e}}rez}}, \bibinfo {author} {\bibfnamefont {A.}~\bibnamefont {Ryou}}, \bibinfo {author} {\bibfnamefont {J.}~\bibnamefont {Simon}}, \bibinfo {author} {\bibfnamefont {A.}~\bibnamefont {Smull}}, \bibinfo {author}
  {\bibfnamefont {M.}~\bibnamefont {Urbanek}}, \bibinfo {author} {\bibfnamefont {R.~J.~M.}\ \bibnamefont {van~de Veerdonk}}, \bibinfo {author} {\bibfnamefont {Z.}~\bibnamefont {Vendeiro}}, \bibinfo {author} {\bibfnamefont {T.~Y.}\ \bibnamefont {Wu}}, \bibinfo {author} {\bibfnamefont {X.}~\bibnamefont {Xie}},\ and\ \bibinfo {author} {\bibfnamefont {B.~J.}\ \bibnamefont {Bloom}},\ }\bibfield  {title} {\bibinfo {title} {{High-fidelity universal gates in the $^{171}$Yb ground state nuclear spin qubit}},\ }\href {http://arxiv.org/abs/2411.11708} {\bibfield  {journal} {\bibinfo  {journal} {arXiv Prepr.}\ }\textbf {\bibinfo {volume} {2411.11708}} (\bibinfo {year} {2024})}\BibitemShut {NoStop}%
\bibitem [{\citenamefont {Madjarov}\ \emph {et~al.}(2020)\citenamefont {Madjarov}, \citenamefont {Covey}, \citenamefont {Shaw}, \citenamefont {Choi}, \citenamefont {Kale}, \citenamefont {Cooper}, \citenamefont {Pichler}, \citenamefont {Schkolnik}, \citenamefont {Williams},\ and\ \citenamefont {Endres}}]{Madjarov2020}%
  \BibitemOpen
  \bibfield  {author} {\bibinfo {author} {\bibfnamefont {I.~S.}\ \bibnamefont {Madjarov}}, \bibinfo {author} {\bibfnamefont {J.~P.}\ \bibnamefont {Covey}}, \bibinfo {author} {\bibfnamefont {A.~L.}\ \bibnamefont {Shaw}}, \bibinfo {author} {\bibfnamefont {J.}~\bibnamefont {Choi}}, \bibinfo {author} {\bibfnamefont {A.}~\bibnamefont {Kale}}, \bibinfo {author} {\bibfnamefont {A.}~\bibnamefont {Cooper}}, \bibinfo {author} {\bibfnamefont {H.}~\bibnamefont {Pichler}}, \bibinfo {author} {\bibfnamefont {V.}~\bibnamefont {Schkolnik}}, \bibinfo {author} {\bibfnamefont {J.~R.}\ \bibnamefont {Williams}},\ and\ \bibinfo {author} {\bibfnamefont {M.}~\bibnamefont {Endres}},\ }\bibfield  {title} {\bibinfo {title} {{High-fidelity entanglement and detection of alkaline-earth Rydberg atoms}},\ }\href {https://doi.org/10.1038/s41567-020-0903-z} {\bibfield  {journal} {\bibinfo  {journal} {Nat. Phys.}\ }\textbf {\bibinfo {volume} {16}},\ \bibinfo {pages} {857} (\bibinfo {year} {2020})}\BibitemShut {NoStop}%
\bibitem [{\citenamefont {Li}\ \emph {et~al.}(2022)\citenamefont {Li}, \citenamefont {Huie}, \citenamefont {Chen}, \citenamefont {DeMarco},\ and\ \citenamefont {Covey}}]{Li2022}%
  \BibitemOpen
  \bibfield  {author} {\bibinfo {author} {\bibfnamefont {L.}~\bibnamefont {Li}}, \bibinfo {author} {\bibfnamefont {W.}~\bibnamefont {Huie}}, \bibinfo {author} {\bibfnamefont {N.}~\bibnamefont {Chen}}, \bibinfo {author} {\bibfnamefont {B.}~\bibnamefont {DeMarco}},\ and\ \bibinfo {author} {\bibfnamefont {J.~P.}\ \bibnamefont {Covey}},\ }\bibfield  {title} {\bibinfo {title} {{Active Cancellation of Servo-Induced Noise on Stabilized Lasers via Feedforward}},\ }\href {https://doi.org/10.1103/PhysRevApplied.18.064005} {\bibfield  {journal} {\bibinfo  {journal} {Phys. Rev. Appl.}\ }\textbf {\bibinfo {volume} {18}},\ \bibinfo {pages} {064005} (\bibinfo {year} {2022})}\BibitemShut {NoStop}%
\bibitem [{\citenamefont {Carolan}\ \emph {et~al.}(2015)\citenamefont {Carolan}, \citenamefont {Harrold}, \citenamefont {Sparrow}, \citenamefont {Mart{\'{i}}n-L{\'{o}}pez}, \citenamefont {Russell}, \citenamefont {Silverstone}, \citenamefont {Shadbolt}, \citenamefont {Matsuda}, \citenamefont {Oguma}, \citenamefont {Itoh}, \citenamefont {Marshall}, \citenamefont {Thompson}, \citenamefont {Matthews}, \citenamefont {Hashimoto}, \citenamefont {O'Brien},\ and\ \citenamefont {Laing}}]{Carolan2015}%
  \BibitemOpen
  \bibfield  {author} {\bibinfo {author} {\bibfnamefont {J.}~\bibnamefont {Carolan}}, \bibinfo {author} {\bibfnamefont {C.}~\bibnamefont {Harrold}}, \bibinfo {author} {\bibfnamefont {C.}~\bibnamefont {Sparrow}}, \bibinfo {author} {\bibfnamefont {E.}~\bibnamefont {Mart{\'{i}}n-L{\'{o}}pez}}, \bibinfo {author} {\bibfnamefont {N.~J.}\ \bibnamefont {Russell}}, \bibinfo {author} {\bibfnamefont {J.~W.}\ \bibnamefont {Silverstone}}, \bibinfo {author} {\bibfnamefont {P.~J.}\ \bibnamefont {Shadbolt}}, \bibinfo {author} {\bibfnamefont {N.}~\bibnamefont {Matsuda}}, \bibinfo {author} {\bibfnamefont {M.}~\bibnamefont {Oguma}}, \bibinfo {author} {\bibfnamefont {M.}~\bibnamefont {Itoh}}, \bibinfo {author} {\bibfnamefont {G.~D.}\ \bibnamefont {Marshall}}, \bibinfo {author} {\bibfnamefont {M.~G.}\ \bibnamefont {Thompson}}, \bibinfo {author} {\bibfnamefont {J.~C.~F.}\ \bibnamefont {Matthews}}, \bibinfo {author} {\bibfnamefont {T.}~\bibnamefont {Hashimoto}}, \bibinfo {author} {\bibfnamefont {J.~L.}\ \bibnamefont {O'Brien}},\
  and\ \bibinfo {author} {\bibfnamefont {A.}~\bibnamefont {Laing}},\ }\bibfield  {title} {\bibinfo {title} {{Universal linear optics}},\ }\href {https://doi.org/10.1126/science.aab3642} {\bibfield  {journal} {\bibinfo  {journal} {Science}\ }\textbf {\bibinfo {volume} {349}},\ \bibinfo {pages} {711} (\bibinfo {year} {2015})}\BibitemShut {NoStop}%
\bibitem [{\citenamefont {Pelucchi}\ \emph {et~al.}(2021)\citenamefont {Pelucchi}, \citenamefont {Fagas}, \citenamefont {Aharonovich}, \citenamefont {Englund}, \citenamefont {Figueroa}, \citenamefont {Gong}, \citenamefont {Hannes}, \citenamefont {Liu}, \citenamefont {Lu}, \citenamefont {Matsuda}, \citenamefont {Pan}, \citenamefont {Schreck}, \citenamefont {Sciarrino}, \citenamefont {Silberhorn}, \citenamefont {Wang},\ and\ \citenamefont {J{\"{o}}ns}}]{Pelucchi2022}%
  \BibitemOpen
  \bibfield  {author} {\bibinfo {author} {\bibfnamefont {E.}~\bibnamefont {Pelucchi}}, \bibinfo {author} {\bibfnamefont {G.}~\bibnamefont {Fagas}}, \bibinfo {author} {\bibfnamefont {I.}~\bibnamefont {Aharonovich}}, \bibinfo {author} {\bibfnamefont {D.}~\bibnamefont {Englund}}, \bibinfo {author} {\bibfnamefont {E.}~\bibnamefont {Figueroa}}, \bibinfo {author} {\bibfnamefont {Q.}~\bibnamefont {Gong}}, \bibinfo {author} {\bibfnamefont {H.}~\bibnamefont {Hannes}}, \bibinfo {author} {\bibfnamefont {J.}~\bibnamefont {Liu}}, \bibinfo {author} {\bibfnamefont {C.-Y.}\ \bibnamefont {Lu}}, \bibinfo {author} {\bibfnamefont {N.}~\bibnamefont {Matsuda}}, \bibinfo {author} {\bibfnamefont {J.-W.}\ \bibnamefont {Pan}}, \bibinfo {author} {\bibfnamefont {F.}~\bibnamefont {Schreck}}, \bibinfo {author} {\bibfnamefont {F.}~\bibnamefont {Sciarrino}}, \bibinfo {author} {\bibfnamefont {C.}~\bibnamefont {Silberhorn}}, \bibinfo {author} {\bibfnamefont {J.}~\bibnamefont {Wang}},\ and\ \bibinfo {author} {\bibfnamefont {K.~D.}\
  \bibnamefont {J{\"{o}}ns}},\ }\bibfield  {title} {\bibinfo {title} {{The potential and global outlook of integrated photonics for quantum technologies}},\ }\href {https://doi.org/10.1038/s42254-021-00398-z} {\bibfield  {journal} {\bibinfo  {journal} {Nat. Rev. Phys.}\ }\textbf {\bibinfo {volume} {4}},\ \bibinfo {pages} {194} (\bibinfo {year} {2021})}\BibitemShut {NoStop}%
\bibitem [{\citenamefont {Wollman}\ \emph {et~al.}(2019)\citenamefont {Wollman}, \citenamefont {Verma}, \citenamefont {Lita}, \citenamefont {Farr}, \citenamefont {Shaw}, \citenamefont {Mirin},\ and\ \citenamefont {{Woo Nam}}}]{Wollman2019}%
  \BibitemOpen
  \bibfield  {author} {\bibinfo {author} {\bibfnamefont {E.~E.}\ \bibnamefont {Wollman}}, \bibinfo {author} {\bibfnamefont {V.~B.}\ \bibnamefont {Verma}}, \bibinfo {author} {\bibfnamefont {A.~E.}\ \bibnamefont {Lita}}, \bibinfo {author} {\bibfnamefont {W.~H.}\ \bibnamefont {Farr}}, \bibinfo {author} {\bibfnamefont {M.~D.}\ \bibnamefont {Shaw}}, \bibinfo {author} {\bibfnamefont {R.~P.}\ \bibnamefont {Mirin}},\ and\ \bibinfo {author} {\bibfnamefont {S.}~\bibnamefont {{Woo Nam}}},\ }\bibfield  {title} {\bibinfo {title} {{Kilopixel array of superconducting nanowire single-photon detectors}},\ }\href {https://doi.org/10.1364/OE.27.035279} {\bibfield  {journal} {\bibinfo  {journal} {Opt. Express}\ }\textbf {\bibinfo {volume} {27}},\ \bibinfo {pages} {35279} (\bibinfo {year} {2019})}\BibitemShut {NoStop}%
\bibitem [{\citenamefont {Oripov}\ \emph {et~al.}(2023)\citenamefont {Oripov}, \citenamefont {Rampini}, \citenamefont {Allmaras}, \citenamefont {Shaw}, \citenamefont {Nam}, \citenamefont {Korzh},\ and\ \citenamefont {McCaughan}}]{Oripov2023}%
  \BibitemOpen
  \bibfield  {author} {\bibinfo {author} {\bibfnamefont {B.~G.}\ \bibnamefont {Oripov}}, \bibinfo {author} {\bibfnamefont {D.~S.}\ \bibnamefont {Rampini}}, \bibinfo {author} {\bibfnamefont {J.}~\bibnamefont {Allmaras}}, \bibinfo {author} {\bibfnamefont {M.~D.}\ \bibnamefont {Shaw}}, \bibinfo {author} {\bibfnamefont {S.~W.}\ \bibnamefont {Nam}}, \bibinfo {author} {\bibfnamefont {B.}~\bibnamefont {Korzh}},\ and\ \bibinfo {author} {\bibfnamefont {A.~N.}\ \bibnamefont {McCaughan}},\ }\bibfield  {title} {\bibinfo {title} {{A superconducting nanowire single-photon camera with 400,000 pixels}},\ }\href {https://doi.org/10.1038/s41586-023-06550-2} {\bibfield  {journal} {\bibinfo  {journal} {Nature}\ }\textbf {\bibinfo {volume} {622}},\ \bibinfo {pages} {730} (\bibinfo {year} {2023})}\BibitemShut {NoStop}%
\bibitem [{\citenamefont {Fleming}\ \emph {et~al.}(2025)\citenamefont {Fleming}, \citenamefont {McCutcheon}, \citenamefont {Wollman}, \citenamefont {Beyer}, \citenamefont {Anant}, \citenamefont {Korzh}, \citenamefont {Allmaras}, \citenamefont {Narv{\'{a}}ez}, \citenamefont {Leedumrongwatthanakun}, \citenamefont {Buller}, \citenamefont {Malik},\ and\ \citenamefont {Shaw}}]{Fleming2025}%
  \BibitemOpen
  \bibfield  {author} {\bibinfo {author} {\bibfnamefont {F.}~\bibnamefont {Fleming}}, \bibinfo {author} {\bibfnamefont {W.}~\bibnamefont {McCutcheon}}, \bibinfo {author} {\bibfnamefont {E.~E.}\ \bibnamefont {Wollman}}, \bibinfo {author} {\bibfnamefont {A.~D.}\ \bibnamefont {Beyer}}, \bibinfo {author} {\bibfnamefont {V.}~\bibnamefont {Anant}}, \bibinfo {author} {\bibfnamefont {B.}~\bibnamefont {Korzh}}, \bibinfo {author} {\bibfnamefont {J.~P.}\ \bibnamefont {Allmaras}}, \bibinfo {author} {\bibfnamefont {L.}~\bibnamefont {Narv{\'{a}}ez}}, \bibinfo {author} {\bibfnamefont {S.}~\bibnamefont {Leedumrongwatthanakun}}, \bibinfo {author} {\bibfnamefont {G.~S.}\ \bibnamefont {Buller}}, \bibinfo {author} {\bibfnamefont {M.}~\bibnamefont {Malik}},\ and\ \bibinfo {author} {\bibfnamefont {M.~D.}\ \bibnamefont {Shaw}},\ }\bibfield  {title} {\bibinfo {title} {{High-efficiency, high-count-rate 2D superconducting nanowire single-photon detector array}},\ }\href {http://arxiv.org/abs/2501.07357} {\bibfield  {journal} {\bibinfo
   {journal} {arXiv Prepr.}\ }\textbf {\bibinfo {volume} {2501.07357}} (\bibinfo {year} {2025})}\BibitemShut {NoStop}%
\bibitem [{\citenamefont {{Scholl, P. and Shaw, A. L. and Finkelstein, R. and Tsai, R. B.-S. and Choi, J. and Endres}}(2023)}]{Scholl2023b}%
  \BibitemOpen
  \bibfield  {author} {\bibinfo {author} {\bibfnamefont {M.}~\bibnamefont {{Scholl, P. and Shaw, A. L. and Finkelstein, R. and Tsai, R. B.-S. and Choi, J. and Endres}}},\ }\bibfield  {title} {\bibinfo {title} {{Erasure-cooling, control, and hyper-entanglement of motion in optical tweezers}},\ }\href@noop {} {\bibfield  {journal} {\bibinfo  {journal} {arXiv Prepr.}\ }\textbf {\bibinfo {volume} {2311.15580}} (\bibinfo {year} {2023})}\BibitemShut {NoStop}%
\bibitem [{\citenamefont {Graham}\ \emph {et~al.}(2023)\citenamefont {Graham}, \citenamefont {Phuttitarn}, \citenamefont {Chinnarasu}, \citenamefont {Song}, \citenamefont {Poole}, \citenamefont {Jooya}, \citenamefont {Scott}, \citenamefont {Scott}, \citenamefont {Eichler},\ and\ \citenamefont {Saffman}}]{Graham2023}%
  \BibitemOpen
  \bibfield  {author} {\bibinfo {author} {\bibfnamefont {T.~M.}\ \bibnamefont {Graham}}, \bibinfo {author} {\bibfnamefont {L.}~\bibnamefont {Phuttitarn}}, \bibinfo {author} {\bibfnamefont {R.}~\bibnamefont {Chinnarasu}}, \bibinfo {author} {\bibfnamefont {Y.}~\bibnamefont {Song}}, \bibinfo {author} {\bibfnamefont {C.}~\bibnamefont {Poole}}, \bibinfo {author} {\bibfnamefont {K.}~\bibnamefont {Jooya}}, \bibinfo {author} {\bibfnamefont {J.}~\bibnamefont {Scott}}, \bibinfo {author} {\bibfnamefont {A.}~\bibnamefont {Scott}}, \bibinfo {author} {\bibfnamefont {P.}~\bibnamefont {Eichler}},\ and\ \bibinfo {author} {\bibfnamefont {M.}~\bibnamefont {Saffman}},\ }\bibfield  {title} {\bibinfo {title} {{Mid-circuit measurements on a neutral atom quantum processor}},\ }\href {http://arxiv.org/abs/2303.10051} {\bibfield  {journal} {\bibinfo  {journal} {arXiv Prepr.}\ }\textbf {\bibinfo {volume} {2303.10051}} (\bibinfo {year} {2023})}\BibitemShut {NoStop}%
\bibitem [{\citenamefont {Singh}\ \emph {et~al.}(2023)\citenamefont {Singh}, \citenamefont {Bradley}, \citenamefont {Anand}, \citenamefont {Ramesh}, \citenamefont {White},\ and\ \citenamefont {Bernien}}]{Singh2022}%
  \BibitemOpen
  \bibfield  {author} {\bibinfo {author} {\bibfnamefont {K.}~\bibnamefont {Singh}}, \bibinfo {author} {\bibfnamefont {C.~E.}\ \bibnamefont {Bradley}}, \bibinfo {author} {\bibfnamefont {S.}~\bibnamefont {Anand}}, \bibinfo {author} {\bibfnamefont {V.}~\bibnamefont {Ramesh}}, \bibinfo {author} {\bibfnamefont {R.}~\bibnamefont {White}},\ and\ \bibinfo {author} {\bibfnamefont {H.}~\bibnamefont {Bernien}},\ }\bibfield  {title} {\bibinfo {title} {{Mid-circuit correction of correlated phase errors using an array of spectator qubits}},\ }\href {https://doi.org/10.1126/science.ade5337} {\bibfield  {journal} {\bibinfo  {journal} {Science}\ }\textbf {\bibinfo {volume} {380}},\ \bibinfo {pages} {1265} (\bibinfo {year} {2023})}\BibitemShut {NoStop}%
\bibitem [{\citenamefont {Nakamura}\ \emph {et~al.}(2024)\citenamefont {Nakamura}, \citenamefont {Kusano}, \citenamefont {Yokoyama}, \citenamefont {Saito}, \citenamefont {Higashi}, \citenamefont {Ozawa}, \citenamefont {Takano}, \citenamefont {Takasu},\ and\ \citenamefont {Takahashi}}]{Nakamura2024}%
  \BibitemOpen
  \bibfield  {author} {\bibinfo {author} {\bibfnamefont {Y.}~\bibnamefont {Nakamura}}, \bibinfo {author} {\bibfnamefont {T.}~\bibnamefont {Kusano}}, \bibinfo {author} {\bibfnamefont {R.}~\bibnamefont {Yokoyama}}, \bibinfo {author} {\bibfnamefont {K.}~\bibnamefont {Saito}}, \bibinfo {author} {\bibfnamefont {K.}~\bibnamefont {Higashi}}, \bibinfo {author} {\bibfnamefont {N.}~\bibnamefont {Ozawa}}, \bibinfo {author} {\bibfnamefont {T.}~\bibnamefont {Takano}}, \bibinfo {author} {\bibfnamefont {Y.}~\bibnamefont {Takasu}},\ and\ \bibinfo {author} {\bibfnamefont {Y.}~\bibnamefont {Takahashi}},\ }\bibfield  {title} {\bibinfo {title} {{A hybrid atom tweezer array of nuclear spin and optical clock qubits}},\ }\href {http://arxiv.org/abs/2406.12247} {\bibfield  {journal} {\bibinfo  {journal} {arXiv Prepr.}\ }\textbf {\bibinfo {volume} {2406.12247}} (\bibinfo {year} {2024})},\ \Eprint {https://arxiv.org/abs/2406.12247} {arXiv:2406.12247} \BibitemShut {NoStop}%
\bibitem [{\citenamefont {Norcia}\ \emph {et~al.}(2023)\citenamefont {Norcia}, \citenamefont {Cairncross}, \citenamefont {Barnes}, \citenamefont {Battaglino}, \citenamefont {Brown}, \citenamefont {Brown}, \citenamefont {Cassella}, \citenamefont {Chen}, \citenamefont {Coxe}, \citenamefont {Crow}, \citenamefont {Epstein}, \citenamefont {Griger}, \citenamefont {Jones}, \citenamefont {Kim}, \citenamefont {Kindem}, \citenamefont {King}, \citenamefont {Kondov}, \citenamefont {Kotru}, \citenamefont {Lauigan}, \citenamefont {Li}, \citenamefont {Lu}, \citenamefont {Megidish}, \citenamefont {Marjanovic}, \citenamefont {McDonald}, \citenamefont {Mittiga}, \citenamefont {Muniz}, \citenamefont {Narayanaswami}, \citenamefont {Nishiguchi}, \citenamefont {Notermans}, \citenamefont {Paule}, \citenamefont {Pawlak}, \citenamefont {Peng}, \citenamefont {Ryou}, \citenamefont {Smull}, \citenamefont {Stack}, \citenamefont {Stone}, \citenamefont {Sucich}, \citenamefont {Urbanek}, \citenamefont {van~de Veerdonk}, \citenamefont
  {Vendeiro}, \citenamefont {Wilkason}, \citenamefont {Wu}, \citenamefont {Xie}, \citenamefont {Zhang},\ and\ \citenamefont {Bloom}}]{Norcia2023}%
  \BibitemOpen
  \bibfield  {author} {\bibinfo {author} {\bibfnamefont {M.~A.}\ \bibnamefont {Norcia}}, \bibinfo {author} {\bibfnamefont {W.~B.}\ \bibnamefont {Cairncross}}, \bibinfo {author} {\bibfnamefont {K.}~\bibnamefont {Barnes}}, \bibinfo {author} {\bibfnamefont {P.}~\bibnamefont {Battaglino}}, \bibinfo {author} {\bibfnamefont {A.}~\bibnamefont {Brown}}, \bibinfo {author} {\bibfnamefont {M.~O.}\ \bibnamefont {Brown}}, \bibinfo {author} {\bibfnamefont {K.}~\bibnamefont {Cassella}}, \bibinfo {author} {\bibfnamefont {C.-A.}\ \bibnamefont {Chen}}, \bibinfo {author} {\bibfnamefont {R.}~\bibnamefont {Coxe}}, \bibinfo {author} {\bibfnamefont {D.}~\bibnamefont {Crow}}, \bibinfo {author} {\bibfnamefont {J.}~\bibnamefont {Epstein}}, \bibinfo {author} {\bibfnamefont {C.}~\bibnamefont {Griger}}, \bibinfo {author} {\bibfnamefont {A.~M.~W.}\ \bibnamefont {Jones}}, \bibinfo {author} {\bibfnamefont {H.}~\bibnamefont {Kim}}, \bibinfo {author} {\bibfnamefont {J.~M.}\ \bibnamefont {Kindem}}, \bibinfo {author} {\bibfnamefont
  {J.}~\bibnamefont {King}}, \bibinfo {author} {\bibfnamefont {S.~S.}\ \bibnamefont {Kondov}}, \bibinfo {author} {\bibfnamefont {K.}~\bibnamefont {Kotru}}, \bibinfo {author} {\bibfnamefont {J.}~\bibnamefont {Lauigan}}, \bibinfo {author} {\bibfnamefont {M.}~\bibnamefont {Li}}, \bibinfo {author} {\bibfnamefont {M.}~\bibnamefont {Lu}}, \bibinfo {author} {\bibfnamefont {E.}~\bibnamefont {Megidish}}, \bibinfo {author} {\bibfnamefont {J.}~\bibnamefont {Marjanovic}}, \bibinfo {author} {\bibfnamefont {M.}~\bibnamefont {McDonald}}, \bibinfo {author} {\bibfnamefont {T.}~\bibnamefont {Mittiga}}, \bibinfo {author} {\bibfnamefont {J.~A.}\ \bibnamefont {Muniz}}, \bibinfo {author} {\bibfnamefont {S.}~\bibnamefont {Narayanaswami}}, \bibinfo {author} {\bibfnamefont {C.}~\bibnamefont {Nishiguchi}}, \bibinfo {author} {\bibfnamefont {R.}~\bibnamefont {Notermans}}, \bibinfo {author} {\bibfnamefont {T.}~\bibnamefont {Paule}}, \bibinfo {author} {\bibfnamefont {K.~A.}\ \bibnamefont {Pawlak}}, \bibinfo {author} {\bibfnamefont
  {L.~S.}\ \bibnamefont {Peng}}, \bibinfo {author} {\bibfnamefont {A.}~\bibnamefont {Ryou}}, \bibinfo {author} {\bibfnamefont {A.}~\bibnamefont {Smull}}, \bibinfo {author} {\bibfnamefont {D.}~\bibnamefont {Stack}}, \bibinfo {author} {\bibfnamefont {M.}~\bibnamefont {Stone}}, \bibinfo {author} {\bibfnamefont {A.}~\bibnamefont {Sucich}}, \bibinfo {author} {\bibfnamefont {M.}~\bibnamefont {Urbanek}}, \bibinfo {author} {\bibfnamefont {R.~J.~M.}\ \bibnamefont {van~de Veerdonk}}, \bibinfo {author} {\bibfnamefont {Z.}~\bibnamefont {Vendeiro}}, \bibinfo {author} {\bibfnamefont {T.}~\bibnamefont {Wilkason}}, \bibinfo {author} {\bibfnamefont {T.-Y.}\ \bibnamefont {Wu}}, \bibinfo {author} {\bibfnamefont {X.}~\bibnamefont {Xie}}, \bibinfo {author} {\bibfnamefont {X.}~\bibnamefont {Zhang}},\ and\ \bibinfo {author} {\bibfnamefont {B.~J.}\ \bibnamefont {Bloom}},\ }\bibfield  {title} {\bibinfo {title} {{Midcircuit Qubit Measurement and Rearrangement in a Yb-171 Atomic Array}},\ }\href
  {https://doi.org/10.1103/PhysRevX.13.041034} {\bibfield  {journal} {\bibinfo  {journal} {Phys. Rev. X}\ }\textbf {\bibinfo {volume} {13}},\ \bibinfo {pages} {041034} (\bibinfo {year} {2023})}\BibitemShut {NoStop}%
\bibitem [{\citenamefont {Hu}\ \emph {et~al.}(2024)\citenamefont {Hu}, \citenamefont {Sinclair}, \citenamefont {Bytyqi}, \citenamefont {Chong}, \citenamefont {Rudelis}, \citenamefont {Ramette}, \citenamefont {Vendeiro},\ and\ \citenamefont {Vuleti{\'{c}}}}]{Hu2024}%
  \BibitemOpen
  \bibfield  {author} {\bibinfo {author} {\bibfnamefont {B.}~\bibnamefont {Hu}}, \bibinfo {author} {\bibfnamefont {J.}~\bibnamefont {Sinclair}}, \bibinfo {author} {\bibfnamefont {E.}~\bibnamefont {Bytyqi}}, \bibinfo {author} {\bibfnamefont {M.}~\bibnamefont {Chong}}, \bibinfo {author} {\bibfnamefont {A.}~\bibnamefont {Rudelis}}, \bibinfo {author} {\bibfnamefont {J.}~\bibnamefont {Ramette}}, \bibinfo {author} {\bibfnamefont {Z.}~\bibnamefont {Vendeiro}},\ and\ \bibinfo {author} {\bibfnamefont {V.}~\bibnamefont {Vuleti{\'{c}}}},\ }\bibfield  {title} {\bibinfo {title} {{Site-selective cavity readout and classical error correction of a 5-bit atomic register}},\ }\href {http://arxiv.org/abs/2408.15329} {\bibfield  {journal} {\bibinfo  {journal} {arXiv Prepr.}\ }\textbf {\bibinfo {volume} {2408.15329}} (\bibinfo {year} {2024})}\BibitemShut {NoStop}%
\bibitem [{\citenamefont {Deist}\ \emph {et~al.}(2022)\citenamefont {Deist}, \citenamefont {Lu}, \citenamefont {Ho}, \citenamefont {Pasha}, \citenamefont {Zeiher}, \citenamefont {Yan},\ and\ \citenamefont {Stamper-Kurn}}]{Deist2022b}%
  \BibitemOpen
  \bibfield  {author} {\bibinfo {author} {\bibfnamefont {E.}~\bibnamefont {Deist}}, \bibinfo {author} {\bibfnamefont {Y.-H.}\ \bibnamefont {Lu}}, \bibinfo {author} {\bibfnamefont {J.}~\bibnamefont {Ho}}, \bibinfo {author} {\bibfnamefont {M.~K.}\ \bibnamefont {Pasha}}, \bibinfo {author} {\bibfnamefont {J.}~\bibnamefont {Zeiher}}, \bibinfo {author} {\bibfnamefont {Z.}~\bibnamefont {Yan}},\ and\ \bibinfo {author} {\bibfnamefont {D.~M.}\ \bibnamefont {Stamper-Kurn}},\ }\bibfield  {title} {\bibinfo {title} {{Mid-Circuit Cavity Measurement in a Neutral Atom Array}},\ }\href {https://doi.org/10.1103/PhysRevLett.129.203602} {\bibfield  {journal} {\bibinfo  {journal} {Phys. Rev. Lett.}\ }\textbf {\bibinfo {volume} {129}},\ \bibinfo {pages} {203602} (\bibinfo {year} {2022})}\BibitemShut {NoStop}%
\bibitem [{\citenamefont {Bluvstein}\ \emph {et~al.}(2023)\citenamefont {Bluvstein}, \citenamefont {Evered}, \citenamefont {Geim}, \citenamefont {Li}, \citenamefont {Zhou}, \citenamefont {Manovitz}, \citenamefont {Ebadi}, \citenamefont {Cain}, \citenamefont {Kalinowski}, \citenamefont {Hangleiter}, \citenamefont {Ataides}, \citenamefont {Maskara}, \citenamefont {Cong}, \citenamefont {Gao}, \citenamefont {Rodriguez}, \citenamefont {Karolyshyn}, \citenamefont {Semeghini}, \citenamefont {Gullans}, \citenamefont {Greiner}, \citenamefont {Vuleti{\'{c}}},\ and\ \citenamefont {Lukin}}]{Bluvstein2023}%
  \BibitemOpen
  \bibfield  {author} {\bibinfo {author} {\bibfnamefont {D.}~\bibnamefont {Bluvstein}}, \bibinfo {author} {\bibfnamefont {S.~J.}\ \bibnamefont {Evered}}, \bibinfo {author} {\bibfnamefont {A.~A.}\ \bibnamefont {Geim}}, \bibinfo {author} {\bibfnamefont {S.~H.}\ \bibnamefont {Li}}, \bibinfo {author} {\bibfnamefont {H.}~\bibnamefont {Zhou}}, \bibinfo {author} {\bibfnamefont {T.}~\bibnamefont {Manovitz}}, \bibinfo {author} {\bibfnamefont {S.}~\bibnamefont {Ebadi}}, \bibinfo {author} {\bibfnamefont {M.}~\bibnamefont {Cain}}, \bibinfo {author} {\bibfnamefont {M.}~\bibnamefont {Kalinowski}}, \bibinfo {author} {\bibfnamefont {D.}~\bibnamefont {Hangleiter}}, \bibinfo {author} {\bibfnamefont {J.~P.~B.}\ \bibnamefont {Ataides}}, \bibinfo {author} {\bibfnamefont {N.}~\bibnamefont {Maskara}}, \bibinfo {author} {\bibfnamefont {I.}~\bibnamefont {Cong}}, \bibinfo {author} {\bibfnamefont {X.}~\bibnamefont {Gao}}, \bibinfo {author} {\bibfnamefont {P.~S.}\ \bibnamefont {Rodriguez}}, \bibinfo {author} {\bibfnamefont
  {T.}~\bibnamefont {Karolyshyn}}, \bibinfo {author} {\bibfnamefont {G.}~\bibnamefont {Semeghini}}, \bibinfo {author} {\bibfnamefont {M.~J.}\ \bibnamefont {Gullans}}, \bibinfo {author} {\bibfnamefont {M.}~\bibnamefont {Greiner}}, \bibinfo {author} {\bibfnamefont {V.}~\bibnamefont {Vuleti{\'{c}}}},\ and\ \bibinfo {author} {\bibfnamefont {M.~D.}\ \bibnamefont {Lukin}},\ }\bibfield  {title} {\bibinfo {title} {{Logical quantum processor based on reconfigurable atom arrays}},\ }\bibfield  {journal} {\bibinfo  {journal} {Nature}\ }\href {https://doi.org/10.1038/s41586-023-06927-3} {10.1038/s41586-023-06927-3} (\bibinfo {year} {2023})\BibitemShut {NoStop}%
\bibitem [{\citenamefont {Tang}\ \emph {et~al.}(2018)\citenamefont {Tang}, \citenamefont {Yu},\ and\ \citenamefont {Dong}}]{Tang2018}%
  \BibitemOpen
  \bibfield  {author} {\bibinfo {author} {\bibfnamefont {Z.-M.}\ \bibnamefont {Tang}}, \bibinfo {author} {\bibfnamefont {Y.-M.}\ \bibnamefont {Yu}},\ and\ \bibinfo {author} {\bibfnamefont {C.-Z.}\ \bibnamefont {Dong}},\ }\bibfield  {title} {\bibinfo {title} {{Determination of static dipole polarizabilities of Yb atom}},\ }\href {https://doi.org/10.1088/1674-1056/27/6/063101} {\bibfield  {journal} {\bibinfo  {journal} {Chinese Phys. B}\ }\textbf {\bibinfo {volume} {27}},\ \bibinfo {pages} {063101} (\bibinfo {year} {2018})}\BibitemShut {NoStop}%
\bibitem [{\citenamefont {Endres}\ \emph {et~al.}(2016)\citenamefont {Endres}, \citenamefont {Bernien}, \citenamefont {Keesling}, \citenamefont {Levine}, \citenamefont {Anschuetz}, \citenamefont {Krajenbrink}, \citenamefont {Senko}, \citenamefont {Vuletic}, \citenamefont {Greiner},\ and\ \citenamefont {Lukin}}]{Endres2016}%
  \BibitemOpen
  \bibfield  {author} {\bibinfo {author} {\bibfnamefont {M.}~\bibnamefont {Endres}}, \bibinfo {author} {\bibfnamefont {H.}~\bibnamefont {Bernien}}, \bibinfo {author} {\bibfnamefont {A.}~\bibnamefont {Keesling}}, \bibinfo {author} {\bibfnamefont {H.}~\bibnamefont {Levine}}, \bibinfo {author} {\bibfnamefont {E.~R.}\ \bibnamefont {Anschuetz}}, \bibinfo {author} {\bibfnamefont {A.}~\bibnamefont {Krajenbrink}}, \bibinfo {author} {\bibfnamefont {C.}~\bibnamefont {Senko}}, \bibinfo {author} {\bibfnamefont {V.}~\bibnamefont {Vuletic}}, \bibinfo {author} {\bibfnamefont {M.}~\bibnamefont {Greiner}},\ and\ \bibinfo {author} {\bibfnamefont {M.~D.}\ \bibnamefont {Lukin}},\ }\bibfield  {title} {\bibinfo {title} {{Atom-by-atom assembly of defect-free one-dimensional cold atom arrays}},\ }\href {https://doi.org/10.1126/science.aah3752} {\bibfield  {journal} {\bibinfo  {journal} {Science}\ }\textbf {\bibinfo {volume} {354}},\ \bibinfo {pages} {1024} (\bibinfo {year} {2016})}\BibitemShut {NoStop}%
\bibitem [{\citenamefont {Norcia}\ \emph {et~al.}(2024)\citenamefont {Norcia}, \citenamefont {Kim}, \citenamefont {Cairncross}, \citenamefont {Stone}, \citenamefont {Ryou}, \citenamefont {Jaffe}, \citenamefont {Brown}, \citenamefont {Barnes}, \citenamefont {Battaglino}, \citenamefont {Bohdanowicz}, \citenamefont {Brown}, \citenamefont {Cassella}, \citenamefont {Chen}, \citenamefont {Coxe}, \citenamefont {Crow}, \citenamefont {Epstein}, \citenamefont {Griger}, \citenamefont {Halperin}, \citenamefont {Hummel}, \citenamefont {Jones}, \citenamefont {Kindem}, \citenamefont {King}, \citenamefont {Kotru}, \citenamefont {Lauigan}, \citenamefont {Li}, \citenamefont {Lu}, \citenamefont {Megidish}, \citenamefont {Marjanovic}, \citenamefont {McDonald}, \citenamefont {Mittiga}, \citenamefont {Muniz}, \citenamefont {Narayanaswami}, \citenamefont {Nishiguchi}, \citenamefont {Paule}, \citenamefont {Pawlak}, \citenamefont {Peng}, \citenamefont {Pudenz}, \citenamefont {Rodr\'{\i}guez~P\'erez}, \citenamefont {Smull},
  \citenamefont {Stack}, \citenamefont {Urbanek}, \citenamefont {van~de Veerdonk}, \citenamefont {Vendeiro}, \citenamefont {Wadleigh}, \citenamefont {Wilkason}, \citenamefont {Wu}, \citenamefont {Xie}, \citenamefont {Zalys-Geller}, \citenamefont {Zhang},\ and\ \citenamefont {Bloom}}]{Norcia2024}%
  \BibitemOpen
  \bibfield  {author} {\bibinfo {author} {\bibfnamefont {M.~A.}\ \bibnamefont {Norcia}}, \bibinfo {author} {\bibfnamefont {H.}~\bibnamefont {Kim}}, \bibinfo {author} {\bibfnamefont {W.~B.}\ \bibnamefont {Cairncross}}, \bibinfo {author} {\bibfnamefont {M.}~\bibnamefont {Stone}}, \bibinfo {author} {\bibfnamefont {A.}~\bibnamefont {Ryou}}, \bibinfo {author} {\bibfnamefont {M.}~\bibnamefont {Jaffe}}, \bibinfo {author} {\bibfnamefont {M.~O.}\ \bibnamefont {Brown}}, \bibinfo {author} {\bibfnamefont {K.}~\bibnamefont {Barnes}}, \bibinfo {author} {\bibfnamefont {P.}~\bibnamefont {Battaglino}}, \bibinfo {author} {\bibfnamefont {T.~C.}\ \bibnamefont {Bohdanowicz}}, \bibinfo {author} {\bibfnamefont {A.}~\bibnamefont {Brown}}, \bibinfo {author} {\bibfnamefont {K.}~\bibnamefont {Cassella}}, \bibinfo {author} {\bibfnamefont {C.-A.}\ \bibnamefont {Chen}}, \bibinfo {author} {\bibfnamefont {R.}~\bibnamefont {Coxe}}, \bibinfo {author} {\bibfnamefont {D.}~\bibnamefont {Crow}}, \bibinfo {author} {\bibfnamefont {J.}~\bibnamefont
  {Epstein}}, \bibinfo {author} {\bibfnamefont {C.}~\bibnamefont {Griger}}, \bibinfo {author} {\bibfnamefont {E.}~\bibnamefont {Halperin}}, \bibinfo {author} {\bibfnamefont {F.}~\bibnamefont {Hummel}}, \bibinfo {author} {\bibfnamefont {A.~M.~W.}\ \bibnamefont {Jones}}, \bibinfo {author} {\bibfnamefont {J.~M.}\ \bibnamefont {Kindem}}, \bibinfo {author} {\bibfnamefont {J.}~\bibnamefont {King}}, \bibinfo {author} {\bibfnamefont {K.}~\bibnamefont {Kotru}}, \bibinfo {author} {\bibfnamefont {J.}~\bibnamefont {Lauigan}}, \bibinfo {author} {\bibfnamefont {M.}~\bibnamefont {Li}}, \bibinfo {author} {\bibfnamefont {M.}~\bibnamefont {Lu}}, \bibinfo {author} {\bibfnamefont {E.}~\bibnamefont {Megidish}}, \bibinfo {author} {\bibfnamefont {J.}~\bibnamefont {Marjanovic}}, \bibinfo {author} {\bibfnamefont {M.}~\bibnamefont {McDonald}}, \bibinfo {author} {\bibfnamefont {T.}~\bibnamefont {Mittiga}}, \bibinfo {author} {\bibfnamefont {J.~A.}\ \bibnamefont {Muniz}}, \bibinfo {author} {\bibfnamefont {S.}~\bibnamefont
  {Narayanaswami}}, \bibinfo {author} {\bibfnamefont {C.}~\bibnamefont {Nishiguchi}}, \bibinfo {author} {\bibfnamefont {T.}~\bibnamefont {Paule}}, \bibinfo {author} {\bibfnamefont {K.~A.}\ \bibnamefont {Pawlak}}, \bibinfo {author} {\bibfnamefont {L.~S.}\ \bibnamefont {Peng}}, \bibinfo {author} {\bibfnamefont {K.~L.}\ \bibnamefont {Pudenz}}, \bibinfo {author} {\bibfnamefont {D.}~\bibnamefont {Rodr\'{\i}guez~P\'erez}}, \bibinfo {author} {\bibfnamefont {A.}~\bibnamefont {Smull}}, \bibinfo {author} {\bibfnamefont {D.}~\bibnamefont {Stack}}, \bibinfo {author} {\bibfnamefont {M.}~\bibnamefont {Urbanek}}, \bibinfo {author} {\bibfnamefont {R.~J.~M.}\ \bibnamefont {van~de Veerdonk}}, \bibinfo {author} {\bibfnamefont {Z.}~\bibnamefont {Vendeiro}}, \bibinfo {author} {\bibfnamefont {L.}~\bibnamefont {Wadleigh}}, \bibinfo {author} {\bibfnamefont {T.}~\bibnamefont {Wilkason}}, \bibinfo {author} {\bibfnamefont {T.-Y.}\ \bibnamefont {Wu}}, \bibinfo {author} {\bibfnamefont {X.}~\bibnamefont {Xie}}, \bibinfo {author}
  {\bibfnamefont {E.}~\bibnamefont {Zalys-Geller}}, \bibinfo {author} {\bibfnamefont {X.}~\bibnamefont {Zhang}},\ and\ \bibinfo {author} {\bibfnamefont {B.~J.}\ \bibnamefont {Bloom}},\ }\bibfield  {title} {\bibinfo {title} {Iterative assembly of ${}^{171}$$\mathrm{Yb}$ atom arrays with cavity-enhanced optical lattices},\ }\href {https://doi.org/10.1103/PRXQuantum.5.030316} {\bibfield  {journal} {\bibinfo  {journal} {PRX Quantum}\ }\textbf {\bibinfo {volume} {5}},\ \bibinfo {pages} {030316} (\bibinfo {year} {2024})}\BibitemShut {NoStop}%
\bibitem [{\citenamefont {Gyger}\ \emph {et~al.}(2024)\citenamefont {Gyger}, \citenamefont {Ammenwerth}, \citenamefont {Tao}, \citenamefont {Timme}, \citenamefont {Snigirev}, \citenamefont {Bloch},\ and\ \citenamefont {Zeiher}}]{Gyger2024}%
  \BibitemOpen
  \bibfield  {author} {\bibinfo {author} {\bibfnamefont {F.}~\bibnamefont {Gyger}}, \bibinfo {author} {\bibfnamefont {M.}~\bibnamefont {Ammenwerth}}, \bibinfo {author} {\bibfnamefont {R.}~\bibnamefont {Tao}}, \bibinfo {author} {\bibfnamefont {H.}~\bibnamefont {Timme}}, \bibinfo {author} {\bibfnamefont {S.}~\bibnamefont {Snigirev}}, \bibinfo {author} {\bibfnamefont {I.}~\bibnamefont {Bloch}},\ and\ \bibinfo {author} {\bibfnamefont {J.}~\bibnamefont {Zeiher}},\ }\bibfield  {title} {\bibinfo {title} {{Continuous operation of large-scale atom arrays in optical lattices}},\ }\href {https://doi.org/10.1103/PhysRevResearch.6.033104} {\bibfield  {journal} {\bibinfo  {journal} {Phys. Rev. Res.}\ }\textbf {\bibinfo {volume} {6}},\ \bibinfo {pages} {033104} (\bibinfo {year} {2024})}\BibitemShut {NoStop}%
\bibitem [{\citenamefont {Periwal}\ \emph {et~al.}(2021)\citenamefont {Periwal}, \citenamefont {Cooper}, \citenamefont {Kunkel}, \citenamefont {Wienand}, \citenamefont {Davis},\ and\ \citenamefont {Schleier-Smith}}]{Periwal2021}%
  \BibitemOpen
  \bibfield  {author} {\bibinfo {author} {\bibfnamefont {A.}~\bibnamefont {Periwal}}, \bibinfo {author} {\bibfnamefont {E.~S.}\ \bibnamefont {Cooper}}, \bibinfo {author} {\bibfnamefont {P.}~\bibnamefont {Kunkel}}, \bibinfo {author} {\bibfnamefont {J.~F.}\ \bibnamefont {Wienand}}, \bibinfo {author} {\bibfnamefont {E.~J.}\ \bibnamefont {Davis}},\ and\ \bibinfo {author} {\bibfnamefont {M.}~\bibnamefont {Schleier-Smith}},\ }\bibfield  {title} {\bibinfo {title} {{Programmable interactions and emergent geometry in an array of atom clouds}},\ }\href {https://doi.org/10.1038/s41586-021-04156-0} {\bibfield  {journal} {\bibinfo  {journal} {Nature}\ }\textbf {\bibinfo {volume} {600}},\ \bibinfo {pages} {630} (\bibinfo {year} {2021})}\BibitemShut {NoStop}%
\bibitem [{\citenamefont {Peters}\ \emph {et~al.}(2024)\citenamefont {Peters}, \citenamefont {Wang}, \citenamefont {Spierings}, \citenamefont {Drucker}, \citenamefont {Hu}, \citenamefont {Chen},\ and\ \citenamefont {Vuleti{\'{c}}}}]{Peters2024}%
  \BibitemOpen
  \bibfield  {author} {\bibinfo {author} {\bibfnamefont {M.~L.}\ \bibnamefont {Peters}}, \bibinfo {author} {\bibfnamefont {G.}~\bibnamefont {Wang}}, \bibinfo {author} {\bibfnamefont {D.~C.}\ \bibnamefont {Spierings}}, \bibinfo {author} {\bibfnamefont {N.}~\bibnamefont {Drucker}}, \bibinfo {author} {\bibfnamefont {B.}~\bibnamefont {Hu}}, \bibinfo {author} {\bibfnamefont {Y.-T.}\ \bibnamefont {Chen}},\ and\ \bibinfo {author} {\bibfnamefont {V.}~\bibnamefont {Vuleti{\'{c}}}},\ }\bibfield  {title} {\bibinfo {title} {{Cavity-enabled real-time observation of individual atomic collisions}},\ }\href {http://arxiv.org/abs/2411.12622} {\bibfield  {journal} {\bibinfo  {journal} {arXiv Prepr.}\ }\textbf {\bibinfo {volume} {2411.12622}} (\bibinfo {year} {2024})}\BibitemShut {NoStop}%
\bibitem [{\citenamefont {Grinkemeyer}\ \emph {et~al.}(2024)\citenamefont {Grinkemeyer}, \citenamefont {Guardado-Sanchez}, \citenamefont {Dimitrova}, \citenamefont {Shchepanovich}, \citenamefont {Mandopoulou}, \citenamefont {Borregaard}, \citenamefont {Vuleti{\'{c}}},\ and\ \citenamefont {Lukin}}]{Grinkemeyer2024}%
  \BibitemOpen
  \bibfield  {author} {\bibinfo {author} {\bibfnamefont {B.}~\bibnamefont {Grinkemeyer}}, \bibinfo {author} {\bibfnamefont {E.}~\bibnamefont {Guardado-Sanchez}}, \bibinfo {author} {\bibfnamefont {I.}~\bibnamefont {Dimitrova}}, \bibinfo {author} {\bibfnamefont {D.}~\bibnamefont {Shchepanovich}}, \bibinfo {author} {\bibfnamefont {G.~E.}\ \bibnamefont {Mandopoulou}}, \bibinfo {author} {\bibfnamefont {J.}~\bibnamefont {Borregaard}}, \bibinfo {author} {\bibfnamefont {V.}~\bibnamefont {Vuleti{\'{c}}}},\ and\ \bibinfo {author} {\bibfnamefont {M.~D.}\ \bibnamefont {Lukin}},\ }\bibfield  {title} {\bibinfo {title} {{Error-Detected Quantum Operations with Neutral Atoms Mediated by an Optical Cavity}},\ }\href {http://arxiv.org/abs/2410.10787} {\bibfield  {journal} {\bibinfo  {journal} {arXiv Prepr.}\ }\textbf {\bibinfo {volume} {2410.10787}} (\bibinfo {year} {2024})}\BibitemShut {NoStop}%
\bibitem [{\citenamefont {Trupke}\ \emph {et~al.}(2007)\citenamefont {Trupke}, \citenamefont {Goldwin}, \citenamefont {Darqui{\'{e}}}, \citenamefont {Dutier}, \citenamefont {Eriksson}, \citenamefont {Ashmore},\ and\ \citenamefont {Hinds}}]{Trupke2007}%
  \BibitemOpen
  \bibfield  {author} {\bibinfo {author} {\bibfnamefont {M.}~\bibnamefont {Trupke}}, \bibinfo {author} {\bibfnamefont {J.}~\bibnamefont {Goldwin}}, \bibinfo {author} {\bibfnamefont {B.}~\bibnamefont {Darqui{\'{e}}}}, \bibinfo {author} {\bibfnamefont {G.}~\bibnamefont {Dutier}}, \bibinfo {author} {\bibfnamefont {S.}~\bibnamefont {Eriksson}}, \bibinfo {author} {\bibfnamefont {J.}~\bibnamefont {Ashmore}},\ and\ \bibinfo {author} {\bibfnamefont {E.~A.}\ \bibnamefont {Hinds}},\ }\bibfield  {title} {\bibinfo {title} {{Atom Detection and Photon Production in a Scalable, Open, Optical Microcavity}},\ }\href {https://doi.org/10.1103/PhysRevLett.99.063601} {\bibfield  {journal} {\bibinfo  {journal} {Phys. Rev. Lett.}\ }\textbf {\bibinfo {volume} {99}},\ \bibinfo {pages} {063601} (\bibinfo {year} {2007})}\BibitemShut {NoStop}%
\bibitem [{\citenamefont {Derntl}\ \emph {et~al.}(2014)\citenamefont {Derntl}, \citenamefont {Schneider}, \citenamefont {Schalko}, \citenamefont {Bittner}, \citenamefont {Schmiedmayer}, \citenamefont {Schmid},\ and\ \citenamefont {Trupke}}]{Derntl2014}%
  \BibitemOpen
  \bibfield  {author} {\bibinfo {author} {\bibfnamefont {C.}~\bibnamefont {Derntl}}, \bibinfo {author} {\bibfnamefont {M.}~\bibnamefont {Schneider}}, \bibinfo {author} {\bibfnamefont {J.}~\bibnamefont {Schalko}}, \bibinfo {author} {\bibfnamefont {A.}~\bibnamefont {Bittner}}, \bibinfo {author} {\bibfnamefont {J.}~\bibnamefont {Schmiedmayer}}, \bibinfo {author} {\bibfnamefont {U.}~\bibnamefont {Schmid}},\ and\ \bibinfo {author} {\bibfnamefont {M.}~\bibnamefont {Trupke}},\ }\bibfield  {title} {\bibinfo {title} {{Arrays of open, independently tunable microcavities}},\ }\href {https://doi.org/10.1364/OE.22.022111} {\bibfield  {journal} {\bibinfo  {journal} {Opt. Express}\ }\textbf {\bibinfo {volume} {22}},\ \bibinfo {pages} {22111} (\bibinfo {year} {2014})}\BibitemShut {NoStop}%
\bibitem [{\citenamefont {Menon}\ \emph {et~al.}(2024)\citenamefont {Menon}, \citenamefont {Glachman}, \citenamefont {Pompili}, \citenamefont {Dibos},\ and\ \citenamefont {Bernien}}]{Menon2024}%
  \BibitemOpen
  \bibfield  {author} {\bibinfo {author} {\bibfnamefont {S.~G.}\ \bibnamefont {Menon}}, \bibinfo {author} {\bibfnamefont {N.}~\bibnamefont {Glachman}}, \bibinfo {author} {\bibfnamefont {M.}~\bibnamefont {Pompili}}, \bibinfo {author} {\bibfnamefont {A.}~\bibnamefont {Dibos}},\ and\ \bibinfo {author} {\bibfnamefont {H.}~\bibnamefont {Bernien}},\ }\bibfield  {title} {\bibinfo {title} {{An integrated atom array-nanophotonic chip platform with background-free imaging}},\ }\href {https://doi.org/10.1038/s41467-024-50355-4} {\bibfield  {journal} {\bibinfo  {journal} {Nat. Commun.}\ }\textbf {\bibinfo {volume} {15}},\ \bibinfo {pages} {6156} (\bibinfo {year} {2024})}\BibitemShut {NoStop}%
\bibitem [{\citenamefont {Shadmany}\ \emph {et~al.}(2024)\citenamefont {Shadmany}, \citenamefont {Kumar}, \citenamefont {Soper}, \citenamefont {Palm}, \citenamefont {Yin}, \citenamefont {Ando}, \citenamefont {Li}, \citenamefont {Taneja}, \citenamefont {Jaffe}, \citenamefont {Schuster},\ and\ \citenamefont {Simon}}]{Shadmany2024}%
  \BibitemOpen
  \bibfield  {author} {\bibinfo {author} {\bibfnamefont {D.}~\bibnamefont {Shadmany}}, \bibinfo {author} {\bibfnamefont {A.}~\bibnamefont {Kumar}}, \bibinfo {author} {\bibfnamefont {A.}~\bibnamefont {Soper}}, \bibinfo {author} {\bibfnamefont {L.}~\bibnamefont {Palm}}, \bibinfo {author} {\bibfnamefont {C.}~\bibnamefont {Yin}}, \bibinfo {author} {\bibfnamefont {H.}~\bibnamefont {Ando}}, \bibinfo {author} {\bibfnamefont {B.}~\bibnamefont {Li}}, \bibinfo {author} {\bibfnamefont {L.}~\bibnamefont {Taneja}}, \bibinfo {author} {\bibfnamefont {M.}~\bibnamefont {Jaffe}}, \bibinfo {author} {\bibfnamefont {D.}~\bibnamefont {Schuster}},\ and\ \bibinfo {author} {\bibfnamefont {J.}~\bibnamefont {Simon}},\ }\bibfield  {title} {\bibinfo {title} {{Cavity QED in a High NA Resonator}},\ }\href {http://arxiv.org/abs/2407.04784} {\bibfield  {journal} {\bibinfo  {journal} {arXiv Prepr.}\ }\textbf {\bibinfo {volume} {2407.04784}} (\bibinfo {year} {2024})}\BibitemShut {NoStop}%
\bibitem [{\citenamefont {Graham}\ \emph {et~al.}(2022)\citenamefont {Graham}, \citenamefont {Song}, \citenamefont {Scott}, \citenamefont {Poole}, \citenamefont {Phuttitarn}, \citenamefont {Jooya}, \citenamefont {Eichler}, \citenamefont {Jiang}, \citenamefont {Marra}, \citenamefont {Grinkemeyer}, \citenamefont {Kwon}, \citenamefont {Ebert}, \citenamefont {Cherek}, \citenamefont {Lichtman}, \citenamefont {Gillette}, \citenamefont {Gilbert}, \citenamefont {Bowman}, \citenamefont {Ballance}, \citenamefont {Campbell}, \citenamefont {Dahl}, \citenamefont {Crawford}, \citenamefont {Blunt}, \citenamefont {Rogers}, \citenamefont {Noel},\ and\ \citenamefont {Saffman}}]{Graham2022}%
  \BibitemOpen
  \bibfield  {author} {\bibinfo {author} {\bibfnamefont {T.~M.}\ \bibnamefont {Graham}}, \bibinfo {author} {\bibfnamefont {Y.}~\bibnamefont {Song}}, \bibinfo {author} {\bibfnamefont {J.}~\bibnamefont {Scott}}, \bibinfo {author} {\bibfnamefont {C.}~\bibnamefont {Poole}}, \bibinfo {author} {\bibfnamefont {L.}~\bibnamefont {Phuttitarn}}, \bibinfo {author} {\bibfnamefont {K.}~\bibnamefont {Jooya}}, \bibinfo {author} {\bibfnamefont {P.}~\bibnamefont {Eichler}}, \bibinfo {author} {\bibfnamefont {X.}~\bibnamefont {Jiang}}, \bibinfo {author} {\bibfnamefont {A.}~\bibnamefont {Marra}}, \bibinfo {author} {\bibfnamefont {B.}~\bibnamefont {Grinkemeyer}}, \bibinfo {author} {\bibfnamefont {M.}~\bibnamefont {Kwon}}, \bibinfo {author} {\bibfnamefont {M.}~\bibnamefont {Ebert}}, \bibinfo {author} {\bibfnamefont {J.}~\bibnamefont {Cherek}}, \bibinfo {author} {\bibfnamefont {M.~T.}\ \bibnamefont {Lichtman}}, \bibinfo {author} {\bibfnamefont {M.}~\bibnamefont {Gillette}}, \bibinfo {author} {\bibfnamefont {J.}~\bibnamefont
  {Gilbert}}, \bibinfo {author} {\bibfnamefont {D.}~\bibnamefont {Bowman}}, \bibinfo {author} {\bibfnamefont {T.}~\bibnamefont {Ballance}}, \bibinfo {author} {\bibfnamefont {C.}~\bibnamefont {Campbell}}, \bibinfo {author} {\bibfnamefont {E.~D.}\ \bibnamefont {Dahl}}, \bibinfo {author} {\bibfnamefont {O.}~\bibnamefont {Crawford}}, \bibinfo {author} {\bibfnamefont {N.~S.}\ \bibnamefont {Blunt}}, \bibinfo {author} {\bibfnamefont {B.}~\bibnamefont {Rogers}}, \bibinfo {author} {\bibfnamefont {T.}~\bibnamefont {Noel}},\ and\ \bibinfo {author} {\bibfnamefont {M.}~\bibnamefont {Saffman}},\ }\bibfield  {title} {\bibinfo {title} {{Multi-qubit entanglement and algorithms on a neutral-atom quantum computer}},\ }\href {https://doi.org/10.1038/s41586-022-04603-6} {\bibfield  {journal} {\bibinfo  {journal} {Nature}\ }\textbf {\bibinfo {volume} {604}},\ \bibinfo {pages} {457} (\bibinfo {year} {2022})}\BibitemShut {NoStop}%
\bibitem [{\citenamefont {Pfister}\ \emph {et~al.}(2016)\citenamefont {Pfister}, \citenamefont {Kaniewski}, \citenamefont {Tomamichel}, \citenamefont {Mantri}, \citenamefont {Schmucker}, \citenamefont {McMahon}, \citenamefont {Milburn},\ and\ \citenamefont {Wehner}}]{Pfister2016}%
  \BibitemOpen
  \bibfield  {author} {\bibinfo {author} {\bibfnamefont {C.}~\bibnamefont {Pfister}}, \bibinfo {author} {\bibfnamefont {J.}~\bibnamefont {Kaniewski}}, \bibinfo {author} {\bibfnamefont {M.}~\bibnamefont {Tomamichel}}, \bibinfo {author} {\bibfnamefont {A.}~\bibnamefont {Mantri}}, \bibinfo {author} {\bibfnamefont {R.}~\bibnamefont {Schmucker}}, \bibinfo {author} {\bibfnamefont {N.}~\bibnamefont {McMahon}}, \bibinfo {author} {\bibfnamefont {G.}~\bibnamefont {Milburn}},\ and\ \bibinfo {author} {\bibfnamefont {S.}~\bibnamefont {Wehner}},\ }\bibfield  {title} {\bibinfo {title} {{A universal test for gravitational decoherence}},\ }\href {https://doi.org/10.1038/ncomms13022} {\bibfield  {journal} {\bibinfo  {journal} {Nat. Commun.}\ }\textbf {\bibinfo {volume} {7}},\ \bibinfo {pages} {13022} (\bibinfo {year} {2016})}\BibitemShut {NoStop}%
\bibitem [{\citenamefont {Borregaard}\ and\ \citenamefont {Pikovski}(2024)}]{Borregaard2024}%
  \BibitemOpen
  \bibfield  {author} {\bibinfo {author} {\bibfnamefont {J.}~\bibnamefont {Borregaard}}\ and\ \bibinfo {author} {\bibfnamefont {I.}~\bibnamefont {Pikovski}},\ }\bibfield  {title} {\bibinfo {title} {{Testing quantum theory on curved space-time with quantum networks}},\ }\href {http://arxiv.org/abs/2406.19533} {\bibfield  {journal} {\bibinfo  {journal} {arXiv Prepr.}\ }\textbf {\bibinfo {volume} {2406.19533}} (\bibinfo {year} {2024})}\BibitemShut {NoStop}%
\bibitem [{\citenamefont {Covey}\ \emph {et~al.}(2025)\citenamefont {Covey}, \citenamefont {Pikovski},\ and\ \citenamefont {Borregaard}}]{Covey2025}%
  \BibitemOpen
  \bibfield  {author} {\bibinfo {author} {\bibfnamefont {J.~P.}\ \bibnamefont {Covey}}, \bibinfo {author} {\bibfnamefont {I.}~\bibnamefont {Pikovski}},\ and\ \bibinfo {author} {\bibfnamefont {J.}~\bibnamefont {Borregaard}},\ }\bibfield  {title} {\bibinfo {title} {{Probing curved spacetime with a distributed atomic processor clock}},\ }\href {http://arxiv.org/abs/2502.12954} {\bibfield  {journal} {\bibinfo  {journal} {arXiv Prepr.}\ }\textbf {\bibinfo {volume} {2502.12954}} (\bibinfo {year} {2025})}\BibitemShut {NoStop}%
\bibitem [{\citenamefont {Wcis{\l}o}\ \emph {et~al.}(2018)\citenamefont {Wcis{\l}o}, \citenamefont {Ablewski}, \citenamefont {Beloy}, \citenamefont {Bilicki}, \citenamefont {Bober}, \citenamefont {Brown}, \citenamefont {Fasano}, \citenamefont {Ciury{\l}o}, \citenamefont {Hachisu}, \citenamefont {Ido}, \citenamefont {Lodewyck}, \citenamefont {Ludlow}, \citenamefont {McGrew}, \citenamefont {Morzy{\'{n}}ski}, \citenamefont {Nicolodi}, \citenamefont {Schioppo}, \citenamefont {Sekido}, \citenamefont {{Le Targat}}, \citenamefont {Wolf}, \citenamefont {Zhang}, \citenamefont {Zjawin},\ and\ \citenamefont {Zawada}}]{Wcislo2018}%
  \BibitemOpen
  \bibfield  {author} {\bibinfo {author} {\bibfnamefont {P.}~\bibnamefont {Wcis{\l}o}}, \bibinfo {author} {\bibfnamefont {P.}~\bibnamefont {Ablewski}}, \bibinfo {author} {\bibfnamefont {K.}~\bibnamefont {Beloy}}, \bibinfo {author} {\bibfnamefont {S.}~\bibnamefont {Bilicki}}, \bibinfo {author} {\bibfnamefont {M.}~\bibnamefont {Bober}}, \bibinfo {author} {\bibfnamefont {R.}~\bibnamefont {Brown}}, \bibinfo {author} {\bibfnamefont {R.}~\bibnamefont {Fasano}}, \bibinfo {author} {\bibfnamefont {R.}~\bibnamefont {Ciury{\l}o}}, \bibinfo {author} {\bibfnamefont {H.}~\bibnamefont {Hachisu}}, \bibinfo {author} {\bibfnamefont {T.}~\bibnamefont {Ido}}, \bibinfo {author} {\bibfnamefont {J.}~\bibnamefont {Lodewyck}}, \bibinfo {author} {\bibfnamefont {A.}~\bibnamefont {Ludlow}}, \bibinfo {author} {\bibfnamefont {W.}~\bibnamefont {McGrew}}, \bibinfo {author} {\bibfnamefont {P.}~\bibnamefont {Morzy{\'{n}}ski}}, \bibinfo {author} {\bibfnamefont {D.}~\bibnamefont {Nicolodi}}, \bibinfo {author} {\bibfnamefont {M.}~\bibnamefont
  {Schioppo}}, \bibinfo {author} {\bibfnamefont {M.}~\bibnamefont {Sekido}}, \bibinfo {author} {\bibfnamefont {R.}~\bibnamefont {{Le Targat}}}, \bibinfo {author} {\bibfnamefont {P.}~\bibnamefont {Wolf}}, \bibinfo {author} {\bibfnamefont {X.}~\bibnamefont {Zhang}}, \bibinfo {author} {\bibfnamefont {B.}~\bibnamefont {Zjawin}},\ and\ \bibinfo {author} {\bibfnamefont {M.}~\bibnamefont {Zawada}},\ }\bibfield  {title} {\bibinfo {title} {{New bounds on dark matter coupling from a global network of optical atomic clocks}},\ }\bibfield  {journal} {\bibinfo  {journal} {Sci. Adv.}\ }\textbf {\bibinfo {volume} {4}},\ \href {https://doi.org/10.1126/sciadv.aau4869} {10.1126/sciadv.aau4869} (\bibinfo {year} {2018})\BibitemShut {NoStop}%
\bibitem [{\citenamefont {Kennedy}\ \emph {et~al.}(2020)\citenamefont {Kennedy}, \citenamefont {Oelker}, \citenamefont {Robinson}, \citenamefont {Bothwell}, \citenamefont {Kedar}, \citenamefont {Milner}, \citenamefont {Marti}, \citenamefont {Derevianko},\ and\ \citenamefont {Ye}}]{Kennedy2020}%
  \BibitemOpen
  \bibfield  {author} {\bibinfo {author} {\bibfnamefont {C.~J.}\ \bibnamefont {Kennedy}}, \bibinfo {author} {\bibfnamefont {E.}~\bibnamefont {Oelker}}, \bibinfo {author} {\bibfnamefont {J.~M.}\ \bibnamefont {Robinson}}, \bibinfo {author} {\bibfnamefont {T.}~\bibnamefont {Bothwell}}, \bibinfo {author} {\bibfnamefont {D.}~\bibnamefont {Kedar}}, \bibinfo {author} {\bibfnamefont {W.~R.}\ \bibnamefont {Milner}}, \bibinfo {author} {\bibfnamefont {G.~E.}\ \bibnamefont {Marti}}, \bibinfo {author} {\bibfnamefont {A.}~\bibnamefont {Derevianko}},\ and\ \bibinfo {author} {\bibfnamefont {J.}~\bibnamefont {Ye}},\ }\bibfield  {title} {\bibinfo {title} {{Precision Metrology Meets Cosmology: Improved Constraints on Ultralight Dark Matter from Atom-Cavity Frequency Comparisons}},\ }\href {https://doi.org/10.1103/PhysRevLett.125.201302} {\bibfield  {journal} {\bibinfo  {journal} {Phys. Rev. Lett.}\ }\textbf {\bibinfo {volume} {125}},\ \bibinfo {pages} {201302} (\bibinfo {year} {2020})}\BibitemShut {NoStop}%
\bibitem [{\citenamefont {Cho}\ \emph {et~al.}(2012)\citenamefont {Cho}, \citenamefont {Lee}, \citenamefont {Lee}, \citenamefont {Ahn}, \citenamefont {Lee}, \citenamefont {Yu}, \citenamefont {Lee},\ and\ \citenamefont {Park}}]{Cho2012}%
  \BibitemOpen
  \bibfield  {author} {\bibinfo {author} {\bibfnamefont {J.~W.}\ \bibnamefont {Cho}}, \bibinfo {author} {\bibfnamefont {H.-g.}\ \bibnamefont {Lee}}, \bibinfo {author} {\bibfnamefont {S.}~\bibnamefont {Lee}}, \bibinfo {author} {\bibfnamefont {J.}~\bibnamefont {Ahn}}, \bibinfo {author} {\bibfnamefont {W.-K.}\ \bibnamefont {Lee}}, \bibinfo {author} {\bibfnamefont {D.-H.}\ \bibnamefont {Yu}}, \bibinfo {author} {\bibfnamefont {S.~K.}\ \bibnamefont {Lee}},\ and\ \bibinfo {author} {\bibfnamefont {C.~Y.}\ \bibnamefont {Park}},\ }\bibfield  {title} {\bibinfo {title} {{Optical repumping of triplet-P states enhances magneto-optical trapping of ytterbium atoms}},\ }\href {https://doi.org/10.1103/PhysRevA.85.035401} {\bibfield  {journal} {\bibinfo  {journal} {Phys. Rev. A}\ }\textbf {\bibinfo {volume} {85}},\ \bibinfo {pages} {035401} (\bibinfo {year} {2012})}\BibitemShut {NoStop}%
\bibitem [{\citenamefont {Porsev}\ \emph {et~al.}(1999)\citenamefont {Porsev}, \citenamefont {Rakhlina},\ and\ \citenamefont {Kozlov}}]{Porsev1999}%
  \BibitemOpen
  \bibfield  {author} {\bibinfo {author} {\bibfnamefont {S.~G.}\ \bibnamefont {Porsev}}, \bibinfo {author} {\bibfnamefont {Y.~G.}\ \bibnamefont {Rakhlina}},\ and\ \bibinfo {author} {\bibfnamefont {M.~G.}\ \bibnamefont {Kozlov}},\ }\bibfield  {title} {\bibinfo {title} {{Electric-dipole amplitudes, lifetimes, and polarizabilities of the low-lying levels of atomic ytterbium}},\ }\href {https://doi.org/10.1103/PhysRevA.60.2781} {\bibfield  {journal} {\bibinfo  {journal} {Phys. Rev. A}\ }\textbf {\bibinfo {volume} {60}},\ \bibinfo {pages} {2781} (\bibinfo {year} {1999})}\BibitemShut {NoStop}%
\bibitem [{\citenamefont {Scazza}(2015)}]{Scazza2015}%
  \BibitemOpen
  \bibfield  {author} {\bibinfo {author} {\bibfnamefont {F.}~\bibnamefont {Scazza}},\ }\emph {\bibinfo {title} {Probing SU(N)-symmetric orbital interactions with ytterbium Fermi gases in optical lattices}},\ \href {http://nbn-resolving.de/urn:nbn:de:bvb:19-181593} {Ph.D. thesis} (\bibinfo {year} {2015})\BibitemShut {NoStop}%
\bibitem [{\citenamefont {Bettermann}(2022)}]{Bettermann2022}%
  \BibitemOpen
  \bibfield  {author} {\bibinfo {author} {\bibfnamefont {O.}~\bibnamefont {Bettermann}},\ }\emph {\bibinfo {title} {Interorbital interactions in ytterbium-171}},\ \href {http://nbn-resolving.de/urn:nbn:de:bvb:19-321710} {Ph.D. thesis} (\bibinfo {year} {2022})\BibitemShut {NoStop}%
\bibitem [{\citenamefont {Li}(2024)}]{FFcircuit}%
  \BibitemOpen
  \bibfield  {author} {\bibinfo {author} {\bibfnamefont {L.}~\bibnamefont {Li}},\ }\href@noop {} {\bibinfo {title} {Laser phase noise canceling (feedforward) circuit for pdh locking}},\ \bibinfo {howpublished} {\url{https://github.com/leaningktower/Phase-Noise-Feedforward}} (\bibinfo {year} {2024})\BibitemShut {NoStop}%
\bibitem [{\citenamefont {Ma}\ \emph {et~al.}(1994)\citenamefont {Ma}, \citenamefont {Jungner}, \citenamefont {Ye},\ and\ \citenamefont {Hall}}]{Ma1994}%
  \BibitemOpen
  \bibfield  {author} {\bibinfo {author} {\bibfnamefont {L.-S.}\ \bibnamefont {Ma}}, \bibinfo {author} {\bibfnamefont {P.}~\bibnamefont {Jungner}}, \bibinfo {author} {\bibfnamefont {J.}~\bibnamefont {Ye}},\ and\ \bibinfo {author} {\bibfnamefont {J.~L.}\ \bibnamefont {Hall}},\ }\bibfield  {title} {\bibinfo {title} {{Delivering the same optical frequency at two places: accurate cancellation of phase noise introduced by an optical fiber or other time-varying path}},\ }\href {https://doi.org/10.1364/OL.19.001777} {\bibfield  {journal} {\bibinfo  {journal} {Opt. Lett.}\ }\textbf {\bibinfo {volume} {19}},\ \bibinfo {pages} {1777} (\bibinfo {year} {1994})}\BibitemShut {NoStop}%
\bibitem [{\citenamefont {Young}\ \emph {et~al.}(2022)\citenamefont {Young}, \citenamefont {Safari}, \citenamefont {Huft}, \citenamefont {Zhang}, \citenamefont {Oh}, \citenamefont {Chinnarasu},\ and\ \citenamefont {Saffman}}]{Young2022}%
  \BibitemOpen
  \bibfield  {author} {\bibinfo {author} {\bibfnamefont {C.~B.}\ \bibnamefont {Young}}, \bibinfo {author} {\bibfnamefont {A.}~\bibnamefont {Safari}}, \bibinfo {author} {\bibfnamefont {P.}~\bibnamefont {Huft}}, \bibinfo {author} {\bibfnamefont {J.}~\bibnamefont {Zhang}}, \bibinfo {author} {\bibfnamefont {E.}~\bibnamefont {Oh}}, \bibinfo {author} {\bibfnamefont {R.}~\bibnamefont {Chinnarasu}},\ and\ \bibinfo {author} {\bibfnamefont {M.}~\bibnamefont {Saffman}},\ }\bibfield  {title} {\bibinfo {title} {{An architecture for quantum networking of neutral atom processors}},\ }\href {http://arxiv.org/abs/2202.01634} {\bibfield  {journal} {\bibinfo  {journal} {arXiv Prepr.}\ }\textbf {\bibinfo {volume} {arXiv:2202}} (\bibinfo {year} {2022})}\BibitemShut {NoStop}%
\end{thebibliography}%

\end{document}